\documentclass[12pt]{iopart}

\usepackage[T1]{fontenc} 

\usepackage{array}
\usepackage{graphicx}
\usepackage{dcolumn}
\usepackage{bm}
\usepackage[numbers]{natbib}
\usepackage{slashed}
\usepackage{contour}
\usepackage{lmodern}
\usepackage{ulem}
\usepackage{amsmath}
\usepackage{mathrsfs}
\usepackage{multirow}
\usepackage{xcolor}
\usepackage{booktabs}
\usepackage{graphicx}
\usepackage[caption=false]{subfig}
\usepackage{feynmf}
\usepackage{tcolorbox}
\usepackage{upgreek}
\usepackage{float}
\usepackage{placeins}
\usepackage[switch*]{lineno}
\usepackage{xspace}
\usepackage{enumitem}
\usepackage{xfrac}
\usepackage{doi}
\usepackage{pifont}
\usepackage{lineno}
\usepackage{nicematrix}
\usepackage[export]{adjustbox}
\usepackage{soul}

\newcommand{\dAUC}{$\Delta\mathrm{AUC}$\xspace}

\newcommand{\topdata}{\texttt{TopData}\xspace}
\newcommand{\jetnet}{\texttt{JetNet}\xspace}
\newcommand{\JetNet}{\texttt{JetNet}\xspace}
\newcommand{\jetclass}{\texttt{JetClass}\xspace}
\newcommand{\JetClass}{\texttt{JetClass}\xspace}
\newcommand{\skiptop}{\texttt{skiptop}\xspace}
\newcommand{\skiptwz}{\texttt{skiptwz}\xspace}
\newcommand{\skipwz}{\texttt{skipwz}\xspace}
\newcommand{\jetnetskiptop}{\texttt{JetNet-skiptop}\xspace}
\newcommand{\jetnetskiptwz}{\texttt{JetNet-skiptwz}\xspace}
\newcommand{\jetnetskipwz}{\texttt{JetNet-skipwz}\xspace}

\usepackage{iopams}  
\begin{document}

\pagebreak
\title[EDL for UQ and OOD Detection in Jet Identification using DNNs]{Evidential Deep Learning for Uncertainty Quantification and Out-of-Distribution Detection in Jet Identification using Deep Neural Networks}

\author{
Ayush Khot$^1$,
Xiwei Wang$^{2}$,
Avik Roy$^{3}$,
Volodymyr Kindratenko$^{2,3,4}$,
Mark S. Neubauer$^{1,2,3}$\footnote{corresponding author}}

\address{$^1$ Department of Physics, University of Illinois Urbana-Champaign, Urbana, IL 61801}
\address{$^2$ The Grainger College of Engineering, Department of Electrical and Computer Engineering, University of Illinois Urbana-Champaign, Urbana, IL 61801}
\address{$^3$ Center for Artificial Intelligence Innovation, National Center for Supercomputing Applications, University of Illinois Urbana-Champaign, Urbana, IL 61801}
\address{$^4$ The Grainger College of Engineering, Siebel School of Computing and Data Science, University of Illinois Urbana-Champaign, Urbana, IL 61801}
\eads{msn@illinois.edu}

\begin{abstract}

Current methods commonly used for uncertainty quantification (UQ) in deep learning (DL) models utilize Bayesian methods which are computationally expensive and time-consuming. In this paper, we provide a detailed study of UQ based on evidential deep learning (EDL) for deep neural network models designed to identify jets in high energy proton-proton collisions at the Large Hadron Collider and explore its utility  in anomaly detection. EDL is a DL approach that treats learning as an evidence acquisition process designed to provide confidence (or epistemic uncertainty) about test data. Using publicly available datasets for jet classification benchmarking, we explore hyperparameter optimizations for EDL applied to the challenge of UQ for jet identification. We also investigate how the uncertainty is distributed for each jet class, how this method can be implemented for the detection of anomalies, how the uncertainty compares with Bayesian ensemble methods, and how the uncertainty maps onto latent spaces for the models. Our studies uncover some pitfalls of EDL applied to anomaly detection and a more effective way to quantify uncertainty from EDL as compared with the foundational EDL setup. These studies illustrate a methodological approach to interpreting EDL in jet classification models, providing new insights on how EDL quantifies uncertainty and detects out-of-distribution data which may lead to improved EDL methods for DL models applied to classification tasks.
\end{abstract}

%
\vspace{2pc}
\noindent{\it Keywords}:  jet classification, machine learning, deep learning, evidential deep learning, uncertainty quantification, anomaly detection

%
\submitto{Mach. Learn.: Sci. Technol.}
%
\maketitle
%
%




\section{Introduction}

Machine Learning (ML) has become an indispensable tool in experimental high-energy physics (HEP), offering significant advancements in analyzing vast amounts of data obtained from complex detector systems. Over time, ML models have grown in complexity from simple regression and classification models into deep neural networks (DNNs) capable of performing sophisticated tasks to advance HEP. Despite the success of DNNs, they are often limited by their lack of explainability~\cite{xAI-review,Neubauer:2022zpn} and ability to provide reliable uncertainties~\cite{shanahan2022snowmass}. Uncertainty quantification (UQ) is crucial since uncertainties quantify the quality of predictive information and enable measurements to be contrasted or accurately combined. UQ also plays a crucial role in search for new physics (NP) signals, whether from specific NP phenomenological models or completely unexpected deviations from the standard model (SM) in the spirit of scientific exploration. The compatibility of extensions of the SM with data observations is constrained by the finite size of datasets as well systematic uncertainties arising from detector performance and signal modeling.

Classification of jets, referred to as \textit{jet
tagging}, is a major application of ML and DL in the field of HEP. Jets are observed as conical sprays of hadronic showers originating from quarks and gluons produced in the high energy collisions at facilities like the Large Hadron Collider (LHC). Historically, the ATLAS and CMS collaborations using jet tagging algorithms in conjunction with classic statistical and ML models such as decision trees, played a pivotal role in jet tagging efforts (see Refs. \cite{aad2016identification, CMS-PAS-JME-09-001, CMS-PAS-JME-13-007} for instance in the context of top quark tagging). More recently, the advent of DNNs has ushered a new era in jet classification algorithms for LHC physics. DNNs, with their ability to model complex, nonlinear relationships within data, have shown superior efficacy over traditional methods~\cite{jetclassdnn}, particularly in scenarios with \textit{boosted jets} where decay products of high-momentum heavy particles are highly collimated within a jet, requiring detailed analysis of jet substructures commonly employed in results using 13 TeV center-of-mass energy collisions at the LHC.~\cite{atlastopodnn, cms2020identification}.

A diverse range of deep learning (DL) models have been developed to optimize jet tagging~\cite{pearkes2017jet, liam2019reports, datta2017how, louppe2019qcd, butter2018deep, komiske2019energy, qu2020jet, macaluso2018pulling, erdmann2019lorentz,egan2017long, bogatskiy2020lorentz, moreno2020jedi, gong2022efficient,bogatskiy2022pelican,qu2022particle}. There have been a variety of approaches to utilize the ability of DNNs to approximate arbitrary non-linear functions in high-dimensional data~\cite{hornik1989multilayer}, and, as such, they have been successfully applied to the field of computer vision. Alternative models for jet tagging have been inspired by the underlying physics like jet clustering history~\cite{louppe2019qcd}, physical symmetries~\cite{butter2018deep} and physics-inspired feature engineering~\cite{erdmann2019lorentz}. These methods have inspired innovative model architectures and feature engineering by integrating or enhancing input feature spaces with physically meaningful quantities~\cite{erdmann2019lorentz,chakraborty2019interpretable,agarwal2021explainable}.

Despite the success of DL models for jet classification, UQ for these models remains a major challenge and an active area of research~\cite{10.21468/SciPostPhys.8.6.090, dorigo2021dealingnuisanceparametersusing, osti_1848067, viren2022solvingsimulationsystematicsaiml}. The black-box-like nature of DNNs obscures physical insight into the inner workings of these highly accurate classification machines, making it challenging to associate accurate and robust measures of uncertainty with these models. Traditional approaches to UQ in the context of DL models often utilize Bayesian inference models~\cite{vadera2020ursabenchcomprehensivebenchmarkingapproximate}, deep ensemble methods~\cite{lakshminarayanan2017simple}, and generative models like variational autoencoders~\cite{kingma2022autoencodingvariationalbayes}.

A comprehensive review of these traditional approaches can be found in Ref.~\cite{UQreview}. Many of these approaches pose significant challenges in terms of training complexity, convergence, and intuitive understanding of the associated uncertainty estimations. Additionally, some of these approaches are tied to specific models and cannot be easily adapted to other architectures. Recent advances in \textit{explainable} artificial intelligence (XAI)~\cite{MILLER20191} have made it possible to build intelligible relationships between an AI model's inputs, architecture, and predictions~\cite{xAI-intro,xAI-review,xAI-review-2}. Additionally, UQ in association with ML models relies on developing robust explanations~\cite{shanahan2022snowmass,seuss2021bridging} which are important for HEP algorithms such as jet tagging that require robust and interpretable models~\cite{grojean2022lessons, Khot_2023} for high-quality physics results. 

Expanding upon our previous work on interpretability of DL-based top quark taggers~\cite{Khot_2023}, we study \textit{evidential deep learning} (EDL) for UQ~\cite{10.5555/3327144.3327239} to develop a model-agnostic, robust, and interpretable approach towards UQ in jet tagging. EDL represents a novel and largely unexplored (in HEP) approach to UQ, offering a method to evaluate the confidence of predictions made by DNN models. By treating the learning process as evidence acquisition and interpreting more evidence as increased predictive confidence, EDL provides a framework for models to express not just predictions but also the certainty of those predictions. It has a significantly lower computational cost than other DNN-based UQ methods like Ensemble or Bayesian networks. Allowing fast UQ, EDL opens up the possibility of application of UQ beyond the standard application of jet tagging in physics analyses. To translate the success of DNNs in jet and event classification into a fast and online jet tagger, recent work has placed emphasis on developing DNN-enabled FPGAs for trigger-level applications at the LHC~\cite{duarte2018fast, iiyama2021distance, heintz2020accelerated}. As resource consumption and latency of FPGAs directly depend on the size of the network to be implemented, it is easier to embed simpler and faster networks on these devices. Hence, methods that quantify interpretable uncertainties without compromising performance can greatly benefit ML applications in both offline and real-time applications, especially for online event selection and jet tagging at current and future high energy colliders.

To demonstrate the application of EDL for UQ in jet tagging, we explore its integration with the Particle Flow Interaction Network (PFIN) model introduced in Ref.~\cite{Khot_2023}. The PFIN model, originally developed to leverage the intricate details of particle flows for improved jet classification, is enhanced through the adoption of EDL to refine its predictive accuracy and provide new capability with regard to UQ. This adaptation represents a significant step towards rendering DNNs more interpretable and reliable for scientific research, particularly in fields where the precise understanding and handling of data uncertainty is required for data-driven discovery. 

In this paper, we compare the uncertainties estimated by EDL with those from Ensemble and Bayesian methods and analyze the uncertainty distributions. The EDL structure and our chosen respective loss function is reviewed in Section~\ref{sec:review-edl}. To compare our results for existing benchmarks and different models, we use three publically available datasets with varying number of jet classes to understand how the uncertainty shifts with different classes. The datasets were developed by the authors of Ref.~\cite{kasieczka2019machine}, Ref.~\cite{Kansal_MPGAN_2021}, and Ref.~\cite{pmlr-v162-qu22b} and are summarized in Section~\ref{sec:data_n_exp}.  The EDL model hyperparameters, comparative Bayesian methods, dataset features, and their respective preprocessing is reviewed in Section~\ref{sec:data_n_exp}. 
The EDL-based uncertainties we analyzed for UQ is presented in Section~\ref{sec:results}. 
We compare EDL uncertainties with those from Ensemble and Bayesian methods in Section~\ref{sec:comp}. 
We analyze and interpret the EDL-based uncertainty in Section~\ref{sec:interp}. 
In Section~\ref{sec:edl4ad}, we explore the utilization of EDL for out-of-distribution detection toward improved anomaly detection methods.  We detail our outlook on EDL and the limitations of this method in Section~\ref{sec:outlook}. Finally, Section~\ref{sec:conclusion} summarizes our findings and illustrates new dimensions to explore in the conjunction of UQ and HEP.
\section{Review of Evidential Deep Learning}
\label{sec:review-edl}

In jet tagging, UQ is crucial due to the complex nature of particle interactions, and the need for accurate, robust and interpretable jet classification. There are two main types of uncertainty: \textit{aleatoric} and \textit{epistemic}. Aleatoric uncertainty describes the noise in the training data, and epistemic uncertainty relates to insufficient training data~\cite{10.5555/3295222.3295309}. Aleatoric uncertainty is often irreducible and can be estimated through neural networks~\cite{Gawlikowski2023}. On the other hand, epistemic uncertainty reduces with more data and is more difficult to approximate.

Evidential Deep Learning (EDL) introduces a novel approach to quantifying epistemic uncertainty, further referred to as uncertainty, in jet tagging. Unlike Ensemble and Bayesian methods, which rely on multiple inferences to approximate uncertainty, EDL directly models the uncertainty through sampling from a learned higher-order distribution.  Grounded in the Dempster-Shafer Theory of Evidence (DST)~\cite{dempster1968generalization} and implemented through Subjective Logic~\cite{10.5555/3031657}, EDL uses a Dirichlet distribution over class probabilities to interpret neural network outputs as subjective opinions, quantifying both confidence and uncertainty in predictions~\cite{10.5555/3327144.3327239}. This approach reduces computational demands by eliminating the need of multiple network evaluations and offers a more detailed understanding of uncertainty, enabling networks to express a spectrum of potential outcomes and their respective confidence levels. This property of EDL is particularly advantageous in fields like particle physics that rely heavily on uncertainty estimation and statistical methods for interpreting large, complex data. In this paper, we present the first detailed study of EDL being applied to experimental HEP.

The foundational EDL approach~\cite{10.5555/3327144.3327239} evaluates the epistemic uncertainty, or uncertainty mass, in classification tasks involving $K$ exclusive class labels. Each class label has a corresponding belief mass $b_k$, $k = 1, ..., K$, and there is an overall uncertainty mass $u$. All of them are non-negative and sum to $1$ as shown in Equation (\ref{eqn:edl-1}):
\begin{align}
    \label{eqn:edl-1}
    \sum_{k=1}^{K}b_k + u = 1,\ ~~0 \leq b_k \leq 1,\ 0 \leq u \leq1,\ k = 1, ..., K
\end{align}

The belief mass $b_k$ of each class $k$ is derived from a new concept, the evidence $e_k$. Evidence quantifies support gathered from data that advocates for categorizing a sample into a specific class. The relationship between $b_k$ and $e_k$ is shown in Equation (\ref{eqn:edl-2}):
\begin{align}
    \label{eqn:edl-2}
    b_k = \frac{e_k}{S},\ ~~S = \sum_{k=1}^{K}(e_k + 1)
\end{align}
The uncertainty mass is then computed as shown in Equation (\ref{eqn:edl-3}):
\begin{align}
    \label{eqn:edl-3}
    u = \frac{K}{S}
\end{align}
The sum $S = \sum_{k=1}^{K}(e_k + 1)$ represents the Dirichlet strength, indicating the overall evidence strength supporting the classification. This is because the Dirichlet distribution, with parameters $\alpha_k = e_k + 1$, represents these belief mass assignments $b_1, b_2, ... , b_K$, which is also called a \textit{subjective opinion}. The probability density function of the Dirichlet distribution with $K$ parameters $[\alpha_1, ..., \alpha_K]$ is given by
\begin{equation}
    \label{eqn:edl-4}
    D(\mathbf{x}|\mathbf{\alpha})=D(x_1, x_2, \dots, x_K| \alpha_1, \alpha_2, \dots, \alpha_K) = \frac{\prod_{i=1}^K x_i^{\alpha_i - 1}}{B(\mathbf{\alpha})},\ x_i \geq 0,\ ~~\sum_{i=1}^{K}x_i = 1
\end{equation}
where the normalizing constant $B(\alpha)$ can be defined in terms of the Gamma function $\Gamma(\cdot)$
\begin{align}
    \label{eqn:edl-5}
    B(\mathbf{\alpha}) = \frac{\prod_{i=1}^K \Gamma(\alpha_i)}{\Gamma(\sum_{i=1}^K \alpha_i)}
\end{align}
For a given subjective opinion, the expected probability of the $k^{th}$ class is derived as the average value from the respective Dirichlet distribution, as shown in Equation (\ref{eqn:edl-6}):
\begin{align}
    \label{eqn:edl-6}
    p_k = \mathbb{E}(x_k) = \frac{\alpha_k}{S} = \frac{e_k + 1}{S}
\end{align}

The final stage in the EDL framework involves determining the evidence $e_k$. This can be accomplished by slightly modifying the outputs of traditional classification neural networks. Typically, classification neural networks utilize a \textsc{Softmax} layer for output, which assigns probabilities to each class. In the EDL approach, the \textsc{Softmax} layer is replaced with a \text{ReLU} activation layer. This ensures that the outputs are non-negative, which is necessary since these outputs are used as the evidence vector for the Dirichlet distribution that models the uncertainties and confidences in predictions. The outputs of the network, denoted as $\mathbf{f}(\mathbf{x}|\Theta)$, directly provide the evidence for the anticipated Dirichlet distribution through 
\begin{equation*}
e_k = f_k(\mathbf{x}|\Theta) ~~~\textrm{and}~~~ \alpha_k = f_k(\mathbf{x}|\Theta) + 1
\end{equation*}
These modifications enable the network to not only predict outcomes but also provide a probabilistic assessment of these predictions, enriching the decision-making process in critical applications such as jet tagging.

To ensure the model learns these opinions, the optimal loss function for the EDL model is composed of two primary components, the reconstruction loss, $\mathcal{L}_{MSE}$, and the Kullback-Leibler (KL) Divergence, $\mathcal{L}_{KL}$. The reconstruction loss $\mathcal{L}_{MSE}$, is calculated as the mean squared error (MSE) between the predicted classification probabilities $\mathbf{\hat{y}_i}$ and actual targets $\mathbf{y}_i$. Contrary to the traditional cross-entropy loss in a classification setting, using the MSE loss metric allows for simultaneous reduction of the prediction error and the variance of the Dirichlet distribution~\cite{10.5555/3327144.3327239}.
\begin{align}
    \label{eqn:edl-8}
    \mathcal{L}_{MSE}(\Theta)_i 
    &=
    \sum_{k=1}^K\mathbb{E} [ \left(y_{ik} - x_{ik}\right)^2 ] 
\end{align}
The second component of the loss function is a KL Divergence term defined as, 
\begin{align}
    \label{eqn:edl-9}
    \mathcal{L}_{KL}(\Theta)_i&=KL[D(\mathbf{x}_i|\mathbf{\tilde{\alpha}_i})\|D(\mathbf{x}_i|\left \langle 1,\cdots,1 \right \rangle)] \\
    &=\log \left (\frac{\Gamma(\sum_{k=1}^K\tilde{\alpha}_{ik})}{\Gamma(K)\prod_{k=1}^K\Gamma(\tilde{\alpha}_{ik})} \right ) + \sum_{k=1}^K (\tilde{\alpha}_{ik}-1) \left [ \psi(\tilde{\alpha}_{ik}) - \psi\left(\sum_{j=1}^K \tilde{\alpha}_{ij}\right) \right] \nonumber
\end{align}
where 
\begin{equation*}
    \mathbf{\tilde{\alpha}_i} = \mathbf{y_i} + (1 - \mathbf{y_i}) \odot \mathbf{\alpha_i}
\end{equation*}
and $\psi(\cdot)$ is the digamma function.

As a key component in EDL to ensure that the model appropriately handles both in-distribution and out-of-distribution input data, Equation~\ref{eqn:edl-9} encourages the network to be more confident about correct predictions while allowing it to generously admit when it fails to do so. For out-of-distribution and hard-to-classify inputs, it ensures that the model outputs high uncertainty, effectively preventing overconfident and potentially erroneous predictions. For in-distribution inputs, it encourages the model to exhibit a clear preference for one class over others by promoting one high evidence value $e_k$ among the possible classes. This helps in sharpening the model's confidence in its predictions when faced with familiar data. This KL Divergence term is strategically integrated into the overall loss function as a regularization term, modulated by an annealing coefficient $\lambda_t$. The overall loss function is given by
\begin{align}
    \label{eqn:edl-10}
    \mathcal{L}(\Theta)=\sum_{i=1}^{N}\mathcal{L}_{MSE}(\Theta)_i+\lambda_t\sum_{i=1}^{N}\mathcal{L}_{KL}(\Theta)_i
\end{align}
The $\lambda_t$ parameter is a hyperparameter of the EDL model which regulates the network's ability to assign uncertainties to model predictions. The authors of Ref.~\cite{10.5555/3327144.3327239} proposed a dynamically scaled choice of $\lambda_t$ to ensure a gradual increase during the training process, defined as $\lambda_t = min (1.0, t/10) \in[0,1]$, where $t$ represents the epoch index. However, since the default choice did not always provide the most optimal solution in the applications we studied, we further adjusted its strength by parameterizing it as $\lambda_t (\zeta) = \zeta \times min (1.0, t/10)$ with $\zeta \in[0,1]$. This scaling allows the influence of KL Divergence term to be limited initially, avoiding overly harsh penalties that could lead to model convergence towards a uniform distribution prematurely. The annealing strategy ensures that as training progresses and the model stabilizes, the regularization effect of the KL Divergence becomes more important, guiding the model towards more accurate UQ.

\section{Dataset and Experimental Setup}
\label{sec:data_n_exp}

\subsection{Datasets}


In this paper, we consider three different datasets for UQ and anomaly detection using EDL: (1) top tagging, (2) JetNet, and (3) JetClass. The data details and cross validation setup for each of the datasets are summarized below:

\begin{itemize}[leftmargin=*, label={}]
    \item (1) \textbf{Top Tagging dataset} (\topdata)~\cite{kasieczka2019machine, topdata}: This dataset consists of 1 million top (signal) jets and 1 million QCD (background) jets generated with \textsc{Pythia8}~\cite{pythia8} with its default tune at 14~TeV center of mass energy for proton-proton collisions. The detector simulation was performed with \textsc{Delphes}~\cite{delphes} and jets were reconstructed using the $anti-k_t$ algorithm~\cite{cacciari2008anti} with a jet radius of $R = 0.8$ using \textsc{FastJet}~\cite{cacciari2012fastjet}. Only jets with transverse momenta within the range of $550$ and $650$~GeV are considered. For each jet, the dataset contains the four momenta of up to 200 constituents with zero-padded entries for missing constituents.
    The top tagging models are trained with transverse momentum ($p_T$), azimuthal angle ($\phi$), and pseudorapidity ($\eta$) of the 60 most energetic particles. As part of data preprocessing, we standardized the constituents' $\eta$ and $\phi$ by subtracting the jet's $\eta$ and $\phi$. The $p_T$ values of the jets constituents are scaled by the inverse of the sum of constituents $p_T$, i.e. $1 / \sum_i p_{T,i}$.
    The dataset is divided into training, validation, and testing sets with a 6:2:2 split and trained in batches of 250. Some characteristic jet features from the dataset are shown in Figure~\ref{fig:topdata-jet-feats}.

\begin{figure}[htbp]
\centering
\subfloat[]{
\includegraphics[width=0.32\textwidth]{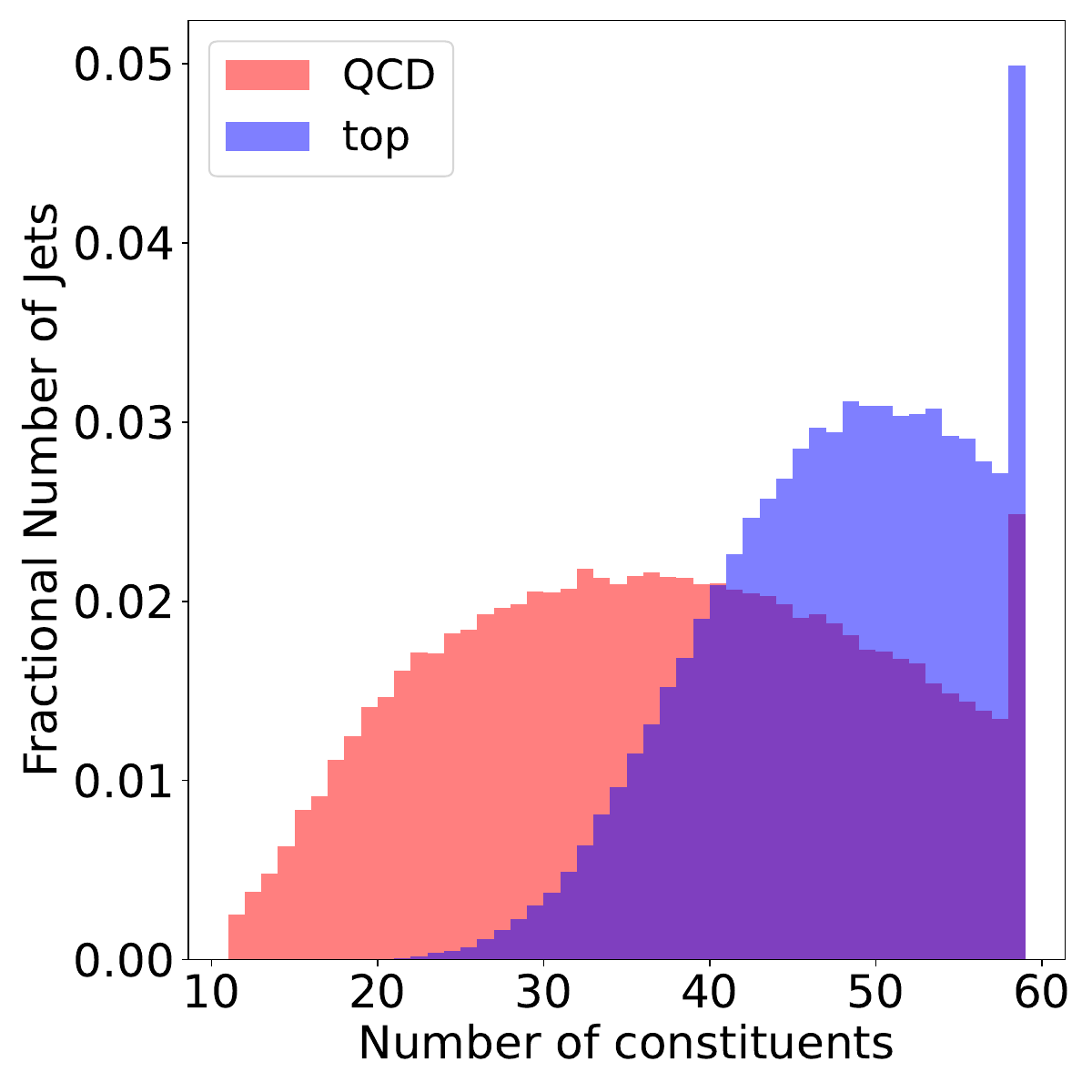}
\label{fig:topdata-Nconst}            
}
\subfloat[]{
\includegraphics[width=0.32\textwidth]{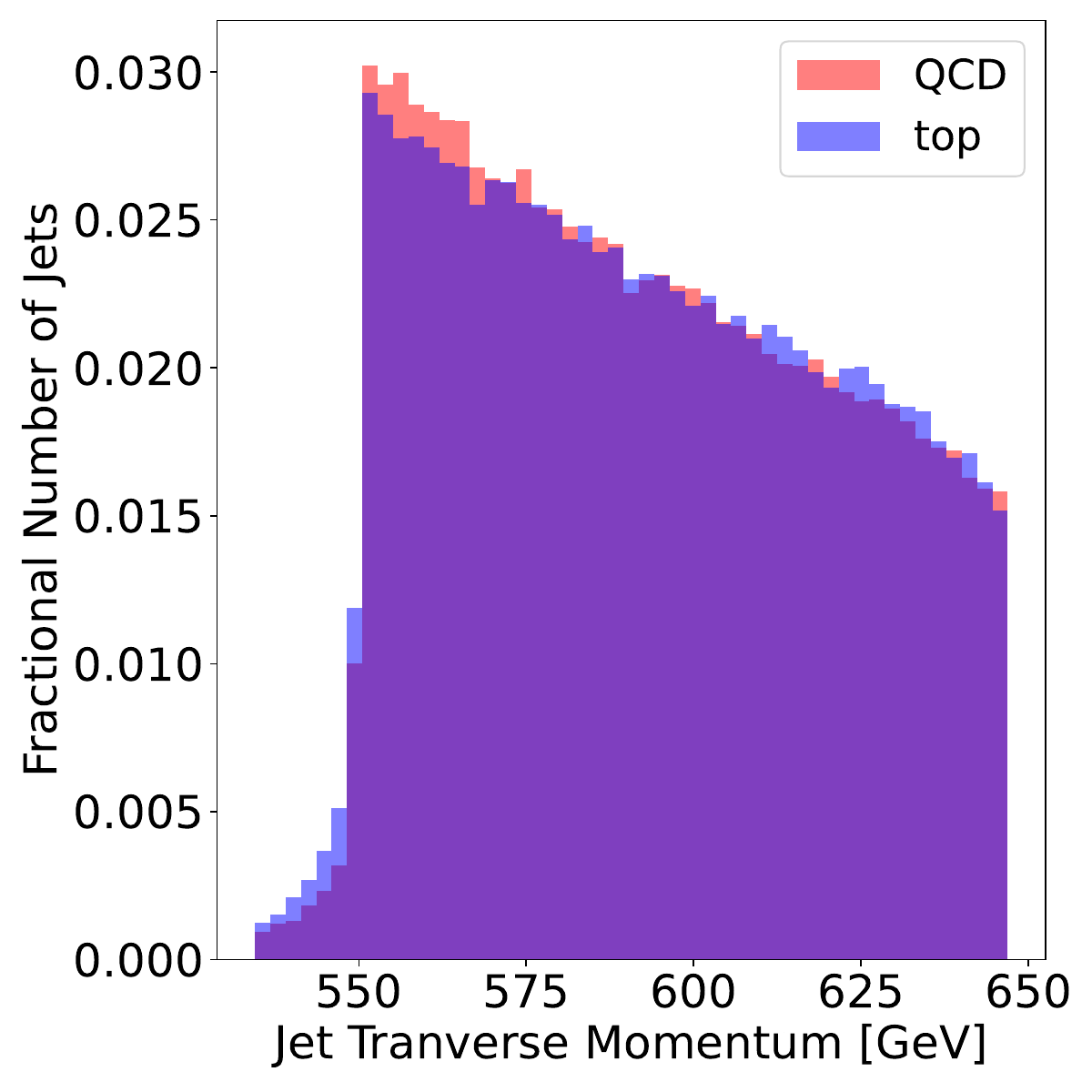}
\label{fig:topdata-jet-pt}            
}
\subfloat[]{
\includegraphics[width=0.32\textwidth]{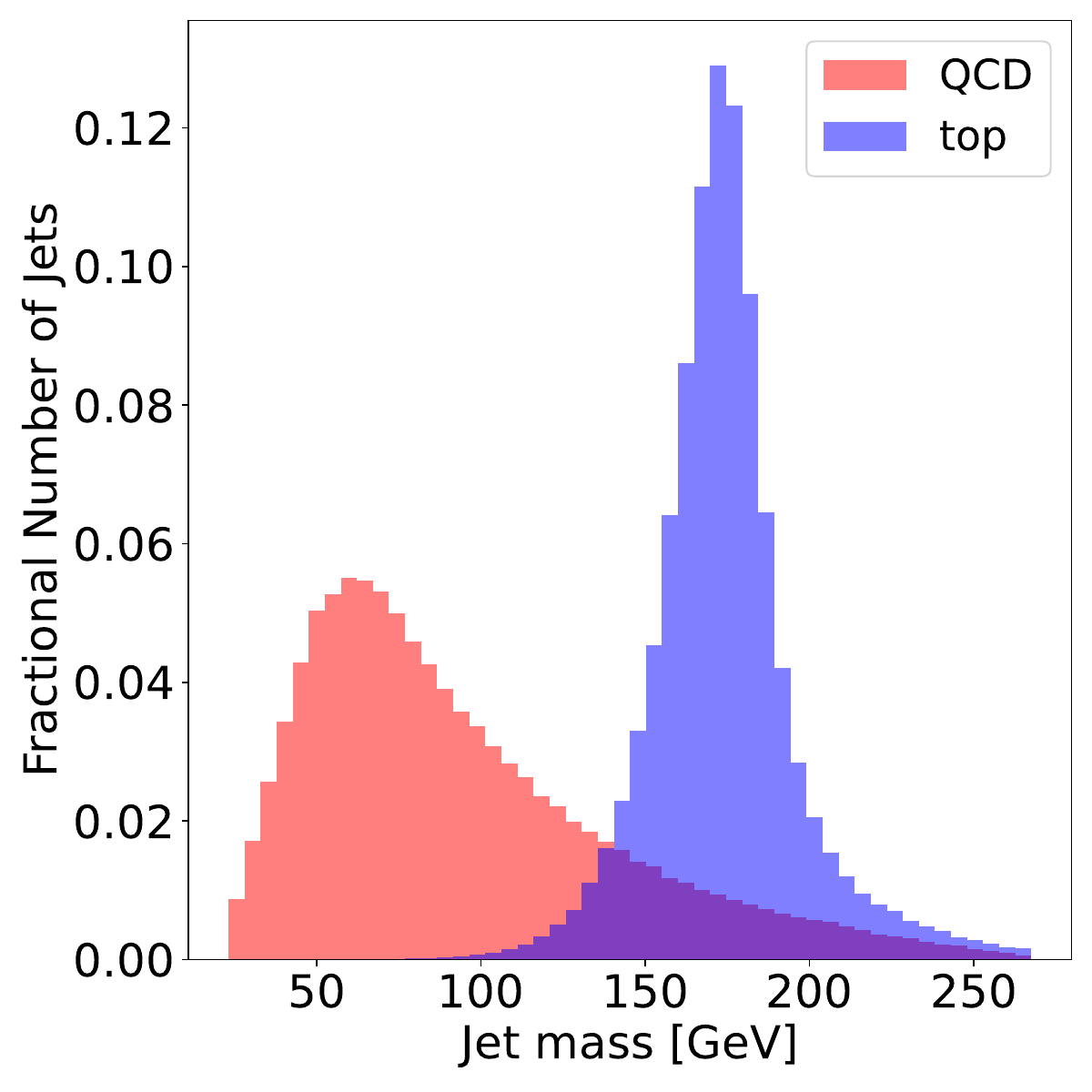}
\label{fig:topdata-jet-m}            
}
\caption{Distribution of \protect\subref{fig:topdata-Nconst} number of constituents,  \protect\subref{fig:topdata-jet-pt} jet transverse momentum ($p_{T,J}$), and  \protect\subref{fig:topdata-jet-m} jet mass ($m_J$) for jets from QCD and top quarks.}
\label{fig:topdata-jet-feats}
\end{figure}
    
    \item (2) \textbf{JetNet dataset} (\jetnet)~\cite{Kansal_MPGAN_2021, kansal_2022_6975118}: This dataset consists of 880k particle jets originating from gluons ($g$), light quarks ($q$), top quarks ($t$), and bosons ($W$ and $Z$). The parton-level events were generated using \textsc{MadGraph5\_aMC@NLO} 2.3.1~\cite{Alwall2014-qv} with its default tune at 13~TeV center of mass energy for proton-proton collisions. These parton-level events are then decayed and showered in \textsc{Pythia8}~\cite{pythia8}. Jets were reconstructed using the $anti-k_t$ algorithm~\cite{cacciari2008anti} with a jet radius of $R = 0.8$ using the \textsc{FastJet} 3.13 and \textsc{FastJet contrib} packages~\cite{cacciari2012fastjet, CACCIARI200657}. Only jets with transverse momenta within the window of $0.8$ and $1.6$~TeV are considered. For each jet, the dataset contains the four momenta of up to 30 constituents with zero-padded entries for missing constituents. Similar to the top tagging dataset, JetNet models are trained with $p_T$, $\phi$, and $\eta$ of jet constituents as input with the same preprocessing. The dataset is divided into training, validation, and testing sets with a 5:3:2 split and trained in batches of 250. Some characteristic jet features from the dataset are shown in Figure~\ref{fig:jetnet-jet-feats}.

\begin{figure}[htbp]
\centering
\subfloat[]{
\includegraphics[width=0.32\textwidth]{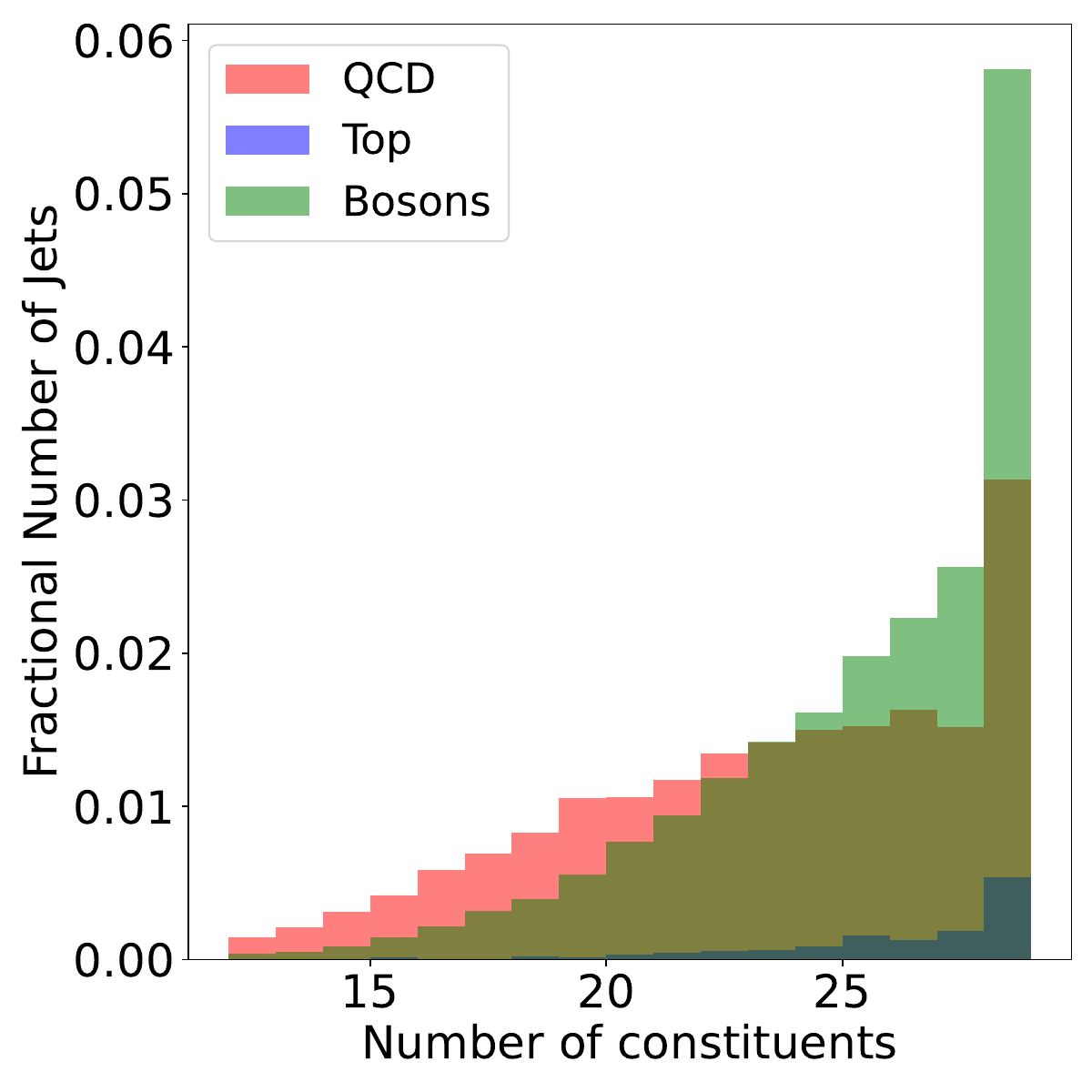}
\label{fig:jetnet-Nconst}            
}
\subfloat[]{
\includegraphics[width=0.32\textwidth]{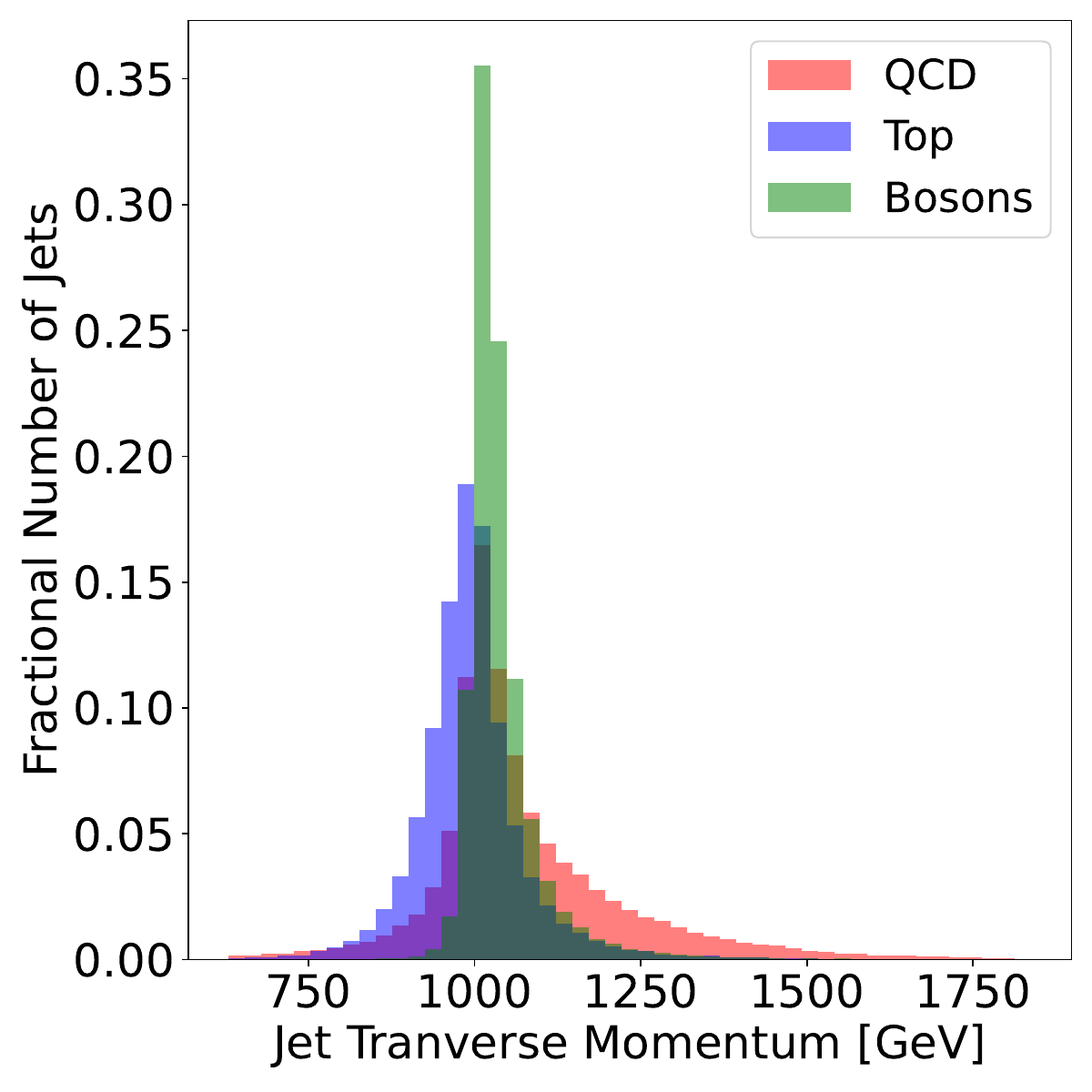}
\label{fig:jetnet-jet-pt}            
}
\subfloat[]{
\includegraphics[width=0.32\textwidth]{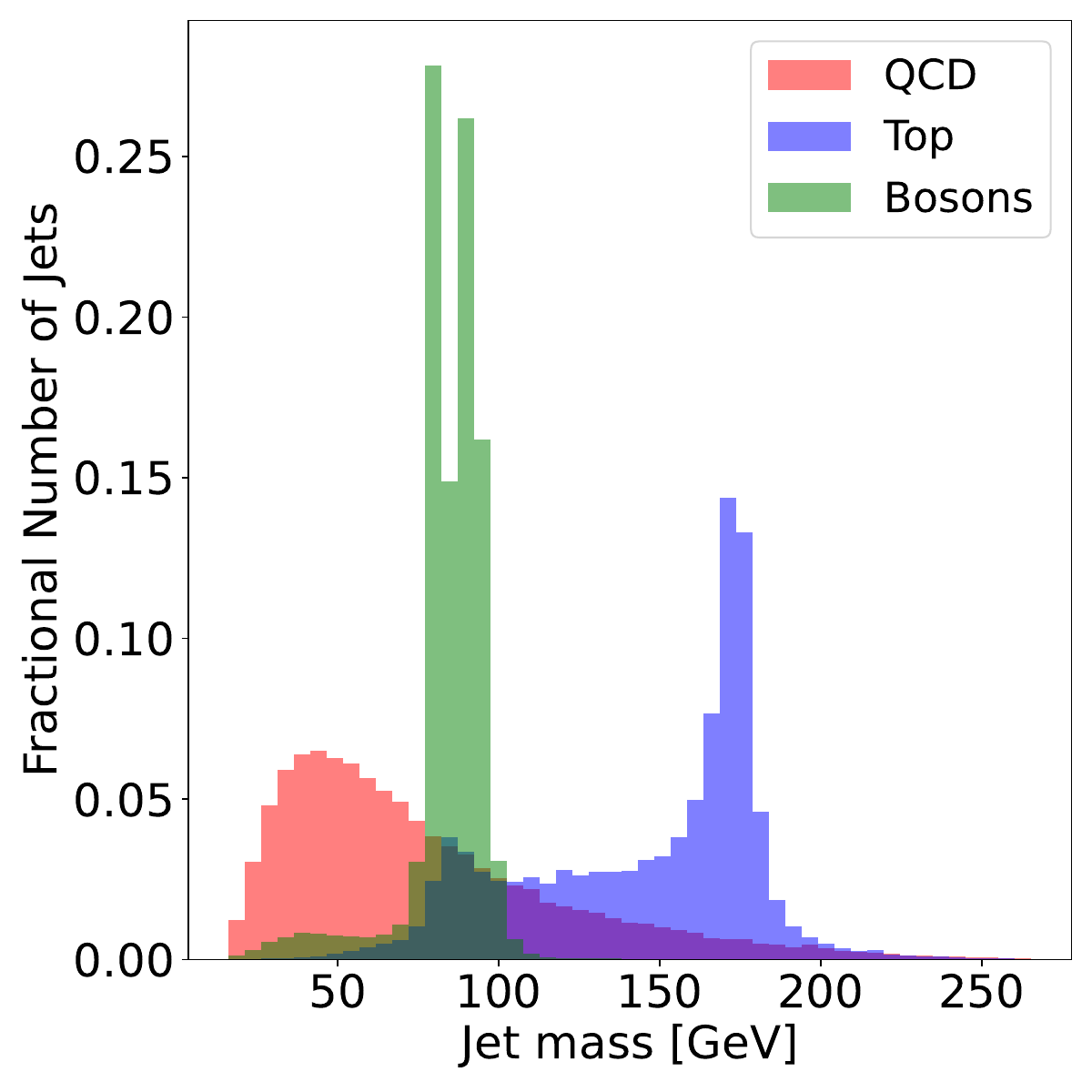}
\label{fig:jetnet-jet-m}            
}
\caption{Distribution of \protect\subref{fig:jetnet-Nconst} number of constituents,  \protect\subref{fig:jetnet-jet-pt} jet transverse momentum ($p_{T,J}$), and  \protect\subref{fig:jetnet-jet-m} jet mass ($m_J$) for QCD ($q,g$), top ($t$), and boson ($W,Z$) jets.}
\label{fig:jetnet-jet-feats}
\end{figure}

    \item (3) \textbf{JetClass dataset} (\jetclass)~\cite{pmlr-v162-qu22b, JetClass}: The dataset consists of 125 million particle jets of ten different types of jets initiated by gluons and quarks ($q/g$), top quarks ($t$), and bosons($W$, $Z$, and $H$). As described in Ref.~\cite{birk2023flow}, jets initiated by a top quark or a Higgs boson are further categorized based on their different decay channels, resulting in the following ten categories: $ q/g$, $t \to bqq'$, $t \to b \ell v$, $Z \to q \bar{q}$, $W \to q q'$, $H \to b \bar{b}$, $H \to c \bar{c}$, $H \to gg$, $H \to 4 q$, and $H \to \ell v q q'$. The jets are extracted from simulated events that are generated with \textsc{MadGraph5\_aMC@NLO}~\cite{Alwall2014-qv}. The parton showering and hadronization was performed with\textsc{Pythia8}~\cite{pythia8} and the detector simulation was performed with \textsc{Delphes}~\cite{delphes}. Jets were reconstructed using the $anti-k_t$ algorithm~\cite{cacciari2008anti} with a jet radius of $R = 0.8$ using the \textsc{FastJet} package~\cite{cacciari2012fastjet}. Only jets with transverse momenta within the range of $550$ and $1000$~GeV and a pseudorapidity $ |\eta^\text{jet}| < 2$ are considered. For each jet, the dataset contains 11 features for each particle, including information on kinematics, particle identification, and trajectory displacement. The particle features include the $p_T$, $\phi$, and $\eta$ of jet constituents, as well as the electric charge. Particle classification is represented using a five-class one-hot encoding to distinguish charged hadrons, neutral hadrons, electrons, muons, and photon. Additionally, the dataset includes measurements of the transverse and longitudinal impact parameters of particle trajectories, reported in mm.
    Each jet contains up to 60 constituents with zero-padded entries for missing constituents. 
    The kinematic variables receive the same data preprocessing as in the other datasets.
    The dataset is divided into training, validation, and testing sets with a 100:5:20 split. In our work, we only use 20M jets for training and 2M jets for validation in batches of 2500 because there is an insubstantial increase in performance for larger training sizes. Some characteristic jet features from the dataset are shown in Figure~\ref{fig:jetclass-jet-feats}. 

\end{itemize}

\begin{figure}[htbp]
\centering
\subfloat[]{
\includegraphics[width=0.32\textwidth]{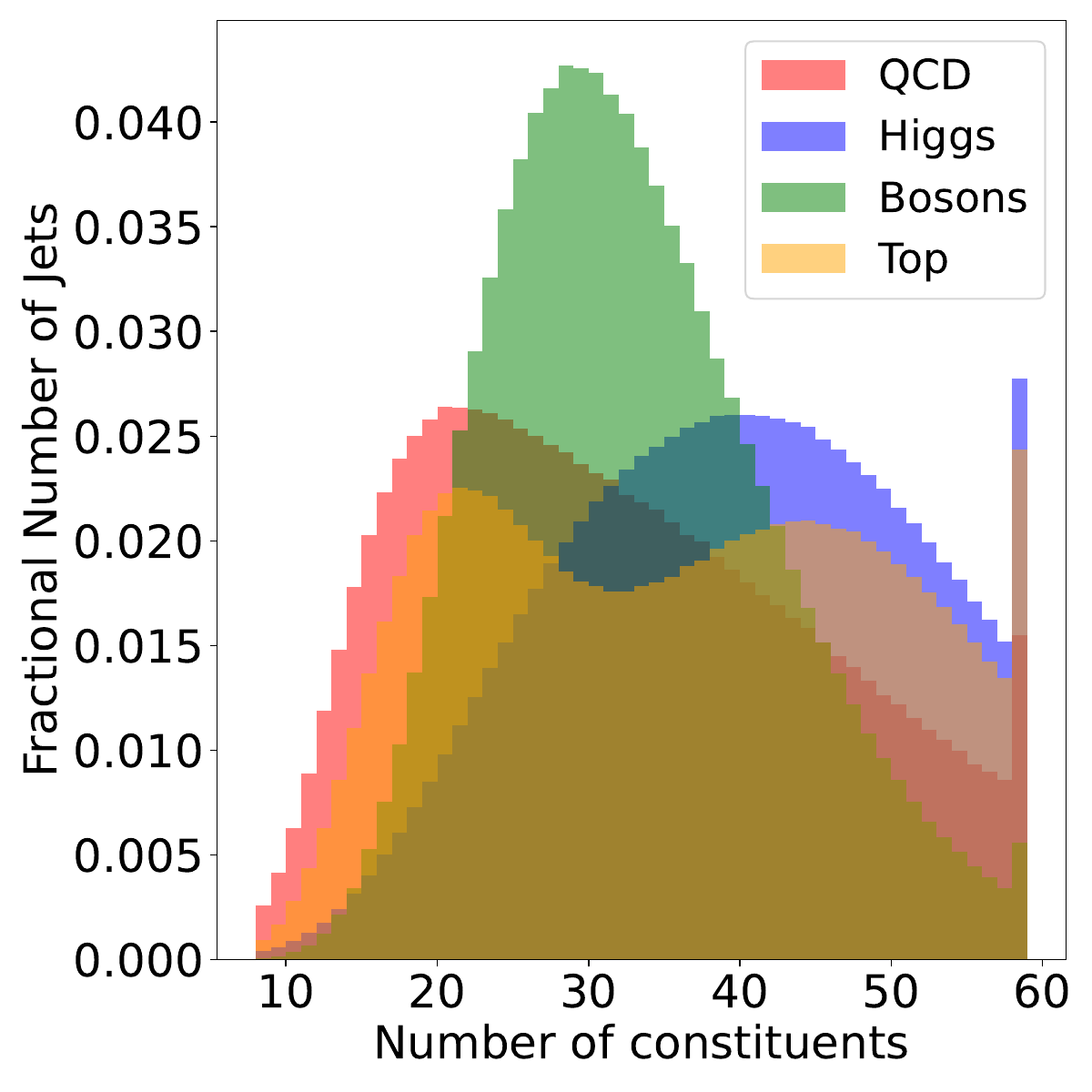}
\label{fig:jetclass-Nconst}            
}
\subfloat[]{
\includegraphics[width=0.32\textwidth]{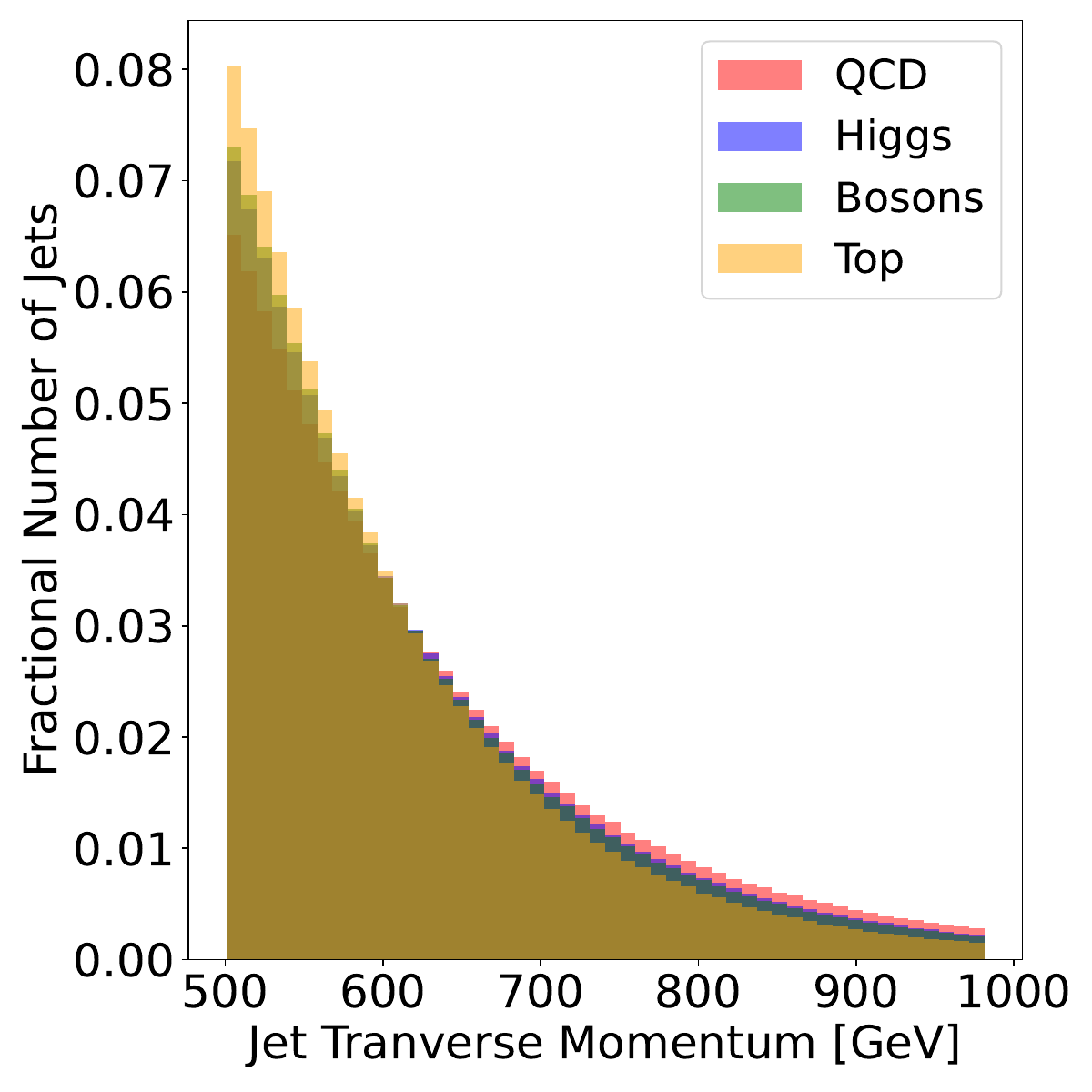}
\label{fig:jetclass-jet-pt}            
}
\subfloat[]{
\includegraphics[width=0.32\textwidth]{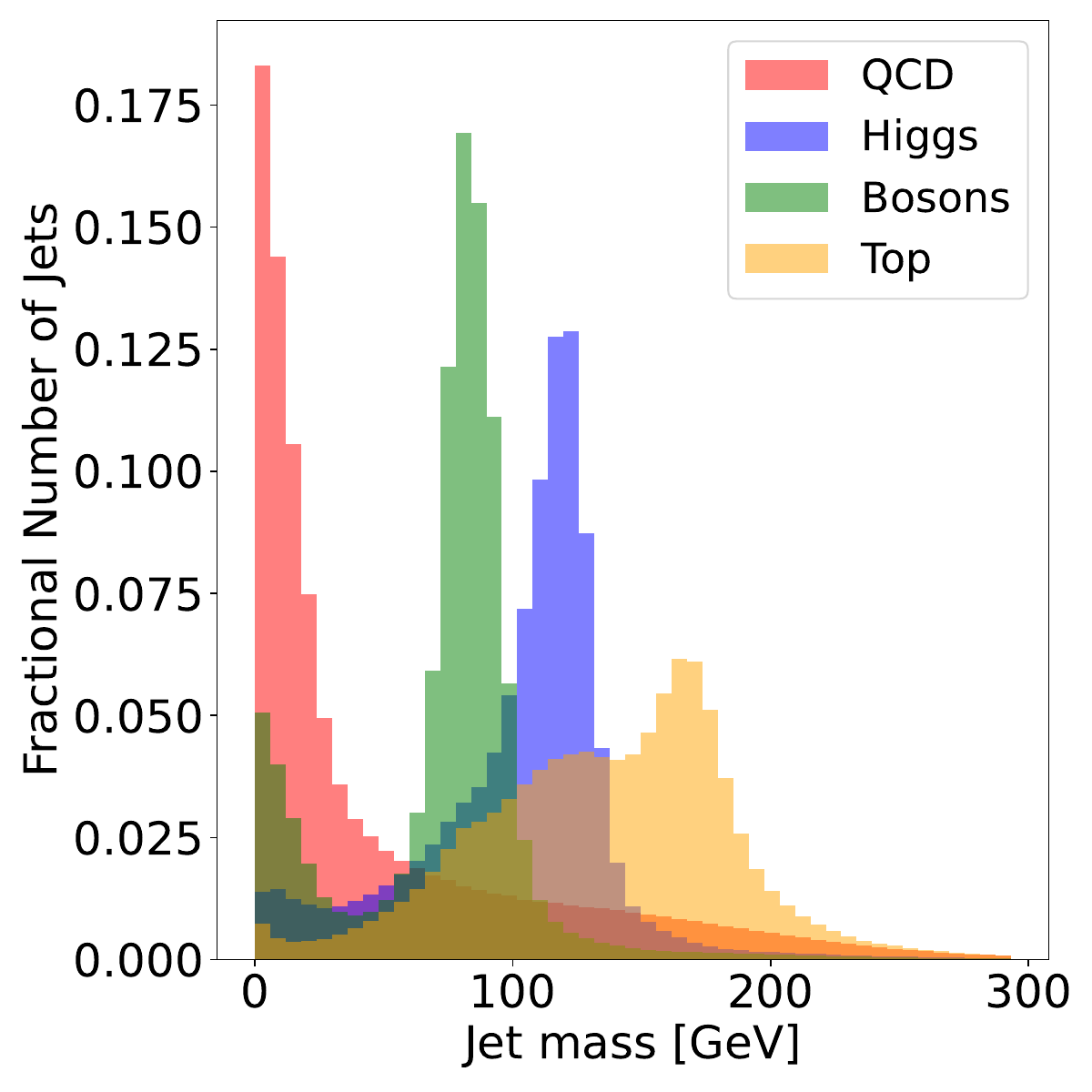}
\label{fig:jetclass-jet-m}            
}
\caption{Distribution of \protect\subref{fig:jetclass-Nconst} number of constituent particles,  \protect\subref{fig:jetclass-jet-pt} jet transverse momentum ($p_{T,J}$), and  \protect\subref{fig:jetclass-jet-m} jet mass ($m_J$) for QCD ($q/g$), Higgs ($H$), boson ($W,Z$) and top ($t$) jets.}
\label{fig:jetclass-jet-feats}
\end{figure}

\subsection{Model}

The DNN tagger model we chose to integrate with the EDL model is the Particle Flow Interaction Network (PFIN)~\cite{Khot_2023}. It is an augmentation of a Particle Flow Network (PFN)~\cite{komiske2019energy} with an Interaction Network (IN)~\cite{IN,moreno2020jedi}. We chose this due to the superior performance of the PFIN model on top tagging and its ability to learn from particle-level interactions in the latent space. These traits make it ideal for EDL to learn from particle-level features and investigate EDL's latent space representation. 

\begin{figure}
    \centering
    \includegraphics[width=1.04\textwidth]{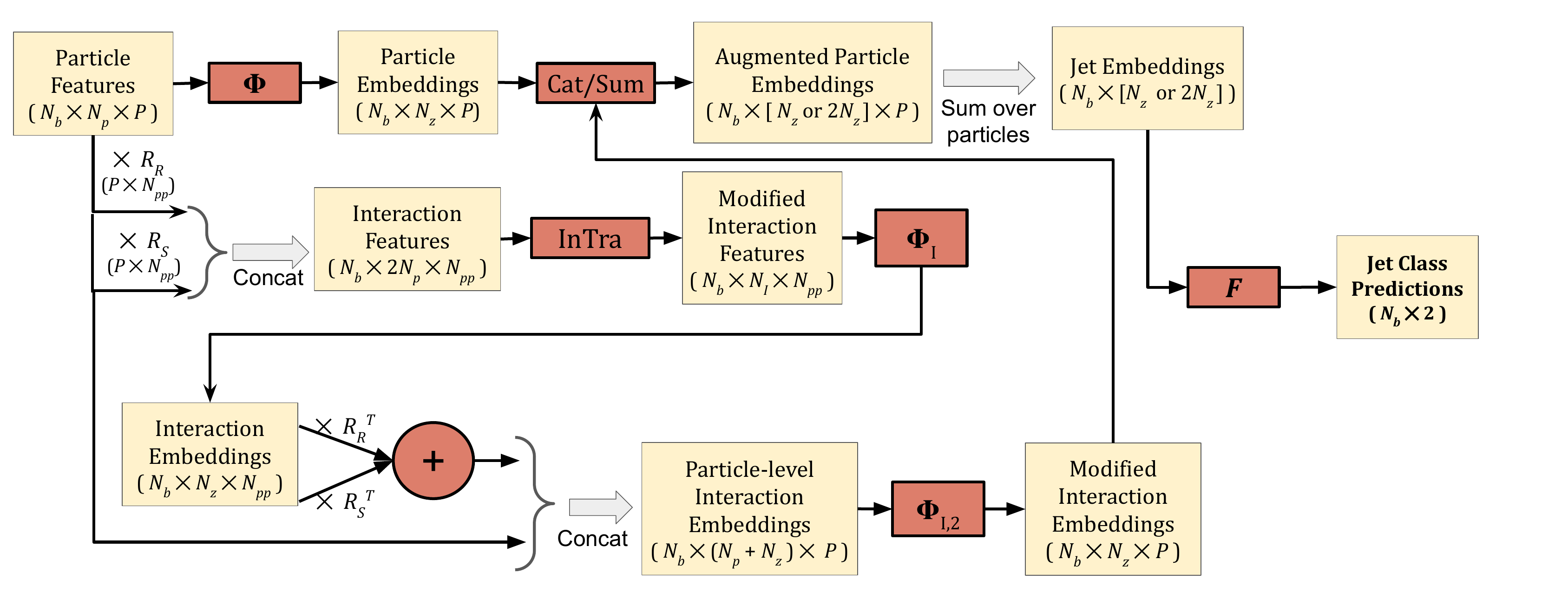}
    \caption{Model architecture and data flow for the PFIN model. $N_b$ represents the batch size. The \texttt{InTra} block computes the pairwise particle interaction features given in Eqn.~\ref{eqn:phys-feats}. The \texttt{Cat/Sum} block creates the augmented particle embeddings by either concatenating or summing the outputs of $\Phi_{I,2}$ with $\Phi$. The $\Phi, \Phi_I, \Phi_{I,2}$, and $F$ networks are implemented as fully connected MLPs with ReLU activation. From Figure 17 in Ref.~\cite{Khot_2023}.}
    \label{fig:PFIN-flow}
\end{figure}

As outlined in Ref.~\cite{Khot_2023}, the dataflow for the PFIN model is illustrated in Figure~\ref{fig:PFIN-flow}. In PFIN, the particle interactions are encapsulated by formulating a fully connected undirected graph with $N_{pp} = \frac{P(P -1)}{2}$ edges where $P$ represents the maximum number of constituent particles the model is trained with. Each particle within this graph is described by a set of $N_p$ attributes. 
We have selected to use $N_p=3$, using the triplet $(p_t, \eta, \phi)$ for each particle in \topdata and \jetnet datasets, following the same preprocessing steps. For the \jetclass dataset, the number of attributes per particle was $N_p = 11$. For each edge in the graph, we combine the features of the two particles involved, resulting in an initial representation of $2N_P$ attributes for every edge. To assist in transforming these node-level features to edge-level attributes, we use two interaction matrices, $R_R$ and $R_S$, each of which has dimensions $P \times N_{pp}$. The edge-level attributes are transformed by the Interaction Transformation (\texttt{InTra}) block to calculate a $N_I = 4$ dimensional representation for each edge by calculating the physics-inspired quantities $\ln\Delta$, $\ln k_T$, $\ln z$, and $\ln m^2$~\cite{qu2022particle, erdmann2019lorentz}, where
\begin{align}
    \label{eqn:phys-feats}
    \Delta &= \sqrt{(\eta_1 - \eta_2)^2 + (\phi_1 - \phi_2)^2} \nonumber \\
    k_T &= \min\left(p_{t,1}, p_{t,2}\right)\Delta \nonumber \\
    z &= \frac{\min\left(p_{t,1}, p_{t,2}\right)}{p_{t,1} + p_{t,2}} \\
    m^2 &= (E_1 + E_2)^2 - || \vec{p}_1 + \vec{p}_2 ||^2 . \nonumber 
\end{align}

The subscripts $1$ and $2$ denote the two particles associated with the edge and each variable within the relations refers to its unpreprocessed value. Since these quantities are symmetric with respect to the particles, the order of the particles does not impact PFIN's dataflow, maintaining the permutation-invariant property of PFN. These interaction features are transformed into $N_z$ dimensional interaction embeddings by the trainable $\Phi_I$ network. These embeddings are propagated back to particle level using the interaction 
matrices, taking into account only those interactions where both particles are involved. These particle-level interaction embeddings are concatenated with the original particle features and further processed into $N_z$ dimensional modified per-particle interaction embeddings through a trainable $\Phi_{I,2}$ network. The embeddings are then combined, either through concatenation or addition, with per-particle embedding from PFN's $\Phi$  network to obtain augmented particle embeddings. These augmented features are then summed over its constituents to obtain the jet-level latent representation. Finally, the $F$ network obtains the output for each of the jet class based on these jet-level latent space features. At the end of the $F$ network, a \textsc{Softmax} layer is used for baseline models to output probabilities while a \textsc{ReLU} layer is used for EDL models to output the Dirichlet parameters. The training for all models is done using the \textsc{Adam} optimizer with minibatches. The model hyperparameters are chosen from the baseline PFIN summation model in Ref.~\cite{Khot_2023}.

\subsection{Baseline methods}

Traditionally, ensemble methods~\cite{lakshminarayanan2017simple} and Monte Carlo (MC) Dropout~\cite{gal2016dropout} have been popular techniques for estimating uncertainty in DNNs. Ensemble methods involve training multiple models on the same task and using their varied outputs to evaluate uncertainty, providing a measure of confidence based on the diversity of the results. On the other hand, MC Dropout leverages dropout layers during both training and inference phases to simulate the effect of  Bayesian inference, thus providing a stochastic basis for uncertainty estimation~\cite{pmlr-v48-gal16}. Both methods are computationally intensive as they require multiple inferences to form a consensus on predictions, reflecting a significant trade-off between accuracy and computational efficiency. We use 10 independent estimates for each prediction in these methods. For the model ensemble, 10 instances of the same model are trained with different seeds to provide 10 independent models. For MC dropout, each sample is passed through the same model 10 times. Given that some of the datasets have more than two classes, minimizing the cross-entropy (CE) loss has been used as the cost function for all Ensemble and MC Dropout models. The maximum of standard deviations of class-wise probability predictions has been used as an estimate of uncertainty for both ensemble and MC dropout methods.

\subsection{Metrics}
\label{sec:metrics}

 The models we trained for this analysis have been evaluated based on two underlying principles: (1) how confident a model is when it correctly predicts the class of a given jet and (2) how well the uncertainty estimate represents the ability of a model to identify misclassified or anomalous jets. Although this paper mostly focuses on the task of uncertainty quantification, the metrics we propose in this section also allow us to assess the performance of anomaly detection (AD) models described in Section~\ref{sec:edl4ad}. To simulate anomalous jets in AD models, we refer to two types of data: in-distribution and out-of-distribution. In-distribution (ID) jets refers to the type of data on which the model is trained, encompassing scenarios and characteristics that the model is expected to handle under normal operating conditions. 
 Conversely, out-of-distribution (OOD) data involves data points, or jet particle types, that are not represented during the training phase. 
 Section~\ref{sec:edl4ad} further explains the creation of ID and OOD datasets for the purposes of this study.
 The following metrics are critical for testing the model's robustness and its ability to handle unexpected or novel situations. 

\begin{itemize}[leftmargin=*, label={}]
    \item \textbf{ID Accuracy}: In our baseline models, ID accuracy is the same as model accuracy, measuring the ability of a model to correctly classify jets. In the case of AD models, this metric represents the accuracy of a model in correctly classifying ID jets, determined by the ratio of correct predictions on ID data to the total number of ID data. This metric ensures that a given model maintains high performance on familiar data and confirms that the enhancement in UQ or AD does not compromise its ability to handle expected scenarios.

\vspace{0.1in}
    \item \textbf{AUROC}: Area Under the Receiver Operating Characteristic Curve (AUROC) is commonly used metric to represent the overall quality of binary classification models. In our context, the AUROC represents how well the uncertainty estimate of a model correlates with an inability of the model to distinguish certain jet classes or identify anomalous jets. In a well-trained classification model, we want the model to be confident, i.e. assign low uncertainties, for correctly classified jets. On the other hand, large uncertainties should be associated with misclassified jets (in a UQ model) or anomalous jets (in an AD model).  Figure~\ref{fig:jetnet-roc} shows a typical ROC constructed from the results of a benchmark EDL model on the \jetnet dataset. The vertical axis represents the fraction of misclassified jets that are assigned an uncertainty greater than a given threshold. The horizontal axis, on the other hand, represents the fraction of correctly classified jets that are assigned an uncertainty greater than a given threshold. The ROC is generated by varying the uncertainty threshold within the range of the observed uncertainties obtained by the model. A higher value of the AUROC would represent the model's superiority in projecting confidence for correctly classified jets while assigning larger uncertainties for incorrectly classified jets. 

    \vspace{0.1in}

    A similar idea can also be constructed in the case of AD models. Figure~\ref{fig:jetnet-skiptop-roc} shows a typical ROC constructed from the results of a benchmark EDL model on the \texttt{JetNet-skiptop} dataset, a variant of the \jetnet dataset that withholds the top jets from the training dataset but reintroduces them as OOD samples in the testing data. In this case, the vertical axis of the ROC represents the OOD detection rate, identified by the fraction of OOD jets assigned an uncertainty larger than the chosen threshold. The horizontal axis represents ID mis-tag rate, which is the fraction of ID jets assigned an uncertainty larger than the chosen threshold. Similar to what is done for UQ models, the ROC is generated by varying the uncertainty threshold within the range of the observed uncertainties obtained by the model. A larger value of the AUROC would imply a model's enhanced ability to tell apart OOD jets.

\begin{figure}[htbp]
\centering
\subfloat[]{
\includegraphics[width=0.49\textwidth]{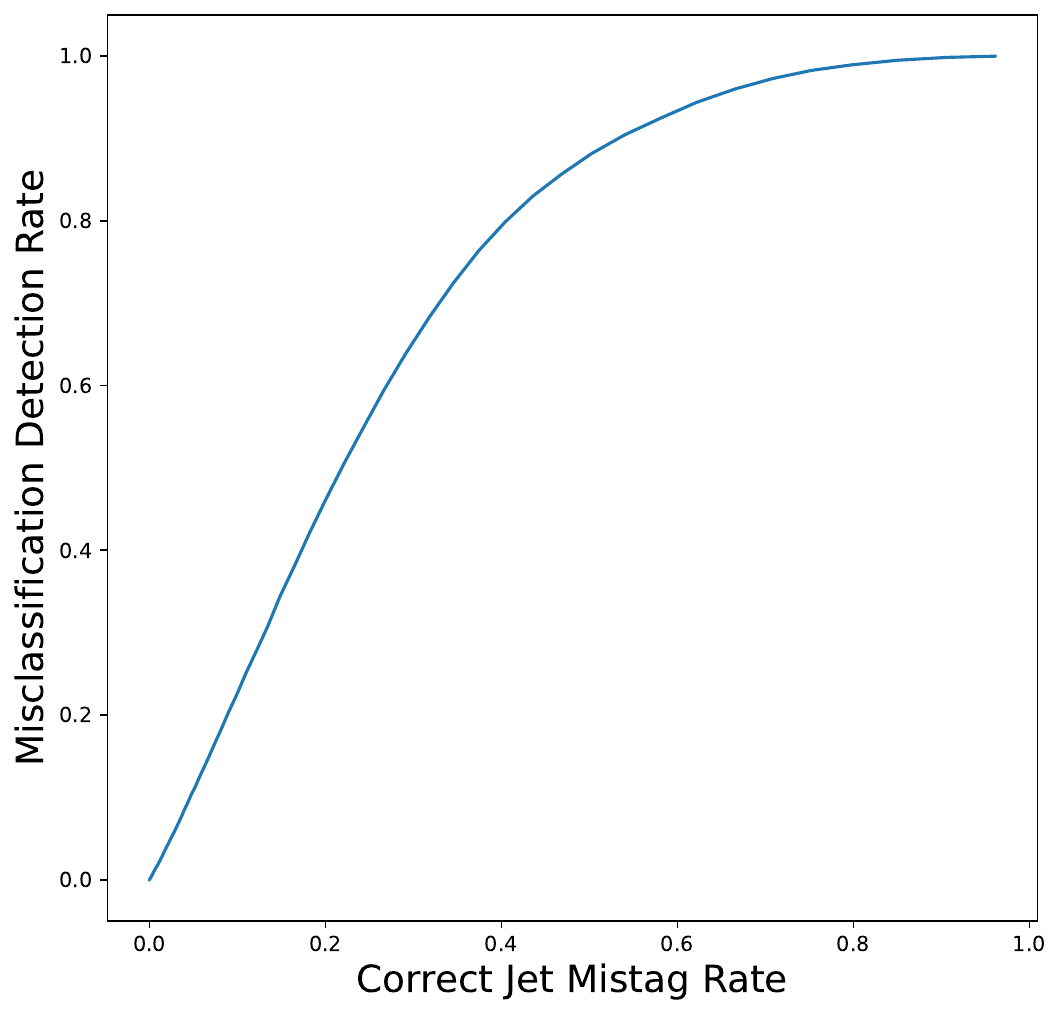}
\label{fig:jetnet-roc}            
}
\subfloat[]{
\includegraphics[width=0.49\textwidth]{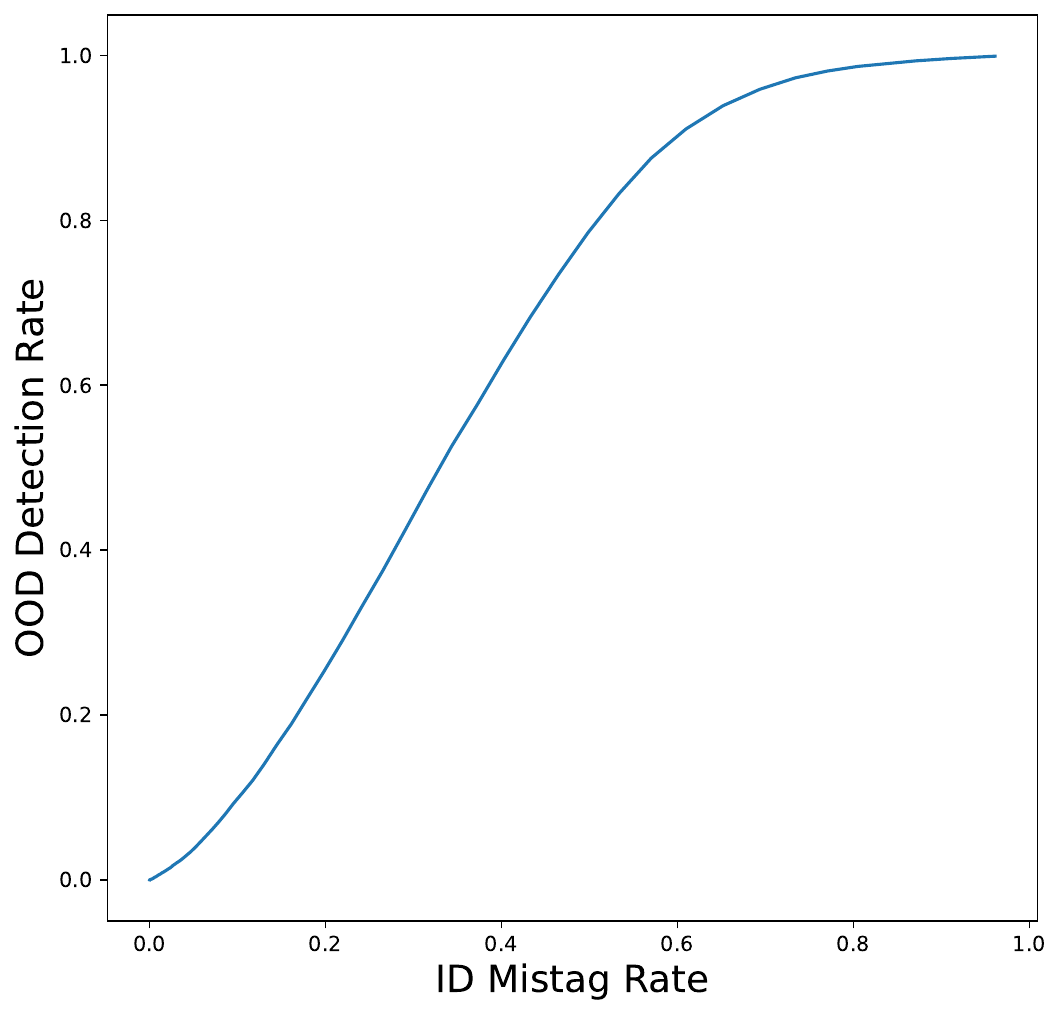}
\label{fig:jetnet-skiptop-roc}            
}
\caption{Receiver operative characteristic (ROC) curves for benchmark EDL models trained on the \protect\subref{fig:jetnet-roc} baseline \JetNet dataset for UQ and \protect\subref{fig:jetnet-skiptop-roc} \texttt{JetNet-skiptop} dataset for AD.}
\label{fig:roc-curve}
\end{figure}

\vspace{0.1in}

\item \textbf{AUROC-STD}: For a Dirichlet distribution with $K$ parameters $[\alpha_1, ..., \alpha_K]$, the Dirichlet standard deviation, D-STD $(\sigma_k)$ for the $k$-th class is given by

\begin{equation}
    \sigma_k= \sqrt{\frac{\alpha_k(S - \alpha_k)}{S^2(S+1)}}
    \label{eqn:d-std}
\end{equation}

where $S = \sum_{k=1}^{K}\alpha_k$ is Dirichlet strength defined in Section~\ref{sec:review-edl}. The quantity $\sigma_k$ as introduced in Eqn.~\ref{eqn:d-std} is a representative of uncertainty associated with the $k$-th class prediction.

\vspace{0.1in}

The Area Under the Receiver Operating Characteristic Curve, using D-STD as uncertainty (AUROC-STD), is similar to AUROC. However, it can only be used on EDL models because only they predict a Dirichlet distribution. We use this metric to compare with AUROC and determine if the D-STD or uncertainties from Ref.~\cite{10.5555/3327144.3327239} are better estimates for UQ and anomaly detection. Since the D-STD predicts uncertainties per class, we use
\begin{equation}
    u_{\text{D-STD}} = \sum_{k=1}^K \sigma_k
\label{eqn:u-dstd}
\end{equation}
as a conservative estimate of total uncertainty associated with the classification. We chose to use linear summation of D-STD as uncertainties are correlated among various jet classes~\cite{Taylor1994}. While quadrature summation was also studied for combining uncertainties, it yielded similar results. The AUROC-STD is then computed using the ROC curve constructed by varying the thresholds on $u_{\text{D-STD}}$. For EDL models, this metric is valuable for assessing the effectiveness of D-STD in representing uncertainty in comparison to Equation~\ref{eqn:edl-3}. 
\end{itemize}

\section{Results on Uncertainty Quantification}
\label{sec:results}
 
In this section, we examine how EDL models perform for UQ on jet classification tasks for the three datasets introduced in Section~\ref{sec:data_n_exp}. Ideally, uncertainties should be high for misclassified jets and low for correctly classified jets in a well-trained model. We examine multiple hyperparameter optimizations for the annealing coefficient $\lambda_t (\zeta)$ in Equation \ref{eqn:edl-10}, comparing fixed or gradually increasing approaches. We observe that gradually increasing $\lambda_t$, as proposed by the authors of Ref.~\cite{10.5555/3327144.3327239}, ensures faster convergence of accuracy. Additionally, we introduce a "Confidence Tuned" variant of the EDL method (EDL-CT) in Section~\ref{sec:jetnet}. This variant initially converges without annealing ($\lambda_t=0$), followed by parameter tuning through retraining with $\lambda_t > 0$. For EDL models, we also examine the use of the D-STD as uncertainty, referenced in Eqn.~\ref{eqn:u-dstd}. 

\subsection{Top tagging dataset}
\label{sec:topdata}

Since \topdata only contains two classes, it is the simplest dataset to investigate the  uncertainty generated from EDL. The performance of EDL model variants in the context of \topdata is given in Table~\ref{tab:baseline}. The model accuracy is found to be very similar for different choices of EDL coefficients, depicting how the introduction of EDL for UQ does not interfere with the decision-making ability of the classifier model for this dataset. It is obseved that higher $\zeta$ values result in a larger AUROC, signifying better discriminative ability between correct and incorrect predictions. The results are summarized in the \topdata column of Table~\ref{tab:baseline}. We find that EDL $\lambda_t (1.0) $ has the largest AUROC and EDL $\lambda_t (0.7)$ exhibits similar performance for the \topdata dataset.

\begin{table}[!ht]
 \centering
 \resizebox{0.99\textwidth}{!}{
 \begin{tabular}{|l||c|c|c||c|c|c||c|c|c|}
    \hline
   & \multicolumn{3}{c||}{\topdata} & \multicolumn{3}{c||}{\jetnet} & \multicolumn{3}{c|}{\jetclass} \\ \hline \hline
    \textbf{Model} & Acc & AUC & STD & Acc & AUC & STD & Acc & AUC & STD  \\ \hline
    EDL $\lambda_t (0)$ & \textbf{0.937} & 0.723 & 0.894 & \textbf{0.803} & 0.550 & 0.792 & \textbf{0.794} & 0.602 & 0.816 \\
    EDL $\lambda_t (0.1)$ & \textbf{0.937} & 0.902 & 0.903 & 0.799 & 0.811 & 0.813 & 0.792 & 0.842 & 0.843 \\
    EDL $\lambda_t (0.5)$ & 0.936 & 0.902 & 0.902 & 0.796 & 0.815 & 0.816 & - & - & - \\
    EDL $\lambda_t (0.7)$ & \textbf{0.937} & \textbf{0.904} & \textbf{0.904} & 0.793 & 0.820 & 0.843 & - & - & - \\
    EDL $\lambda_t (1.0)$ & \textbf{0.937} & \textbf{0.904} & \textbf{0.904} & 0.790 & 0.822 & 0.823 & 0.776 & \textbf{0.847} & \textbf{0.847} \\
    EDL-CT $\lambda_t^{\texttt{CT}} (0.1)$ & - & - & - & 0.801 & 0.814 & 0.815 & - & - & - \\
    EDL-CT $\lambda_t^{\texttt{CT}} (0.5)$ & - & - & - & 0.788 & 0.831 & 0.832 & - & - & - \\
    EDL-CT $\lambda_t^{\texttt{CT}} (0.7)$ & - & - & - & 0.776 & \textbf{0.843} & \textbf{0.843} & - & - & - \\ \hline \hline
    Ensemble & 0.937 & 0.890 & - & 0.806 & 0.772 & - & 0.805 & 0.782 & -  \\ \hline
    MC Dropout & 0.933 & 0.887 & - & 0.797 & 0.743 & - & 0.793 & 0.745 & -  \\
    \hline
  \end{tabular}
  }
  \caption{ID Accuracy (Acc), AUROC (AUC), and AUROC-STD (STD) of the EDL and Ensemble methods on \topdata, \jetnet, and \jetclass datasets. Within \topdata and \jetnet models, the Ensemble model has 970k parameters, while all other models have 97k parameters. The \jetclass Ensemble model has 994k parameters, while all other \jetclass models have 99k parameters. For each dataset, the entries marked in \textbf{bold} represent the EDL model with the highest central value for the corresponding metric. These measurements have uncertainties of $\mathcal{O}(0.001)$.}
  \label{tab:baseline}
\end{table}

\begin{figure}[htbp]

\centering
    \subfloat[]{
    \includegraphics[width=0.3059\textwidth]{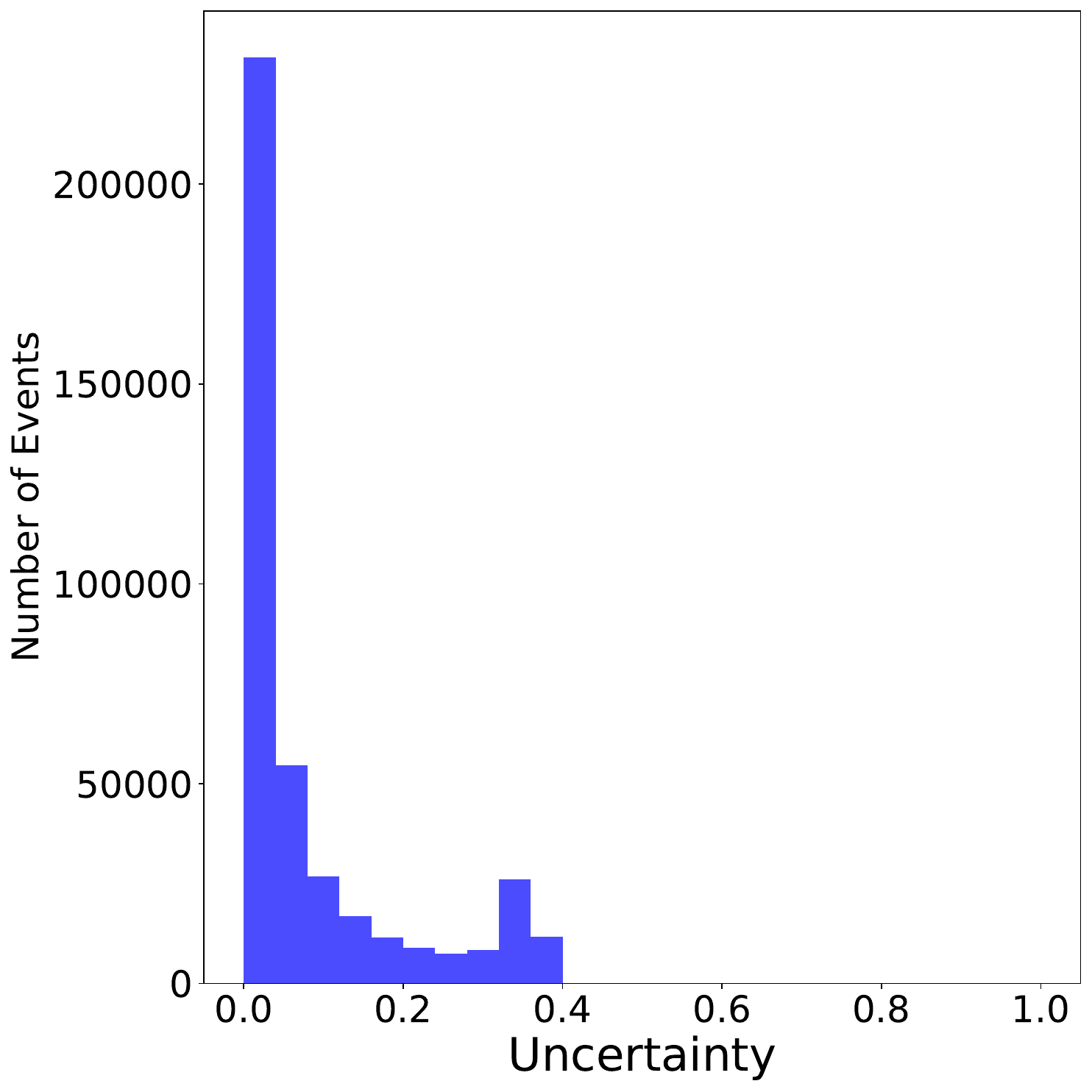}
    \label{fig:topdata_baseline_0.1_unc_total}
    }
    \subfloat[]{
    \includegraphics[width=0.3059\textwidth]{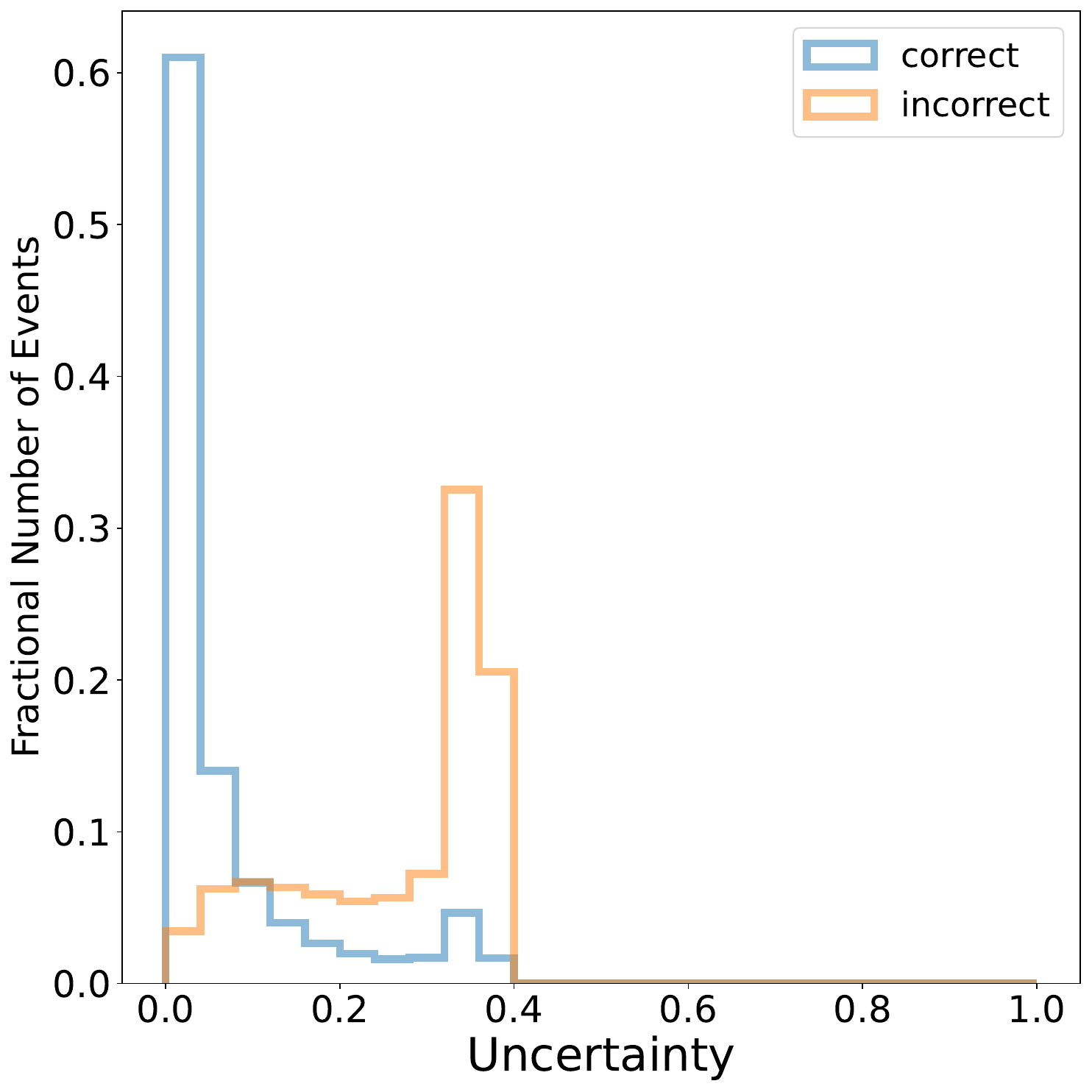}
    \label{fig:topdata_baseline_0.1_unc}
    }
    \hspace*{-0.2cm}
    \subfloat[]{
    \raisebox{-0.15cm}{
    \includegraphics[width=0.3591\textwidth]{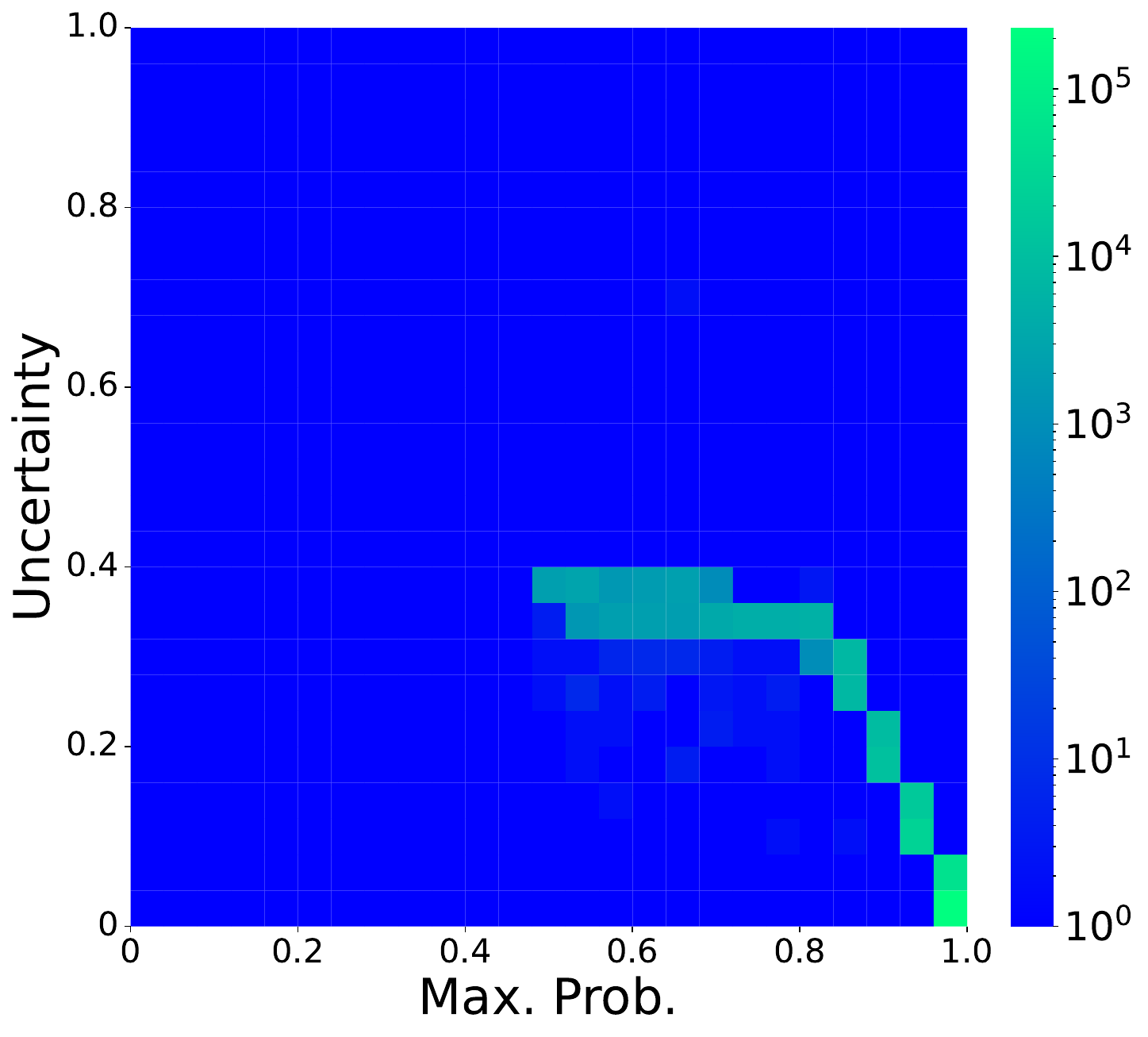}
    \label{fig:topdata_baseline_0.1_up}
    }
    }

    \subfloat[]{
    \includegraphics[width=0.3059\textwidth]{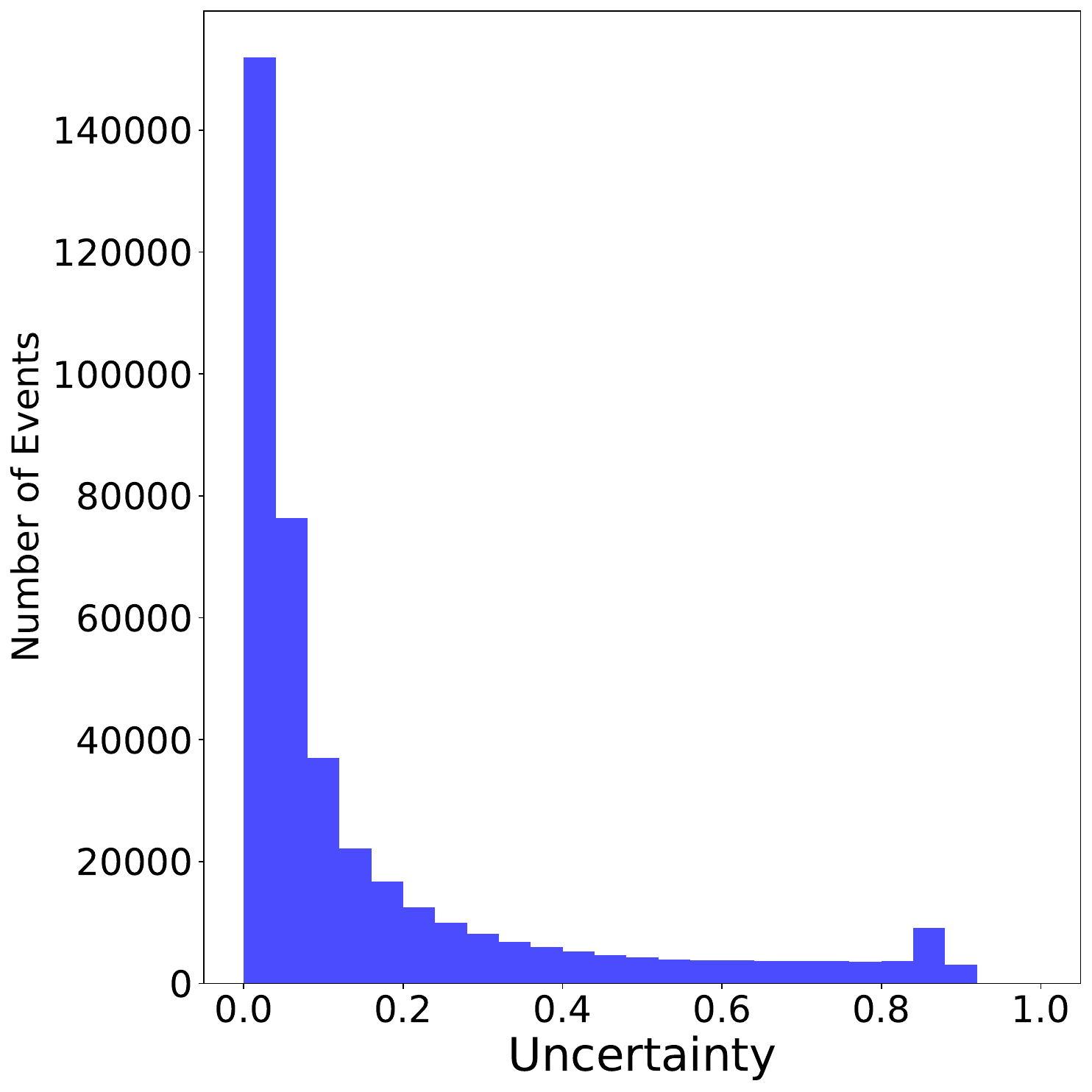}
    \label{fig:topdata_baseline_0.7_unc_total}
    }
    \subfloat[]{
    \includegraphics[width=0.3059\textwidth]{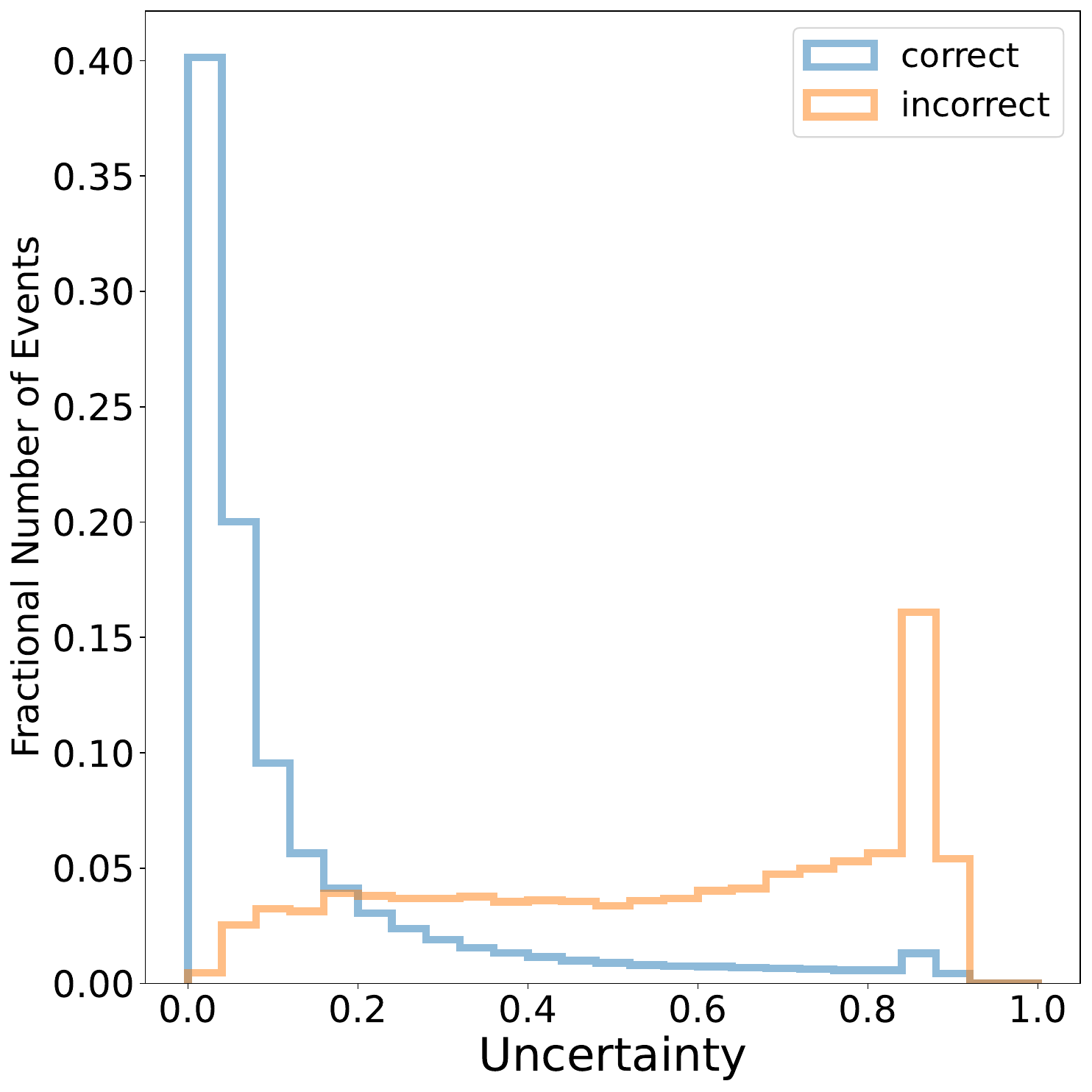}
    \label{fig:topdata_baseline_0.7_unc}
    }
    \hspace*{-0.2cm}
    \subfloat[]{
    \raisebox{-0.15cm}{
    \includegraphics[width=0.3591\textwidth]{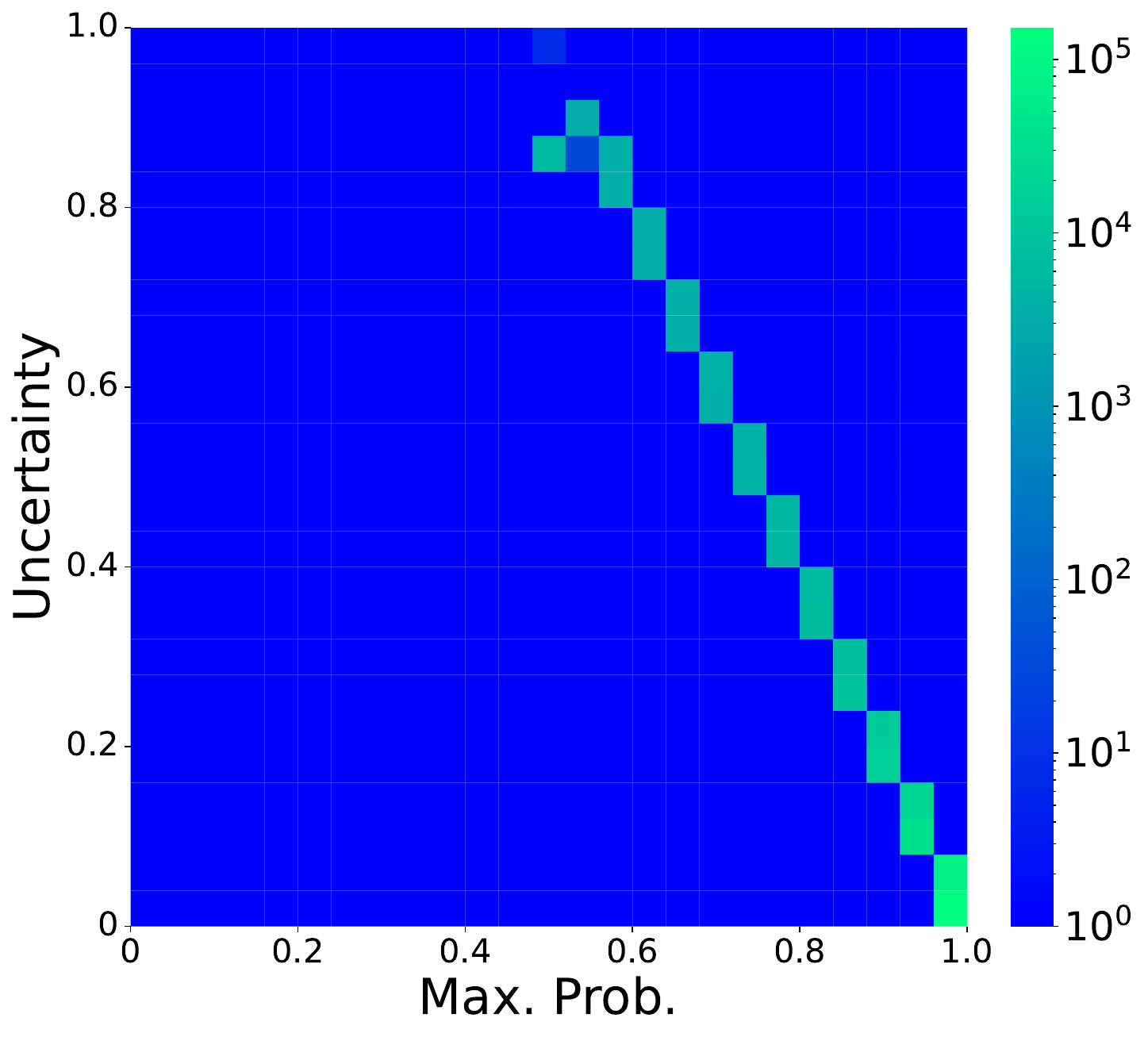}
    \label{fig:topdata_baseline_0.7_up}
    }
    }
    
    \caption{For the \topdata dataset, on each row, from left to right, \protect\subref{fig:topdata_baseline_0.1_unc_total},\protect\subref{fig:topdata_baseline_0.7_unc_total} uncertainty distribution, \protect\subref{fig:topdata_baseline_0.1_unc},\protect\subref{fig:topdata_baseline_0.7_unc} logarithmic uncertainty distributions separated by correct and incorrect jets,
    and \protect\subref{fig:topdata_baseline_0.1_up},\protect\subref{fig:topdata_baseline_0.7_up}
    2D histogram of maximum probability versus uncertainty, 
    for baseline EDL $\lambda_t (0.1) $ (top row) and $\lambda_t (0.7) $ (bottom row).
    } 
    \label{fig:topdata_unc}
\end{figure}

Figure~\ref{fig:topdata_unc} shows the impact of the choice of $\zeta$ as a hyperparameter for the choice of the model. As shown in Figures~\ref{fig:topdata_baseline_0.1_unc_total}~and~\ref{fig:topdata_baseline_0.7_unc_total}, the distribution of the total uncertainty as obtained from these models shows a strong dependence on the choice of the regularization scale of the EDL model. Respectively choosing $\zeta = 0.1$ and $\zeta = 0.7$ for these two models, there are two distinct peaks in the uncertainty distribution. Figure~\ref{fig:topdata_baseline_0.1_unc}~and~\ref{fig:topdata_baseline_0.7_unc}  provide the uncertainty distributions separated for the correctly and incorrectly classified jets generated by the same models. In both instances, smaller uncertainties are attributed to correctly classified jets while the misclassified jets tend to be assigned larger uncertainties.  

A general trend with EDL models is that as the $\zeta$ parameter increases in $\lambda_t (\zeta)$, there are more high-uncertainty jets, which is shown in Figure~\ref{fig:topdata_baseline_0.1_unc}~and~\ref{fig:topdata_baseline_0.7_unc}. This corresponds to higher uncertainties in both correctly classified and misclassified jets. The cause can be attributed to the loss function. The $\zeta$ parameter is used to regulate the magnitude of the KL-divergence loss, which diverges away from a uniform Dirichlet distribution when misclassification takes place.  When $\zeta=0$, the Dirichlet parameters of the correct label keep increasing whenever the prediction is correct to minimize the loss resulting in an overly confident prediction. However, as $\zeta$ increases, the regularizing KL-divergence term takes more priority, penalizing the divergences from the "I do not know" state. Then, the EDL model with larger $\zeta$ will have smaller Dirichlet parameters and high uncertainties as opposed to an EDL model with $\lambda_t (0)$. This can also be seen in the distribution of uncertainty as a function of the largest assigned probability (i.e. Max. Prob) in Figures~\ref{fig:topdata_baseline_0.1_up}~and~\ref{fig:topdata_baseline_0.7_up}. As seen in Figure~\ref{fig:topdata_baseline_0.1_up}, the uncertainty distribution hits a plateau close to the value of~0.4 as an artifact of the training with a weaker constraint on the DL-divergence term in the EDL loss function. 
On the other hand, EDL $ \lambda_t (0.7)$ in Figure~\ref{fig:topdata_baseline_0.7_up} conforms with the general expectations from a well-trained uncertainty-aware classifier, that is (a) a general inverse relationship between Max. Prob and uncertainty and (b) a high concentration of correctly classified events in the low uncertainty bins. Since EDL $\lambda_t (0.7)$ has the highest AUROC of any EDL model, this log-linear relationship indicates better misclassification prediction for this simple binary classification dataset.

Since the EDL model predicts the parameters of a Dirichlet distribution, we can also examine the D-STD as a measure of uncertainty in the top tagging dataset. As stated previously, the AUROC-STD is AUROC but with D-STD uncertainty. As shown in the \topdata column of Table~\ref{tab:baseline}, there is no significant difference between the AUROC and AUROC-STD scores for $\zeta > 0$.
\subsection{JetNet dataset}
\label{sec:jetnet}

In contrast to the binary classification of the \topdata dataset, \JetNet has five distinct classes of jets, giving a more comprehensive overview of how the EDL uncertainty behaves in a multiclass scenario. The \JetNet dataset contains the following jets with their corresponding class labels: quarks (0), gluons (1), top quarks (2), $W$ bosons (3), and $Z$ bosons (4). As shown in the \JetNet column in Table~\ref{tab:baseline}, EDL models with higher $\zeta$ tend to have marginally lower accuracy but higher AUROC and AUROC-STD. This implies that EDL models with higher $\zeta$ make more incorrect predictions but tend to assign commensurately larger uncertainties to them.  

\begin{figure}[htbp]

\centering

    \subfloat[]{
        \includegraphics[width=0.3059\textwidth]{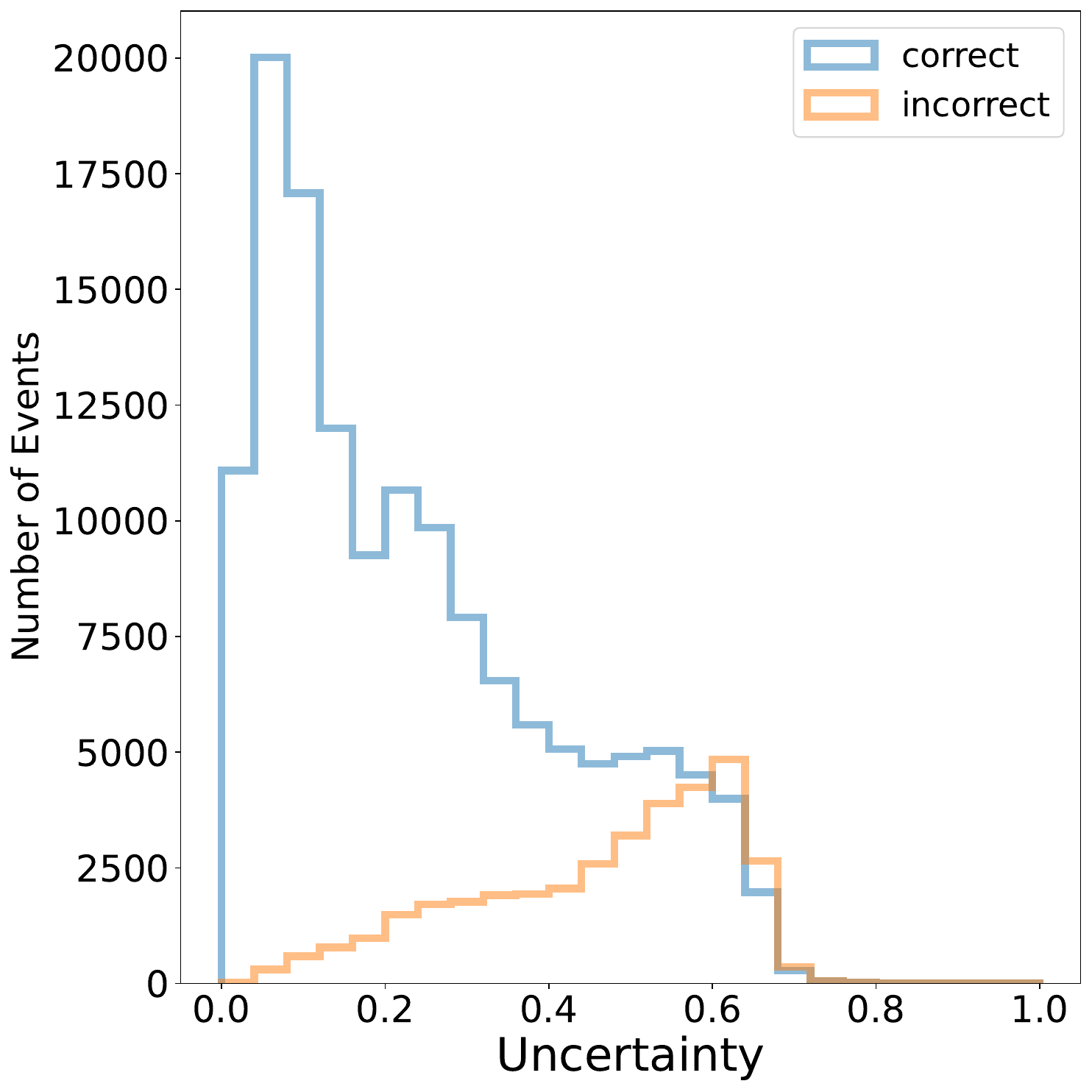}
        \label{fig:jetnet_0.1_baseline_us}
    }
    \subfloat[]{
        \includegraphics[width=0.3059\textwidth]{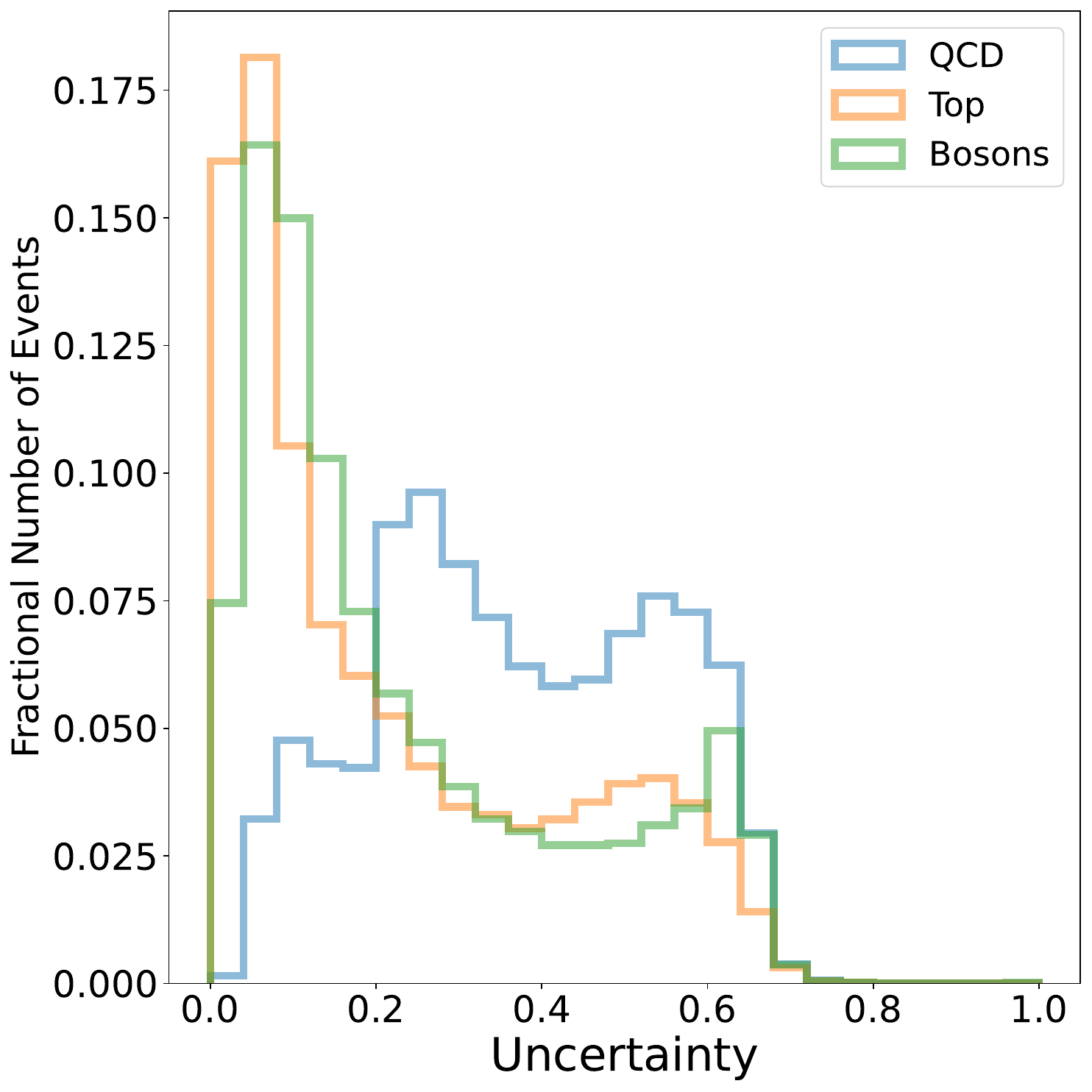}
        \label{fig:jetnet_0.1_baseline_unc_class}
    }
    \hspace*{-0.2cm}
    \subfloat[]{
        \raisebox{-0.15cm}{
        \includegraphics[width=0.3591\textwidth]{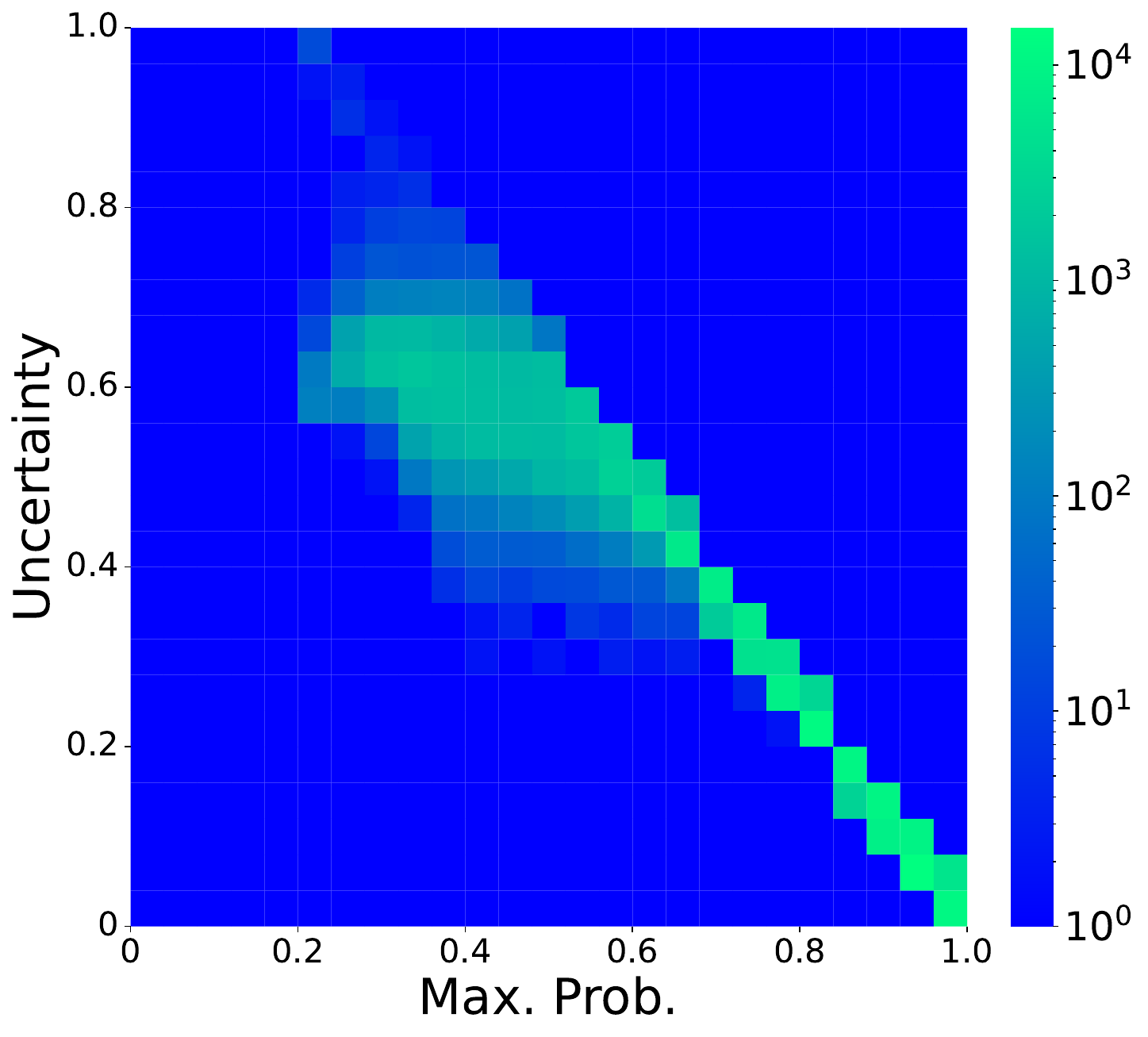}
        }
        \label{fig:jetnet_0.1_baseline_up}
    }

    \subfloat[]{
        \includegraphics[width=0.3059\textwidth]{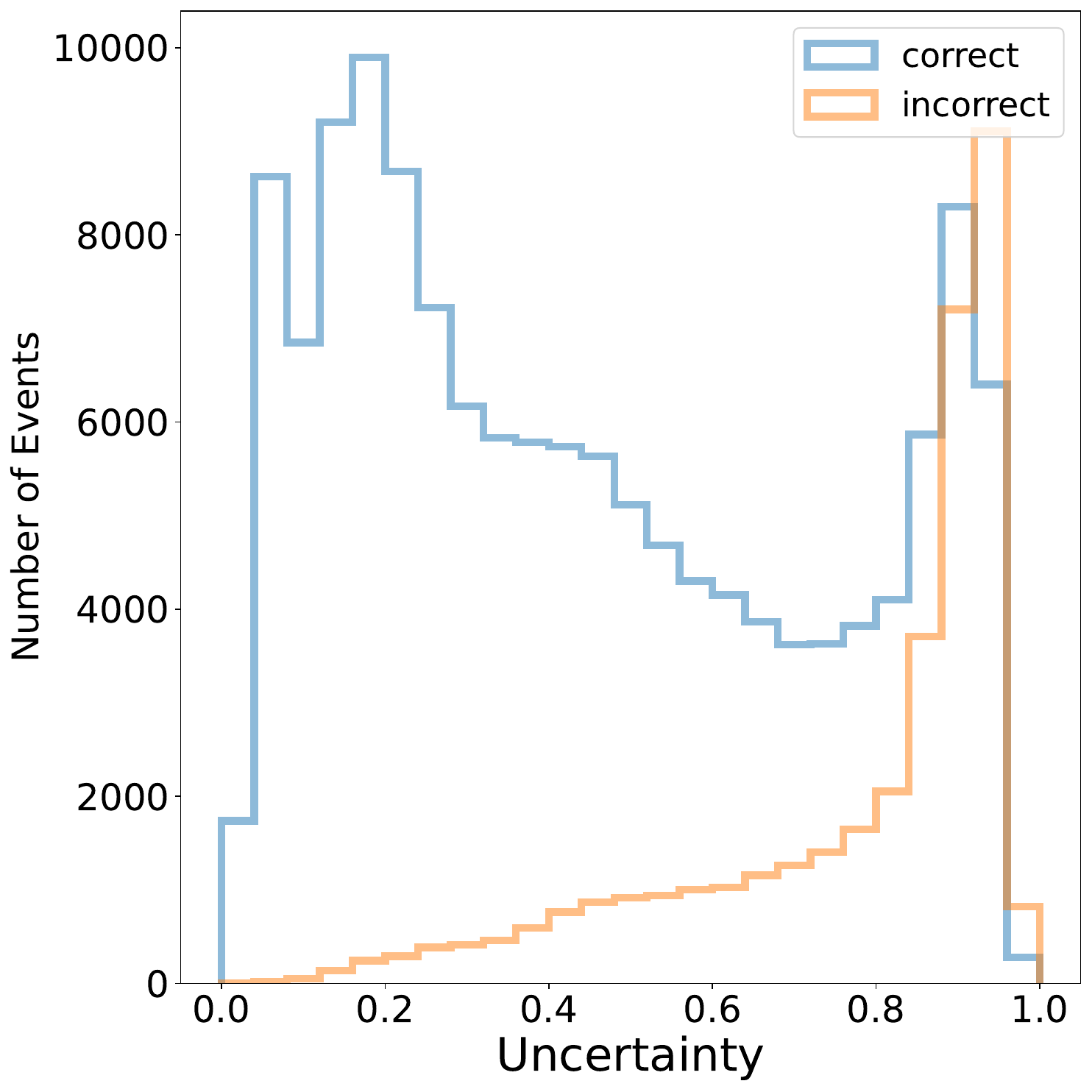}
        \label{fig:jetnet_0.7_baseline_us}
    }
    \subfloat[]{
        \includegraphics[width=0.3059\textwidth]{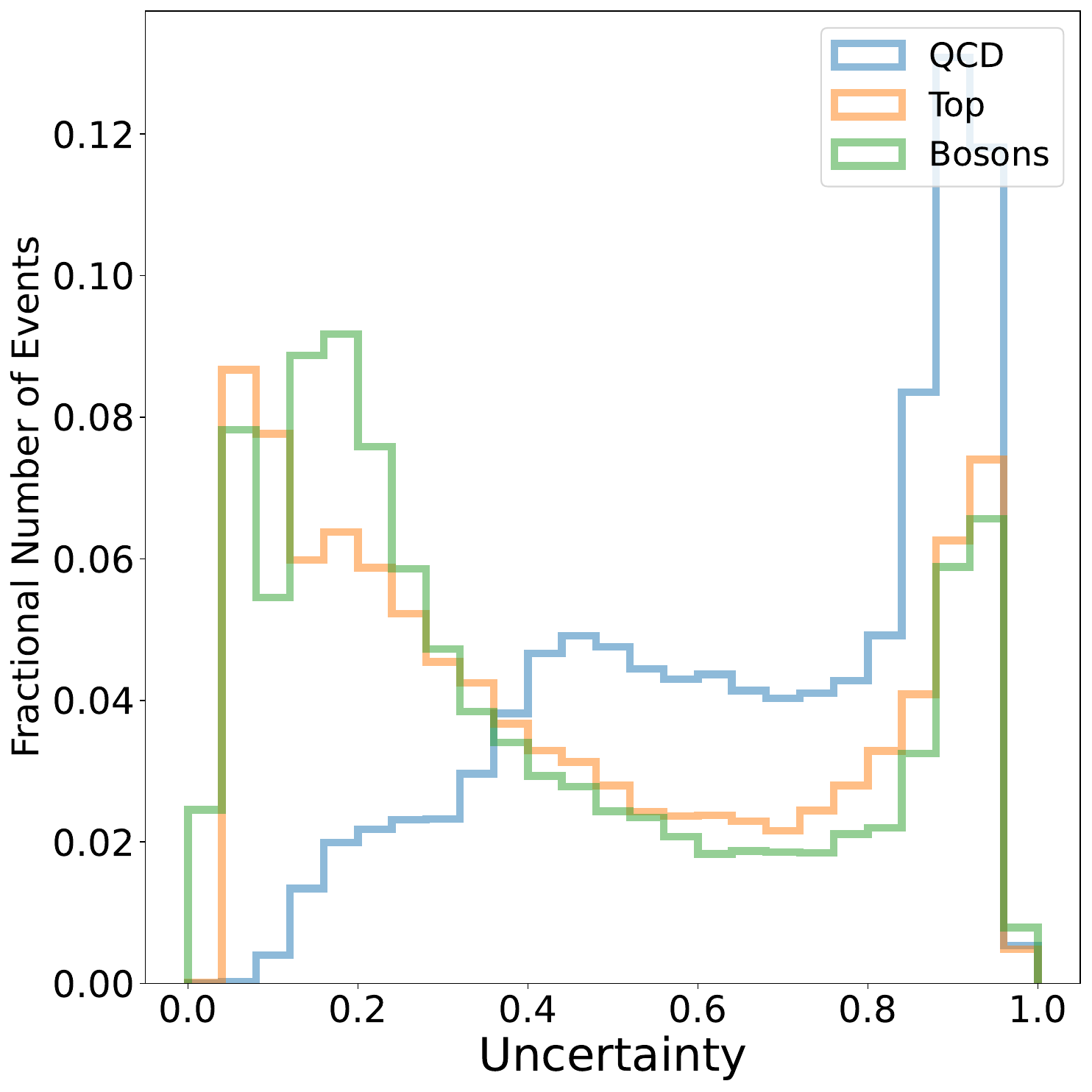}
        \label{fig:jetnet_0.7_baseline_unc_class}
    }
    \hspace*{-0.2cm}
    \subfloat[]{
        \raisebox{-0.15cm}{
        \includegraphics[width=0.3591\textwidth]{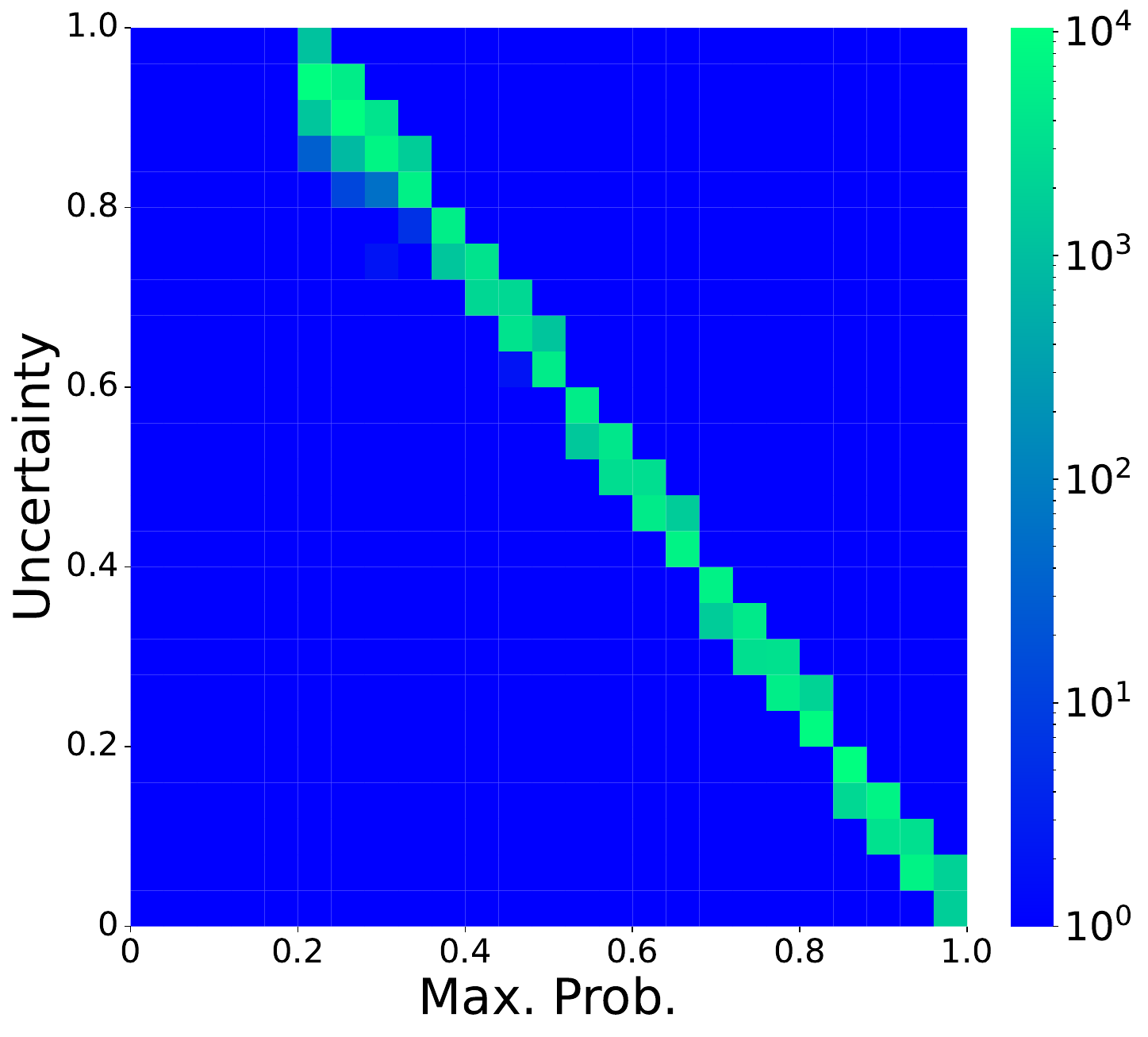}
        }
        \label{fig:jetnet_0.7_baseline_up}
    }
   
    \caption{For the \jetnet dataset, on each row, from left to right, \protect\subref{fig:jetnet_0.1_baseline_us},\protect\subref{fig:jetnet_0.7_baseline_us} uncertainty distribution, separated by correctly and incorrectly classified jets, \protect\subref{fig:jetnet_0.1_baseline_unc_class},\protect\subref{fig:jetnet_0.7_baseline_unc_class} uncertainty distribution for correctly classified jets, separated by initiating particle jet type, and \protect\subref{fig:jetnet_0.1_baseline_up},\protect\subref{fig:jetnet_0.7_baseline_up}
    2D histogram of maximum probability versus uncertainty for baseline EDL $\lambda_t (0.1) $ (top row) and $\lambda_t (0.7) $ (bottom row).
    }
    \label{fig:jetnet_unc}
\end{figure}

To understand why ID accuracy decreases and AUROC increases as $\zeta$ increases, we examine the uncertainties of baseline \JetNet EDL $\lambda_t (0.1)$ and $\lambda_t (0.7)$ in Figures~\ref{fig:jetnet_0.1_baseline_us}~and~\ref{fig:jetnet_0.7_baseline_us}, respectively. Similarly to our observations for the EDL models applied to the \topdata dataset, the range of uncertainties and proportion of high uncertainty jets for \JetNet EDL models grows as the $\zeta$ increases. The uncertainties for EDL $\lambda_t (0.1)$ still have a bimodal distribution associated with correctly classified jets at low uncertainties and misclassified jets at high uncertainties. But for EDL $\lambda_t (0.7) $, there are a large number of correctly classified jets with higher uncertainties. 

\begin{figure}[htbp]

\centering

    \subfloat[]{
        \includegraphics[width=0.5\textwidth]{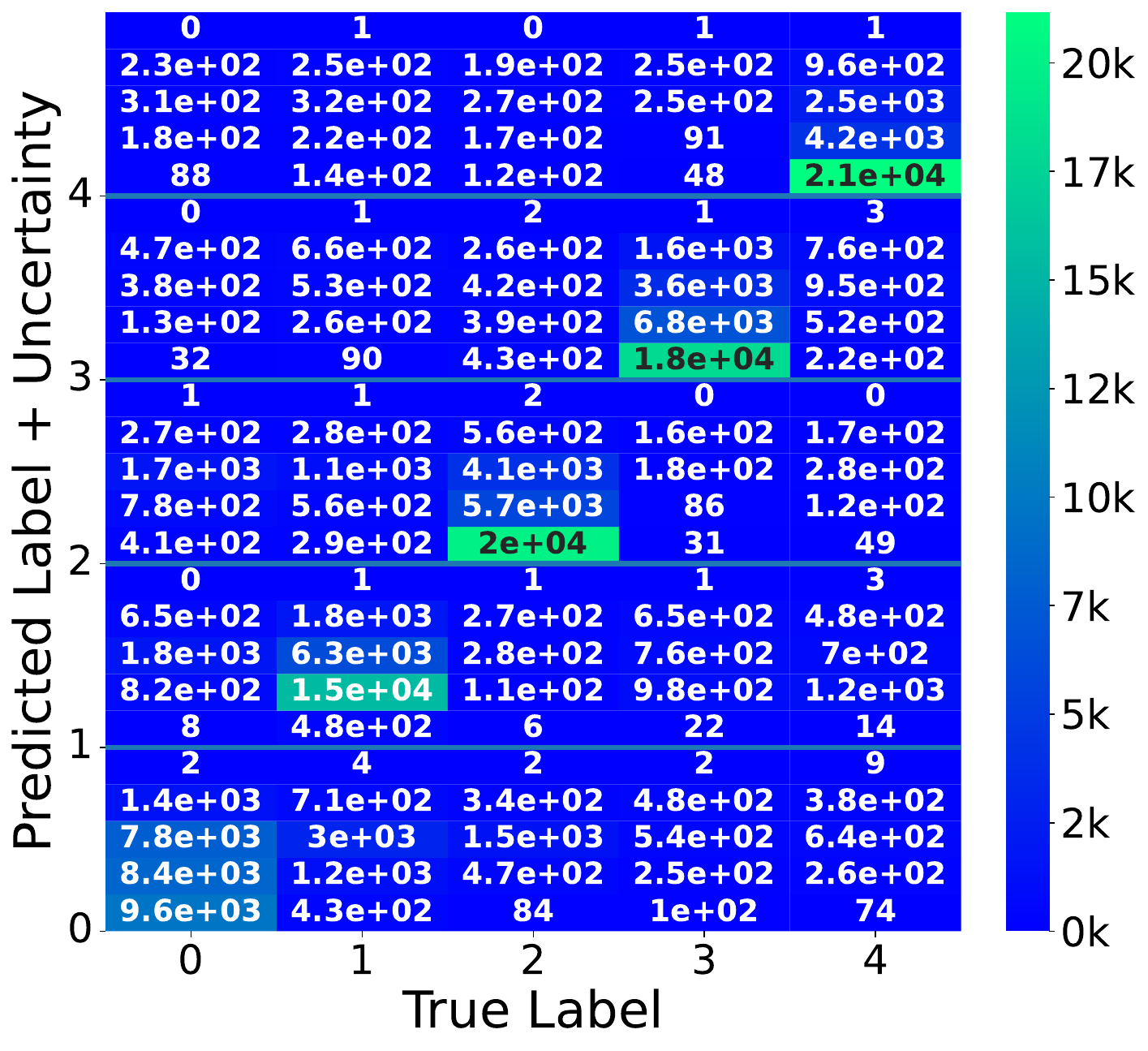}
    \label{fig:jetnet_0.1_baseline_lpu}
    }
    \subfloat[]{
        \includegraphics[width=0.5\textwidth]{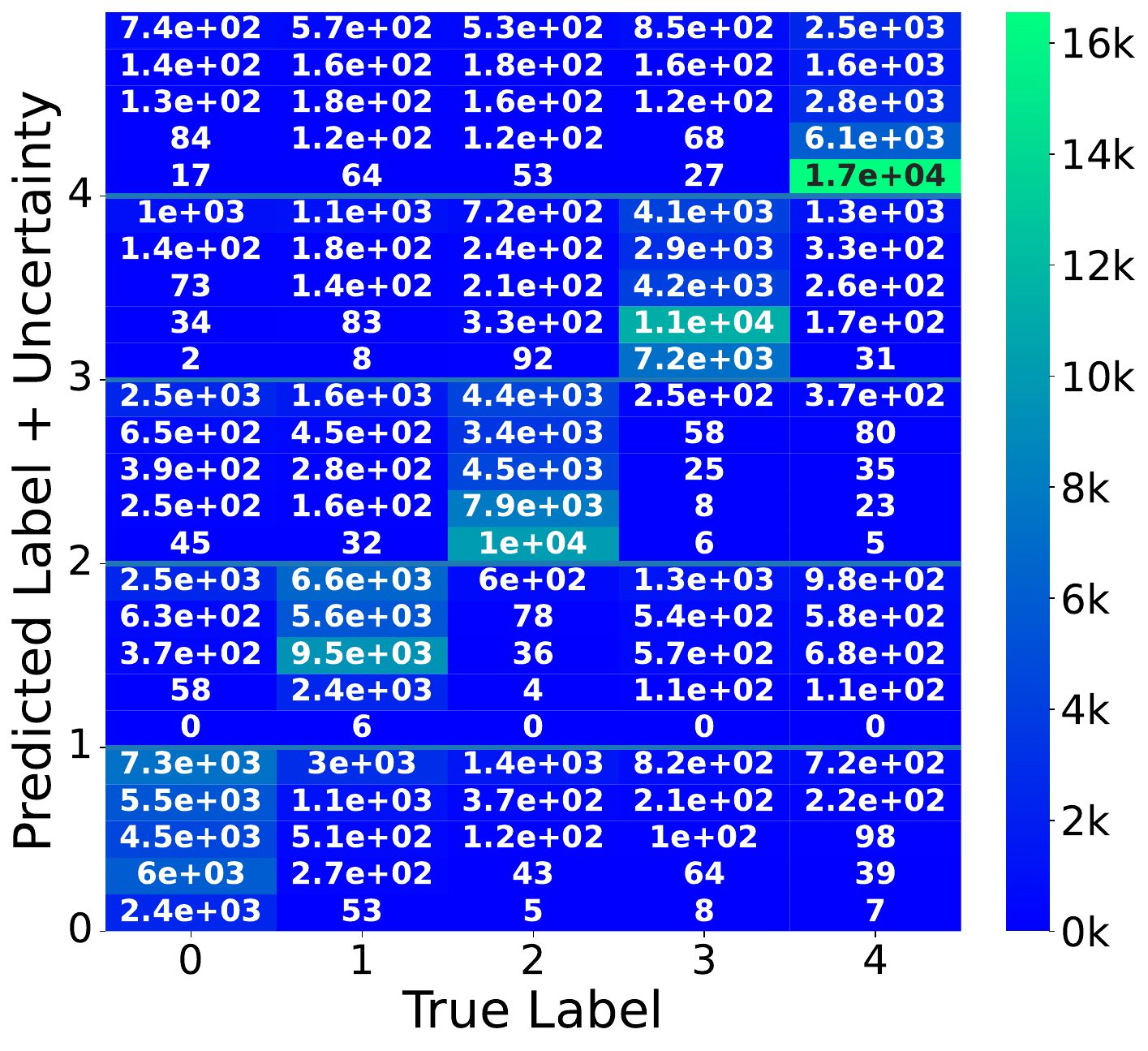}
    \label{fig:jetnet_0.7_baseline_lpu}
    }
    
    \caption{Uncertainty Aware Confusion Matrix for, respectively, baseline \jetnet EDL \protect\subref{fig:jetnet_0.1_baseline_lpu} $\lambda_t (0.1) $ and \protect\subref{fig:jetnet_0.7_baseline_lpu} $\lambda_t (0.7) $.
    }
    \label{fig:jetnet_uacm}
\end{figure}

We visualize the uncertainties for each label and prediction through the Uncertainty Aware Confusion Matrix (UACM), as displayed in Figure~\ref{fig:jetnet_uacm}. The UACM is an extension of the traditional confusion matrix that incorporates uncertainty information for each prediction. The $y$-axis represents a binned distribution of predicted label plus uncertainty, which has a maximum of one, so it can display the general uncertainty distributions for correctly classified and misclassified jets. For both choices of $\zeta$, correctly classified quark and gluon jets with respective labels of 0 and 1 tend to have higher uncertainties.

As depicted in Figures~\ref{fig:jetnet_0.1_baseline_unc_class}~and~\ref{fig:jetnet_0.7_baseline_unc_class}, the high-uncertainty, correctly-classified jets are dominated by QCD jets, while the heavier jets usually have lower uncertainties. This gives us an interesting insight into how the EDL models behave when two or more classes within the training dataset have similar physical characteristics. It is well known that jets initiated by quarks ($q$) and gluons ($g$) have very similar characteristics (being from the fragmentation of particles with color charge) and are generally regarded as \textit{hard-to-tell-apart} (HTA)~\cite{Gallicchio_2013}. In fact, many LHC physics analyses either combine them together as a single jet class of light or QCD jets or employ sophisticated taggers developed specifically for $q$/$g$ separation~\cite{PhysRevLett.107.172001,ATLAS:2023dyu}. This challenge of telling apart quark and gluon jets from their observed characteristics is large uncertainties assigned to these jets for higher values of $\zeta$ even when the model learns to correctly classify them. By increasing $\zeta$, the model penalizes divergences from the "I do not know" state. In both models, as shown in Figures~\ref{fig:jetnet_0.1_baseline_lpu}~and~\ref{fig:jetnet_0.7_baseline_lpu}, the relationship between uncertainty and maximum probability is similar as found in case of EDL models applied the \topdata dataset. However, unlike what we observed for the \topdata dataset, the performance of the model does not necessarily improve with larger $\zeta$. Models with large $\zeta$ show high uncertainty association for incorrectly classified jets at the expense of reduced confidence in correctly classified jets.

In an effort to circumnavigate this issue of large penalties for HTA jets, we introduce an alternative (hybrid) training paradigm. We refer EDL models trained with this paradigm as EDL-CT models. For the first 30 epochs, this model was trained with $\lambda_t = 0$  and the EDL regularization is restored with a nonzero constant after 30 epochs:
\begin{equation}
    \lambda_t^{\texttt{CT}}(\zeta) = 
    \begin{cases}
        0,  \text{for first 30 epochs of training} \\
        \zeta,  \text{for the remaining epochs}
    \end{cases}
\end{equation}

\begin{figure}[htbp]

\centering

    \subfloat[]{
        \includegraphics[width=0.3059\textwidth]{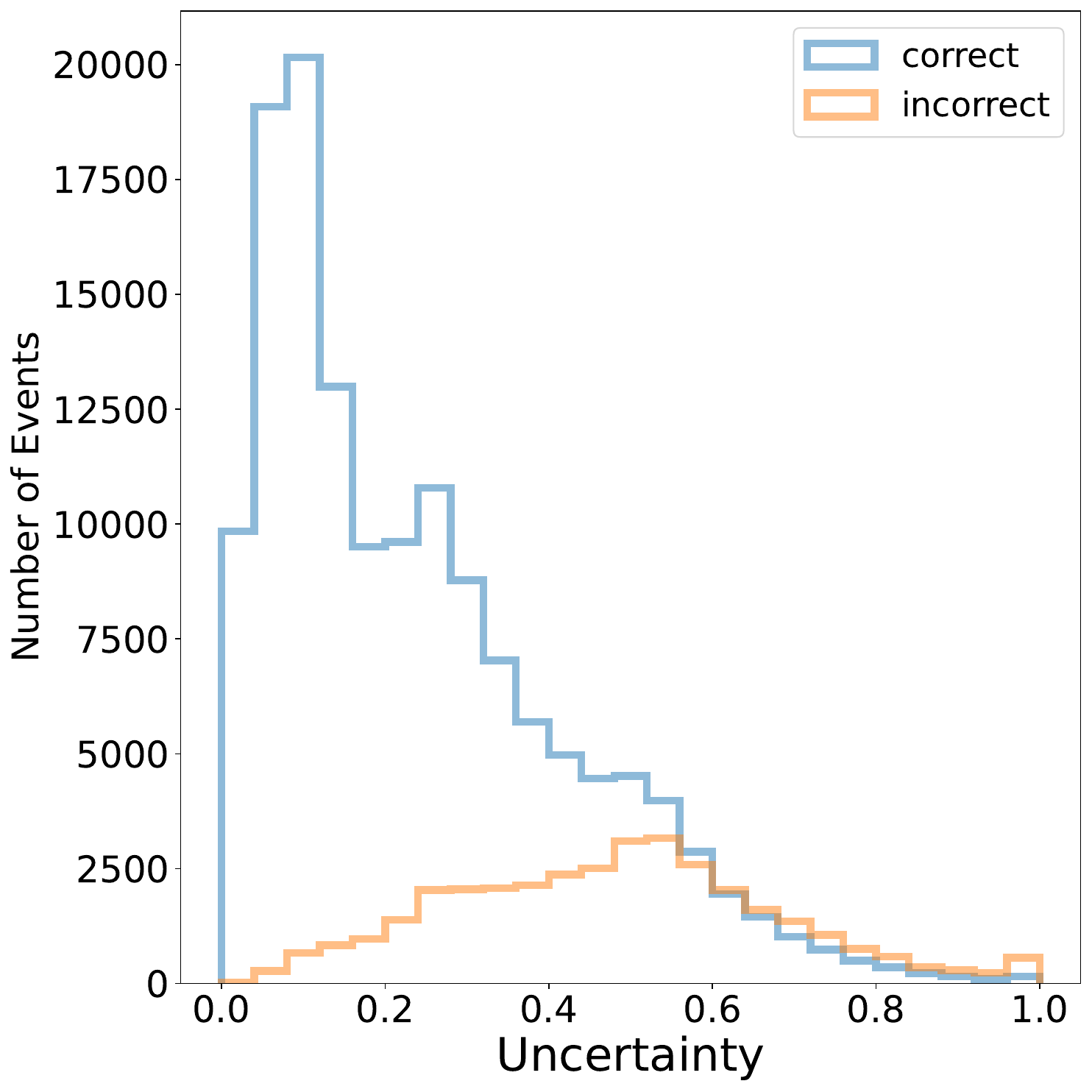}
        \label{fig:jetnet_0.1_baseline_us_oc}
    }
    \subfloat[]{
        \includegraphics[width=0.3059\textwidth]{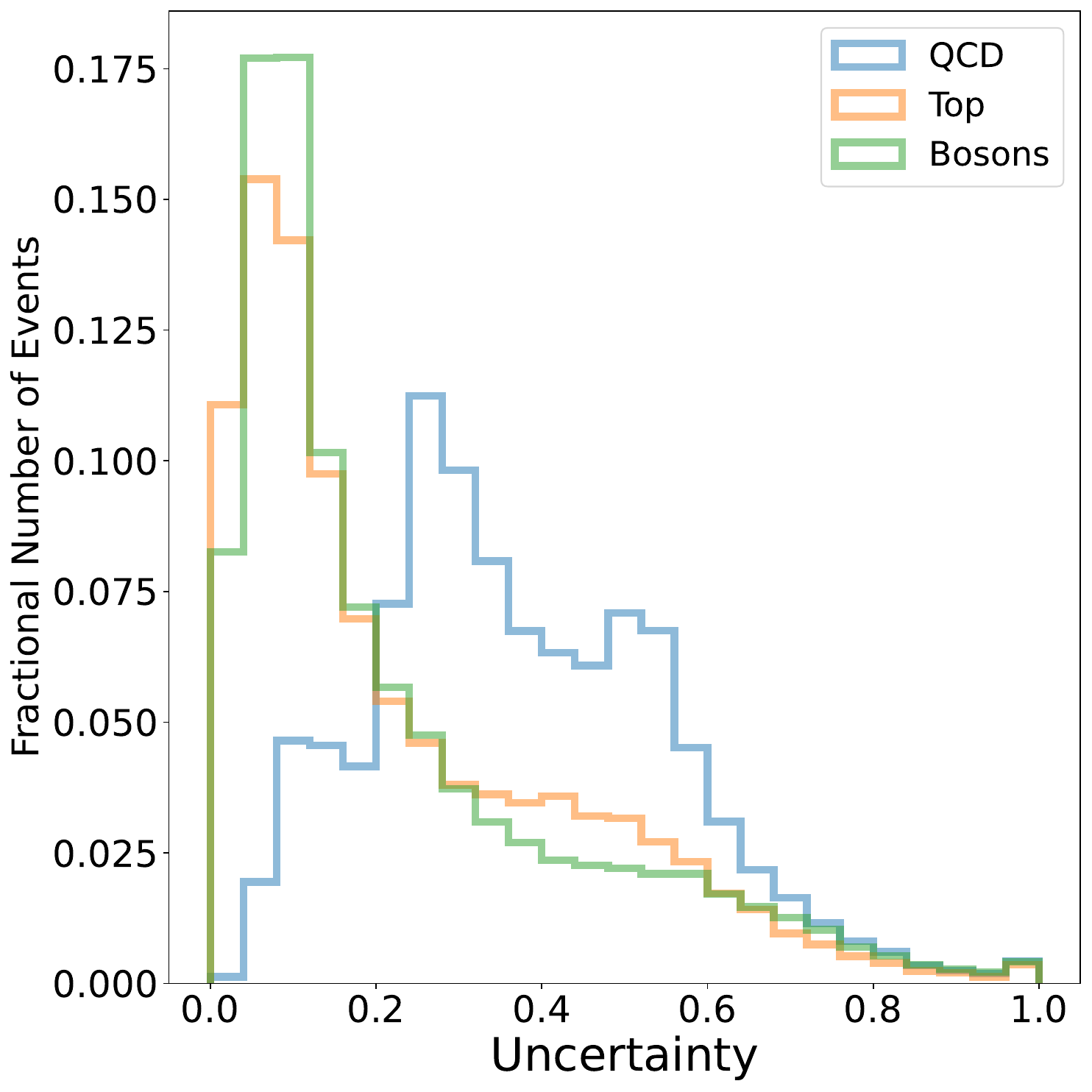}
        \label{fig:jetnet_0.1_baseline_unc_class_oc}
    }
    \hspace*{-0.2cm}
    \subfloat[]{
        \raisebox{-0.15cm}{
        \includegraphics[width=0.3591\textwidth]{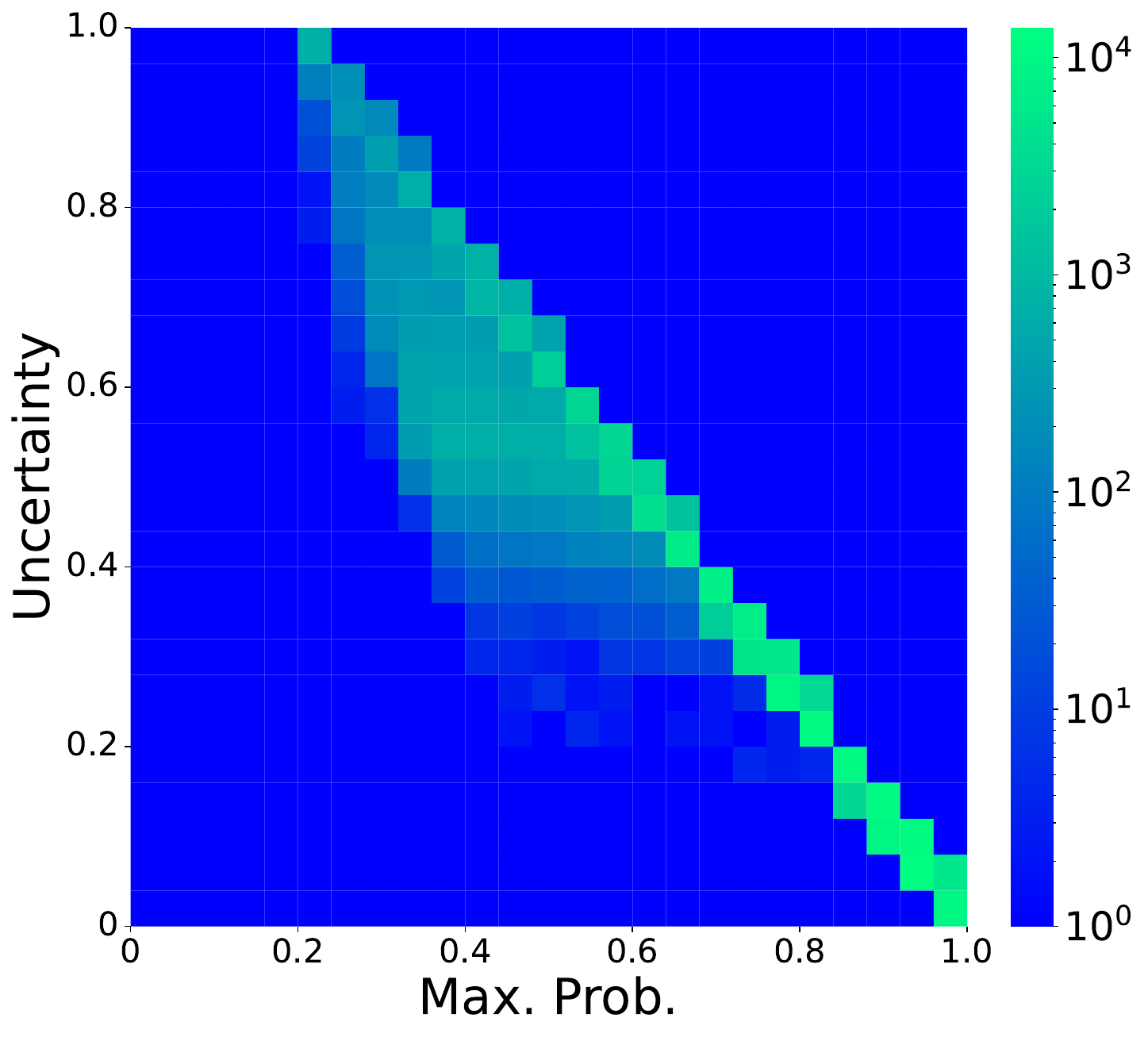}
        }
        \label{fig:jetnet_0.1_baseline_up_oc}
    }

    \subfloat[]{
        \includegraphics[width=0.3059\textwidth]{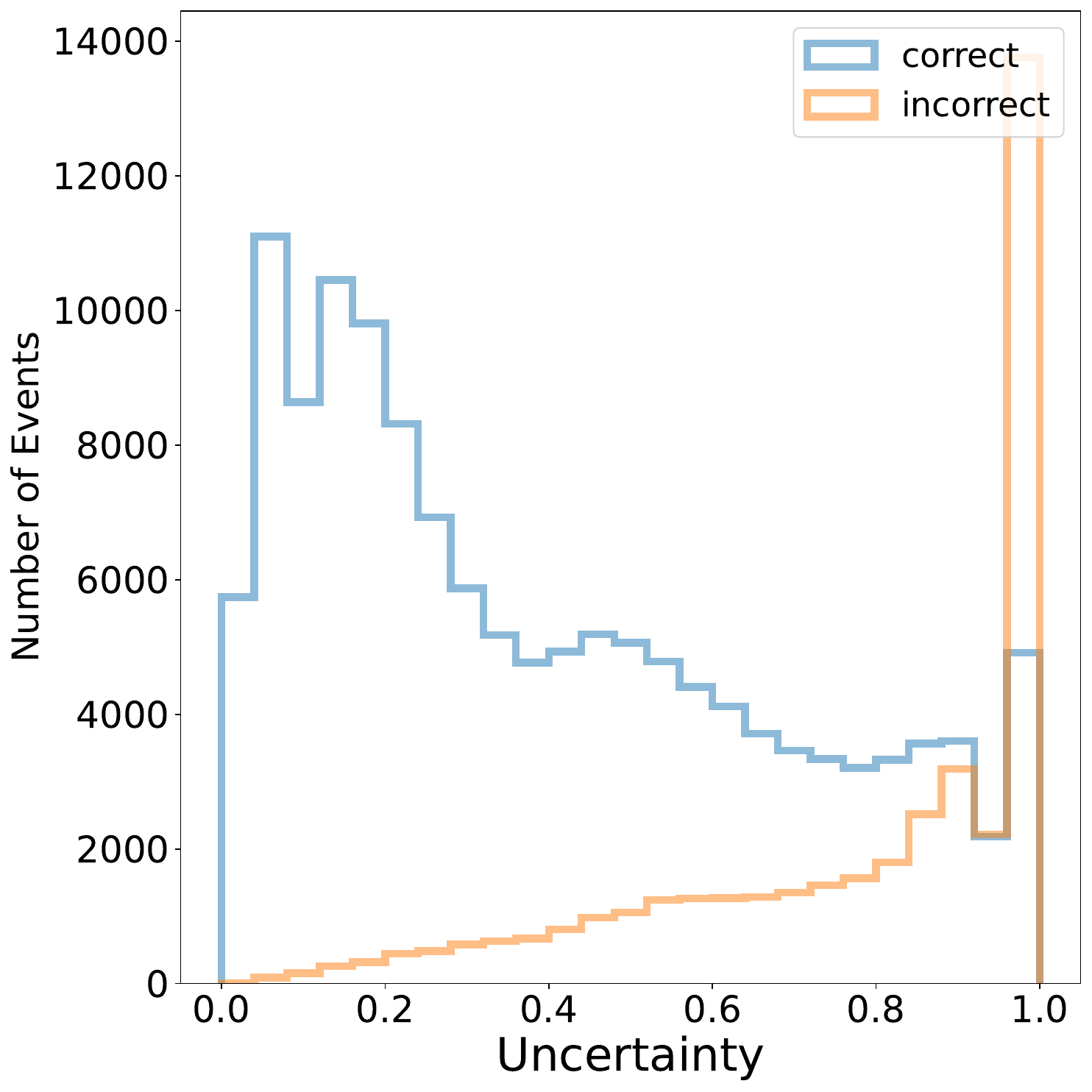}
        \label{fig:jetnet_0.7_baseline_us_oc}
    }
    \subfloat[]{
        \includegraphics[width=0.3059\textwidth]{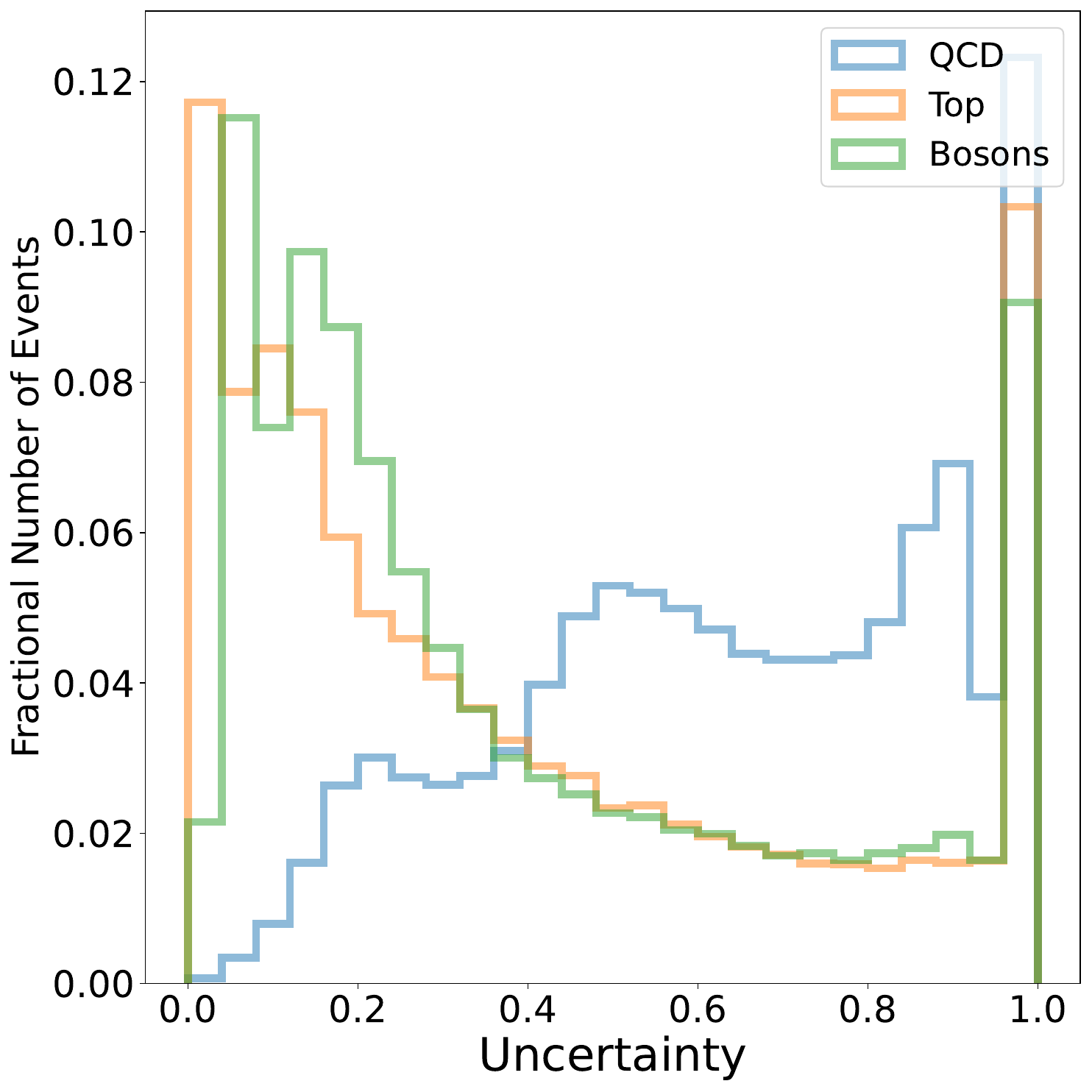}
        \label{fig:jetnet_0.7_baseline_unc_class_oc}
    }
    \hspace*{-0.2cm}
    \subfloat[]{
        \raisebox{-0.15cm}{
        \includegraphics[width=0.3591\textwidth]{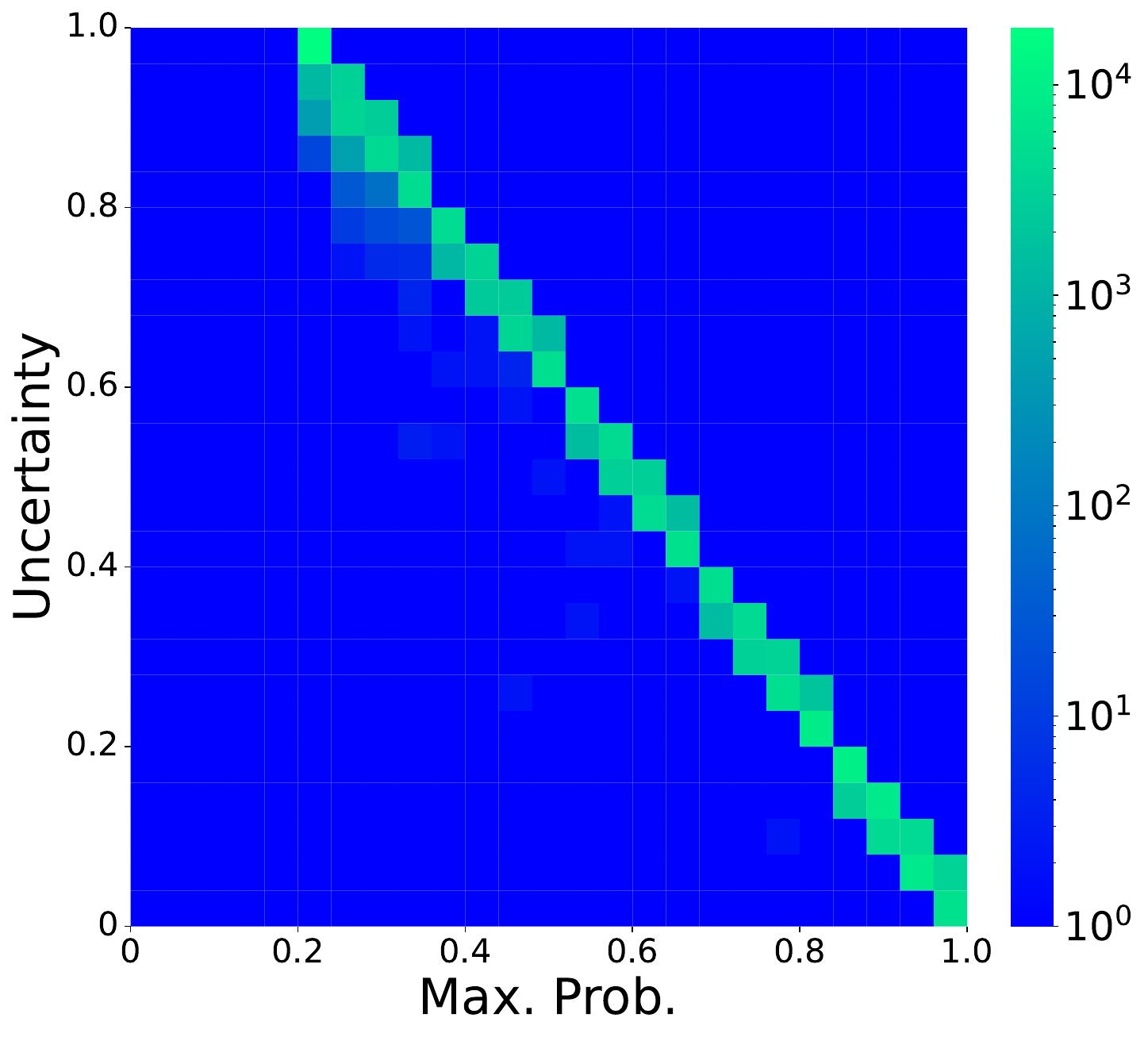}
        }
        \label{fig:jetnet_0.7_baseline_up_oc}
    }
    
    \caption{For the \jetnet dataset, on each row, from left to right, \protect\subref{fig:jetnet_0.1_baseline_us_oc},\protect\subref{fig:jetnet_0.7_baseline_us_oc} uncertainty distribution, separated by correctly and incorrectly classified jets, \protect\subref{fig:jetnet_0.1_baseline_unc_class_oc},\protect\subref{fig:jetnet_0.7_baseline_unc_class_oc} uncertainty distribution for correctly classified jets, separated by initiating particle jet type, and \protect\subref{fig:jetnet_0.1_baseline_up_oc},\protect\subref{fig:jetnet_0.7_baseline_up_oc}
     2D histogram of maximum probability versus uncertainty for baseline EDL-CT $\lambda_t^{\texttt{CT}} (0.1) $ (top row) and $\lambda_t^{\texttt{CT}} (0.7) $ (bottom row).
    }
    \label{fig:jetnet_ocmodel}
\end{figure}

As shown in Table~\ref{tab:baseline}, the EDL-CT model with $\lambda_t^{\texttt{CT}}(0.1)$ has an accuracy comparable with the EDL model with $\lambda_t(0)$ while its AUROC is much higher than the EDL model with $\lambda_t(0.1)$. This is an encouraging result, since it shows that this EDL-CT model can retain its classification performance while large uncertainty assignments correlate with misclassification more strongly. The method of confidence tuning results in smoother uncertainty distributions for correctly classified jets, as seen in Figures~\ref{fig:jetnet_0.1_baseline_us_oc}~and~\ref{fig:jetnet_0.7_baseline_us_oc}. Both choices for EDL-CT models tend to show a softer uncertainty assignment for the $q/g$ jets while most misclassified jets have larger uncertainties. Similar to what we have observed before, a larger choice of $\zeta$ makes the model more conservative: its uncertainties are better calibrated at the expense of model accuracy.

\begin{figure}[htbp]

\centering

    \subfloat[]{
        \includegraphics[width=0.5\textwidth]{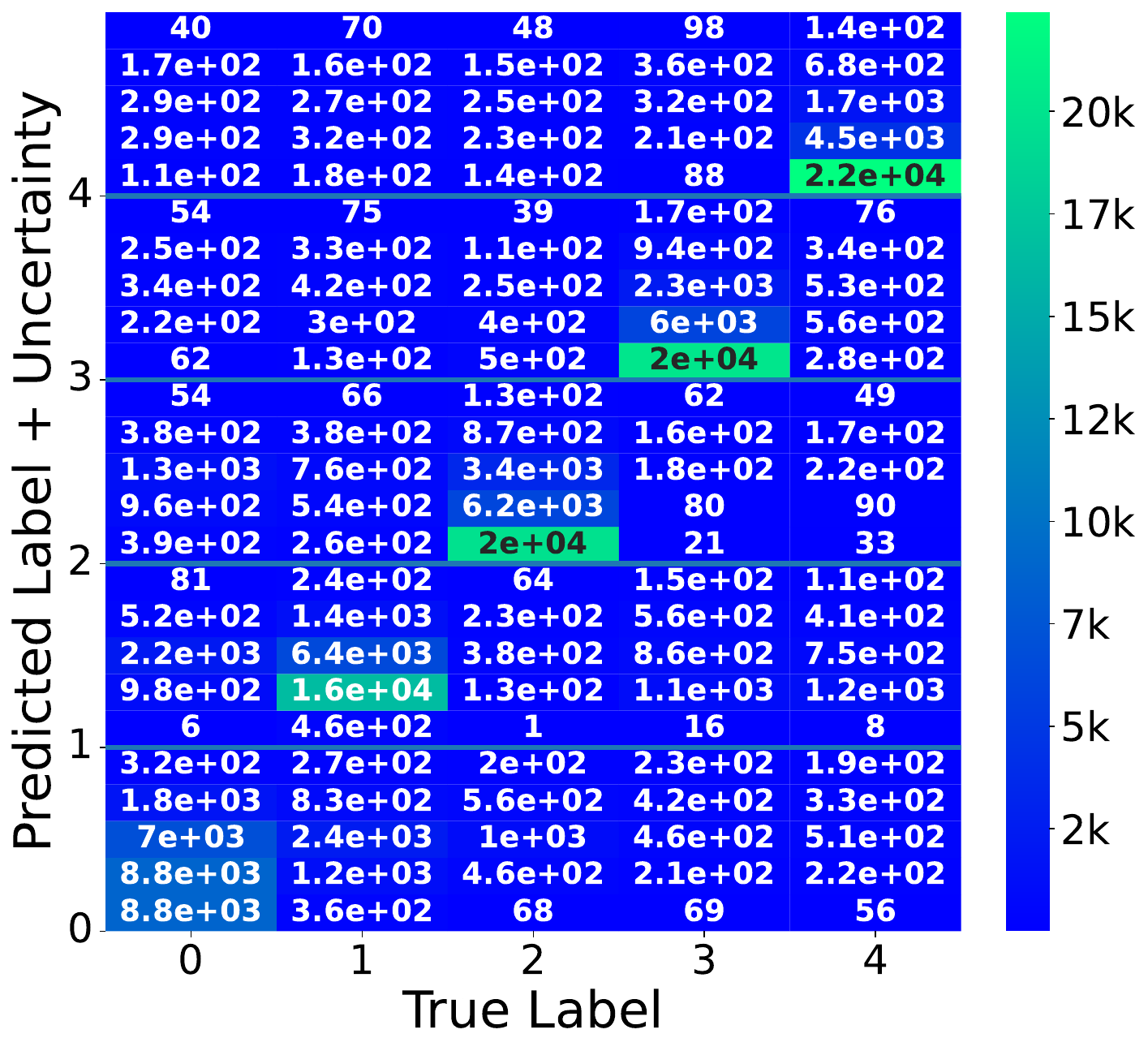}
    \label{fig:jetnet_0.1_baseline_lpu_oc}
    }
    \subfloat[]{
        \includegraphics[width=0.5\textwidth]{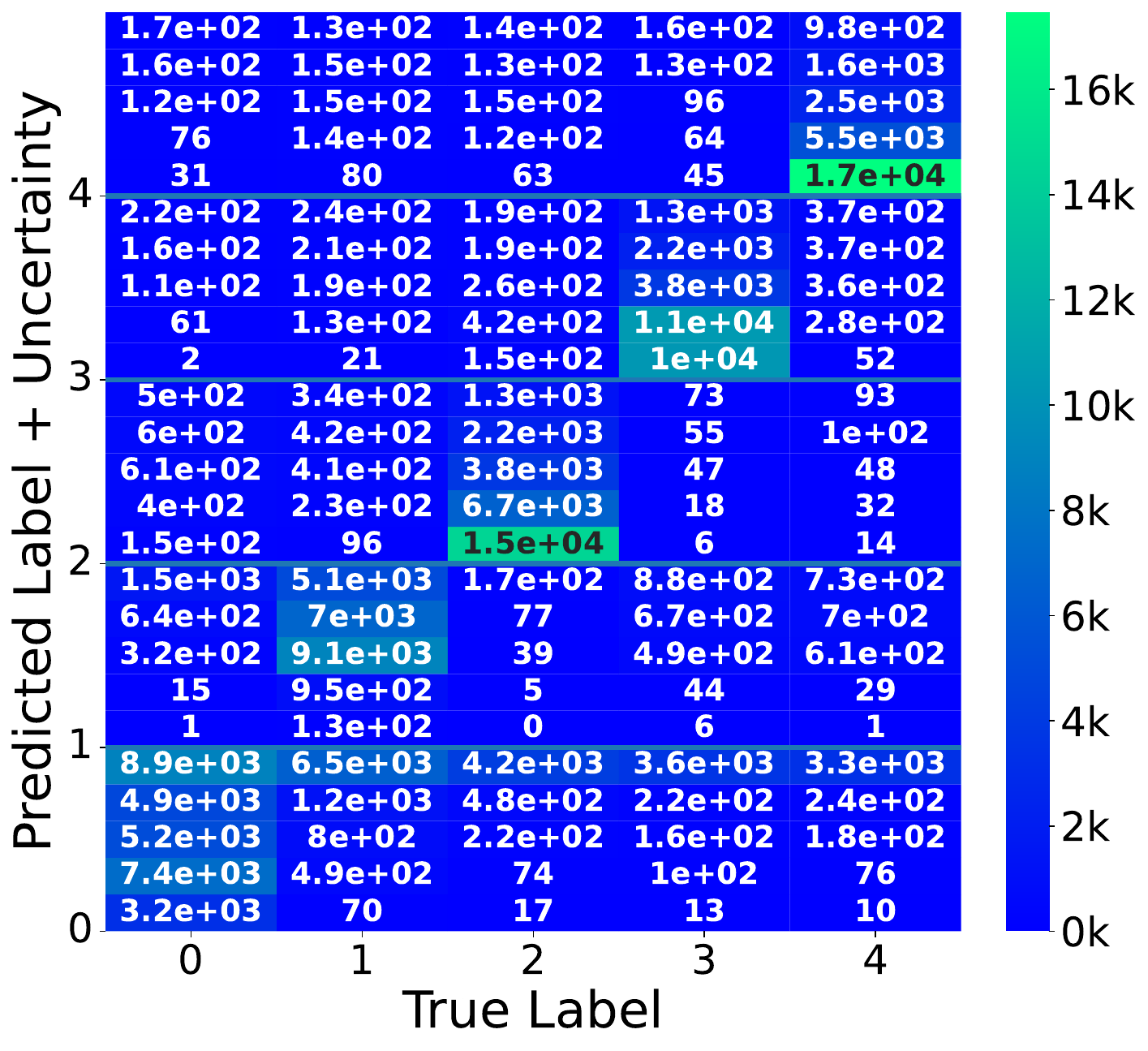}
    \label{fig:jetnet_0.7_baseline_lpu_oc}
    }
    
    \caption{Uncertainty Aware Confusion Matrix for, respectively, baseline \jetnet EDL-CT \protect\subref{fig:jetnet_0.1_baseline_lpu_oc} $\lambda_t^{\texttt{CT}} (0.1) $ and \protect\subref{fig:jetnet_0.7_baseline_lpu_oc} $\lambda_t^{\texttt{CT}} (0.7) $.
    }
    \label{fig:jetnet_uacm_oc}
\end{figure}

\subsection{JetClass dataset}
\label{sec:jetclass}

This dataset is much larger than the \topdata and \JetNet datasets and further subdivides jet classes by particle structure, allowing us to fully explore the extent of EDL-based uncertainty quantification on jet tagging. The classes of the dataset and their indices are: $ q/g$ (0), $H \to b \bar{b}$ (1), $H \to c \bar{c}$ (2), $H \to gg$ (3), $H \to 4 q$ (4), $H \to \ell \nu q q'$ (5), $Z \to q \bar{q}$ (6), $W \to q q'$ (7), $t \to bqq'$ (8), $t \to b \ell \nu$ (9). Building upon our experience with the smaller datasets,  we only trained the \JetClass EDL models with two different choices of non-zero annealing coefficients: $\lambda_t (0.1) $ and $\lambda_t (1.0) $, to illustrate the impact of smaller and larger values of $\zeta$.\footnote{Training the PFIN model for the JetClass dataset showed a somewhat enhanced sensitivity to model initialization, sometimes requiring multiple iterations to reach a converging state.}

\begin{figure}[htbp]

\centering
    \includegraphics[width=1.0\textwidth]{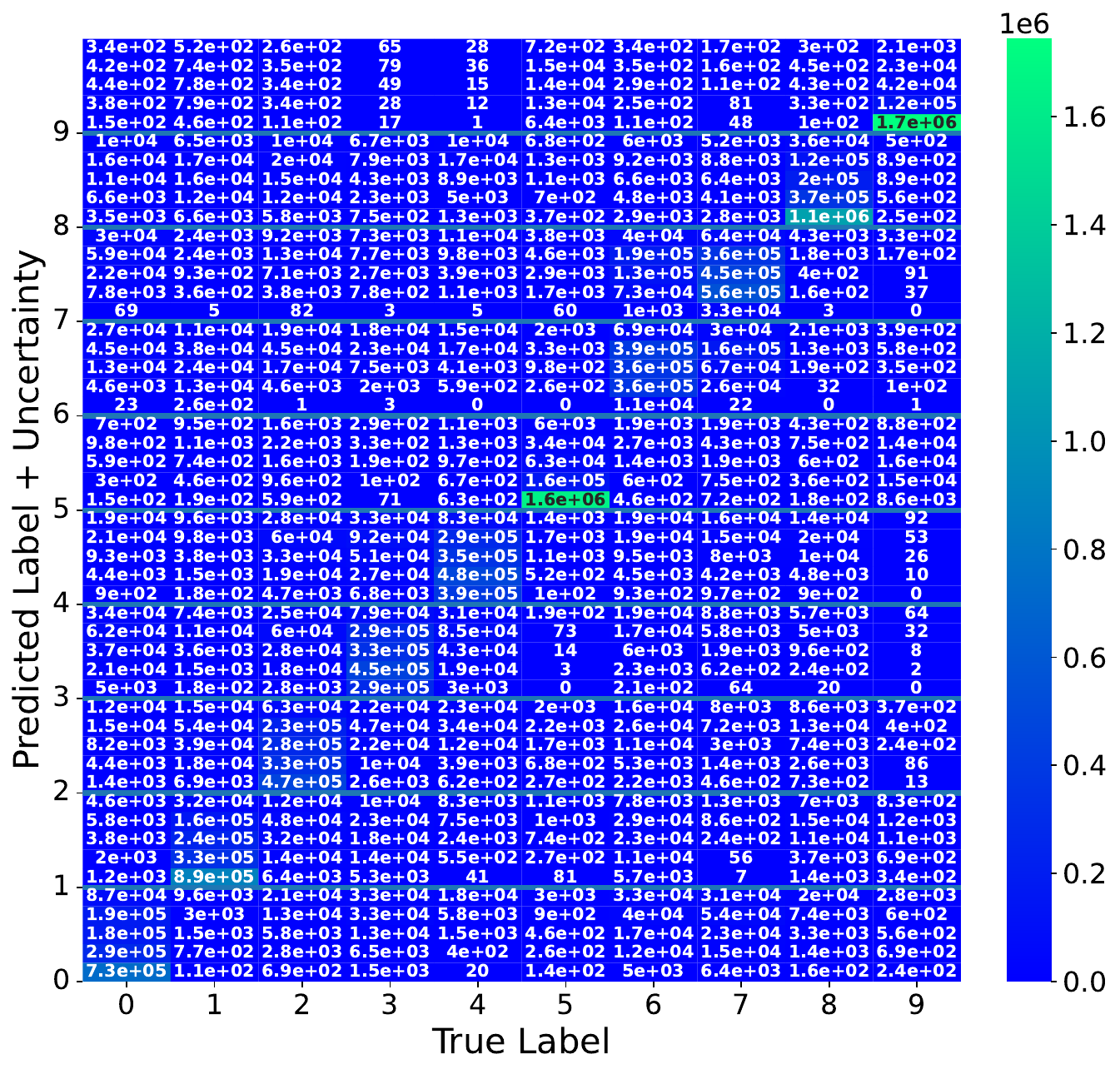}
    \label{fig:jetclass_0.1_baseline_lpu}
    
    \caption{Uncertainty Aware Confusion Matrix for baseline \JetClass EDL $\lambda_t (0.1) $.
    }
    \label{fig:jetclass_uacm_0.1}
\end{figure}

\begin{figure}[htbp]

\centering

    \subfloat[]{
        \includegraphics[width=0.3059\textwidth]{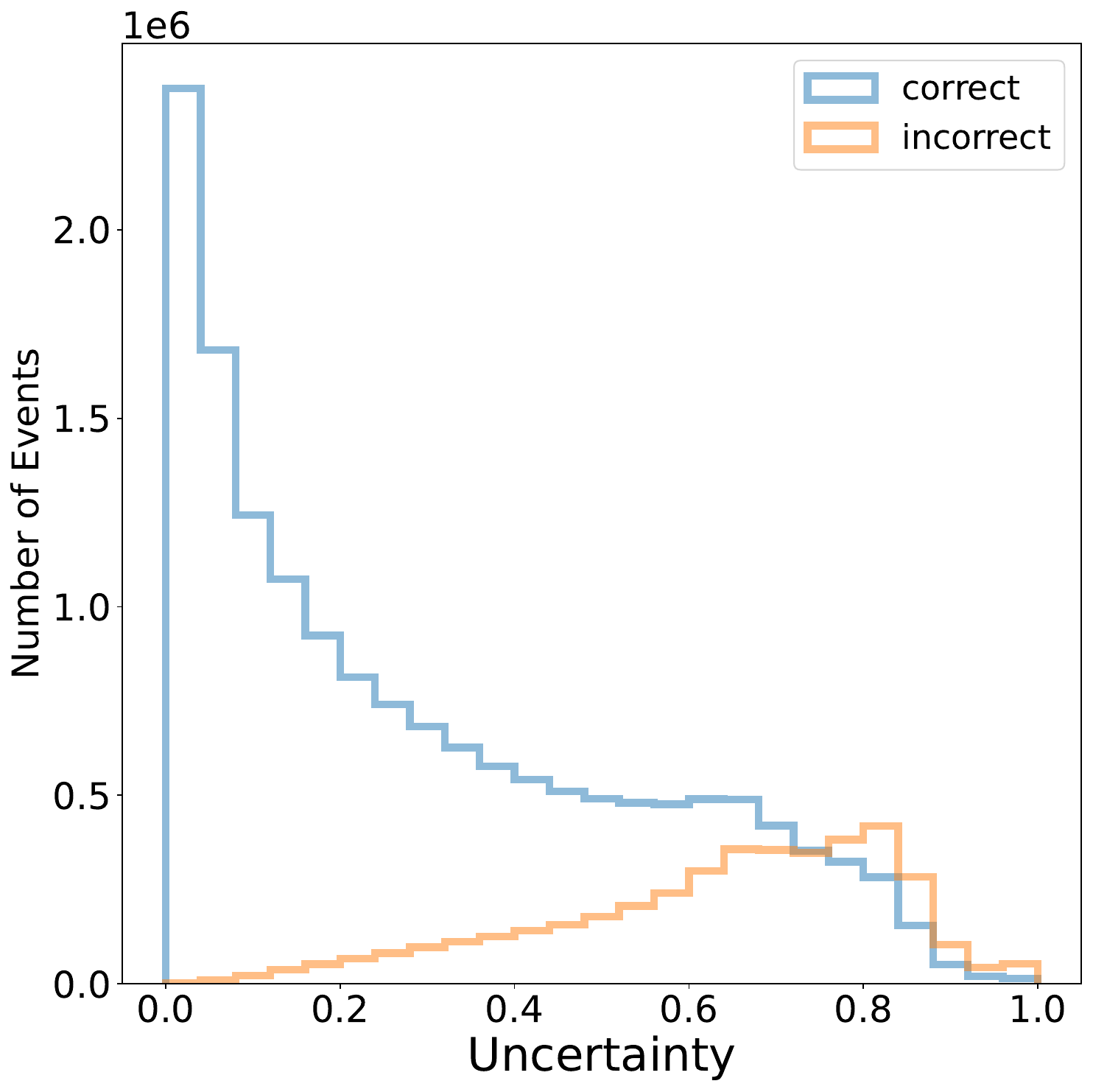}
        \label{fig:jetclass_0.1_baseline_us}
    }
    \subfloat[]{
        \includegraphics[width=0.3059\textwidth]{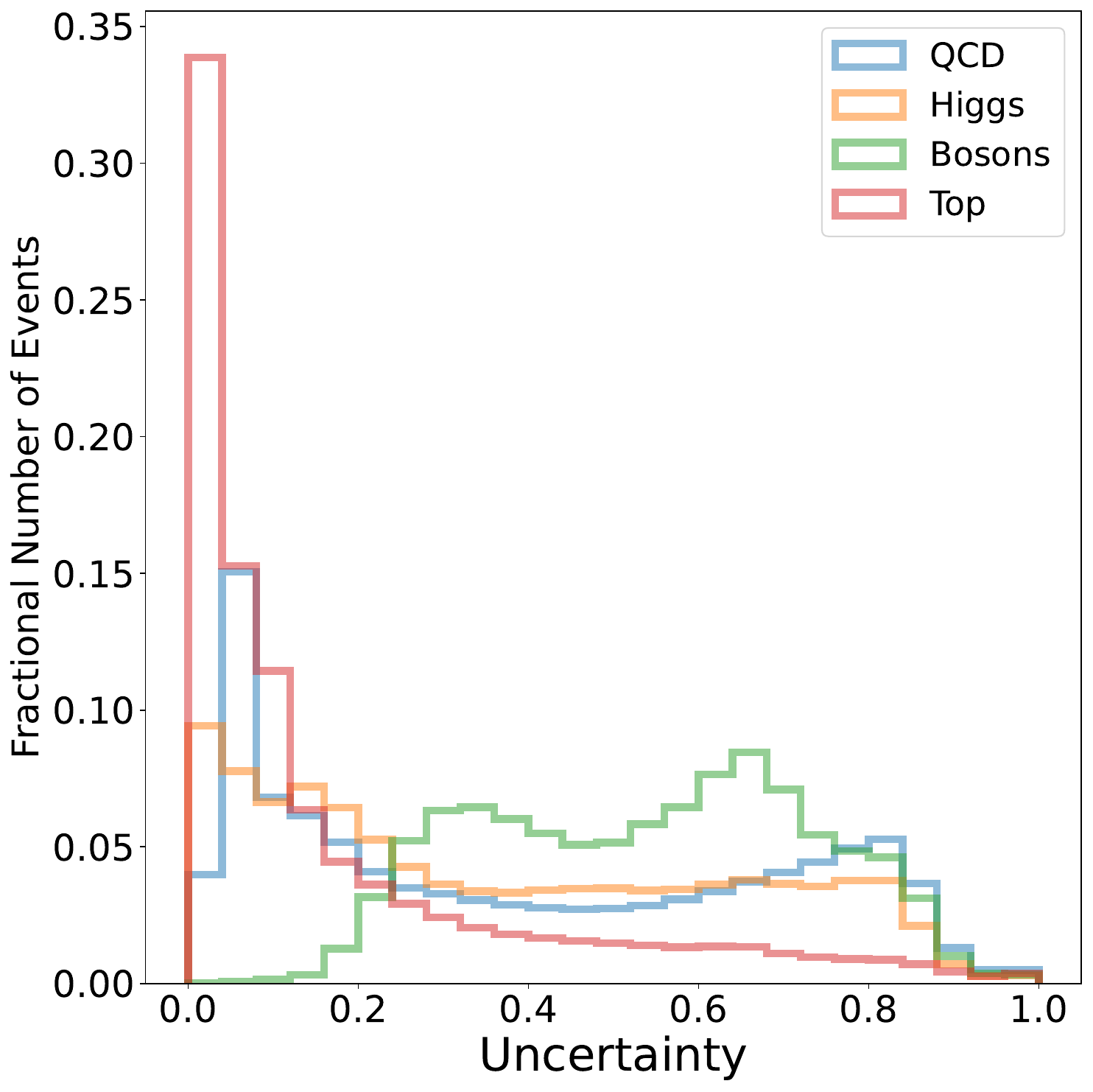}
        \label{fig:jetclass_0.1_baseline_unc_class}
    }
    \hspace*{-0.2cm}
    \subfloat[]{
        \raisebox{-0.15cm}{
        \includegraphics[width=0.3591\textwidth]{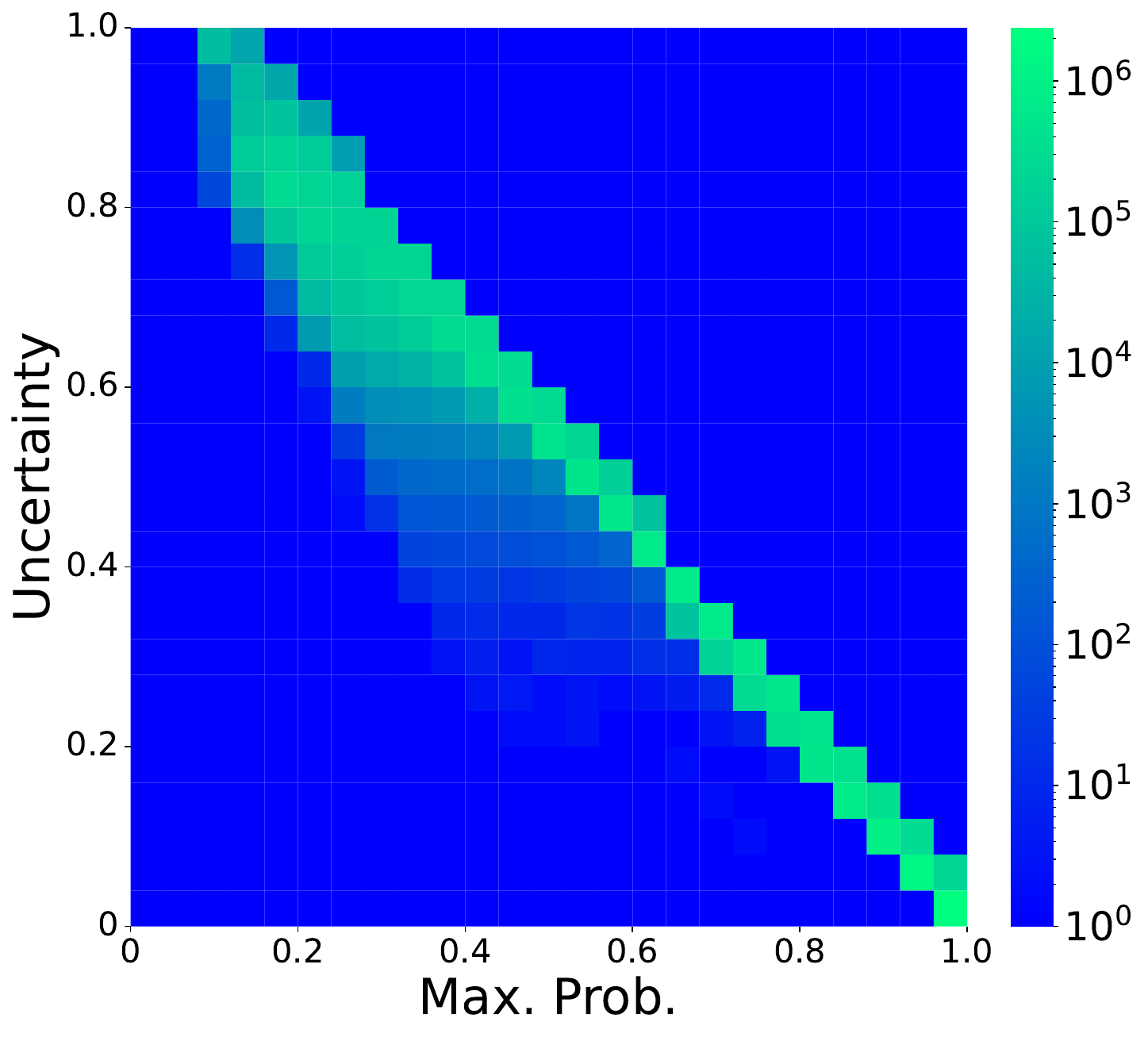}
        }
        \label{fig:jetclass_0.1_baseline_up}
    }

    \subfloat[]{
        \includegraphics[width=0.3059\textwidth]{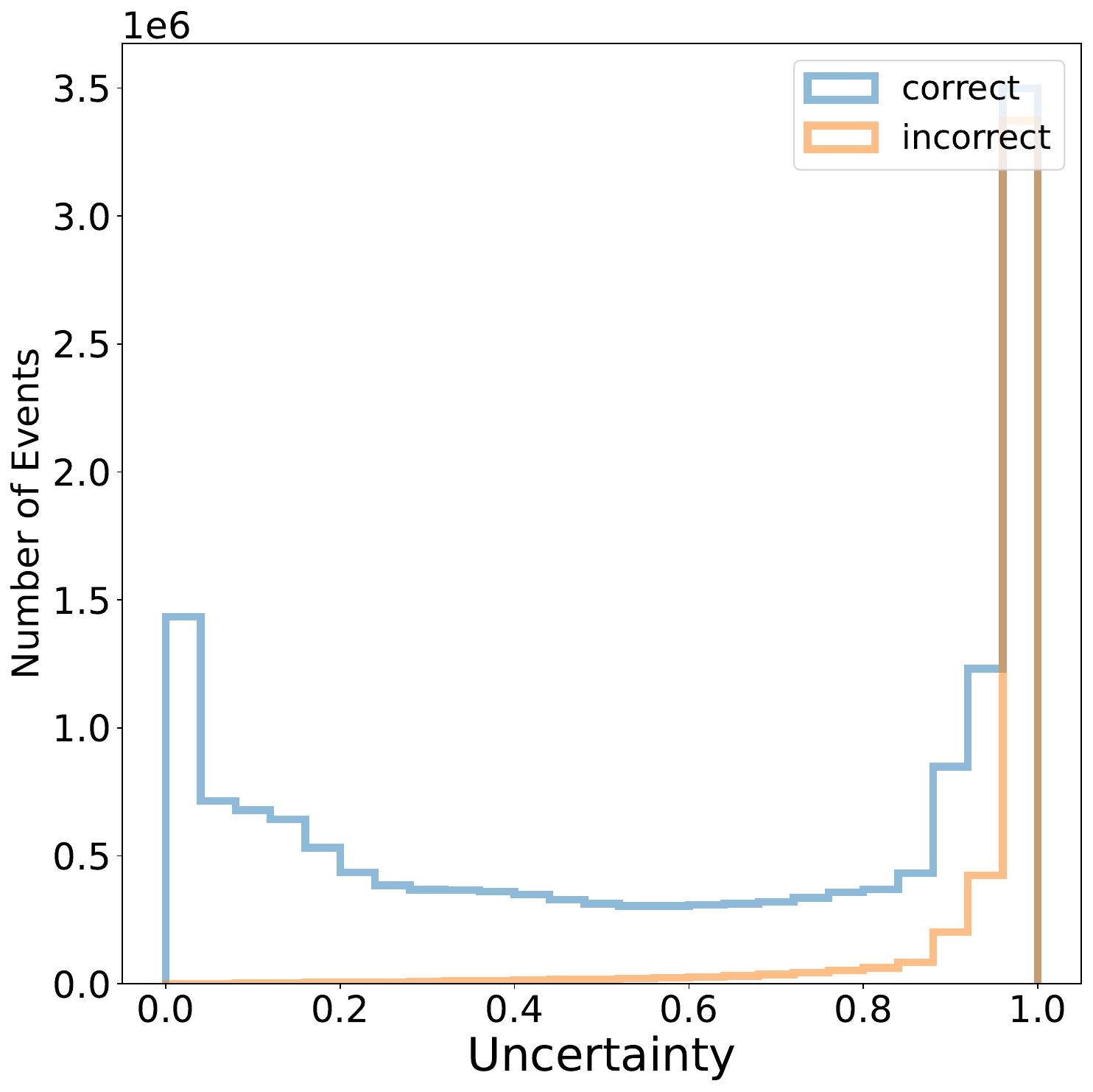}
        \label{fig:jetclass_1.0_baseline_us}
    }
    \subfloat[]{
        \includegraphics[width=0.3059\textwidth]{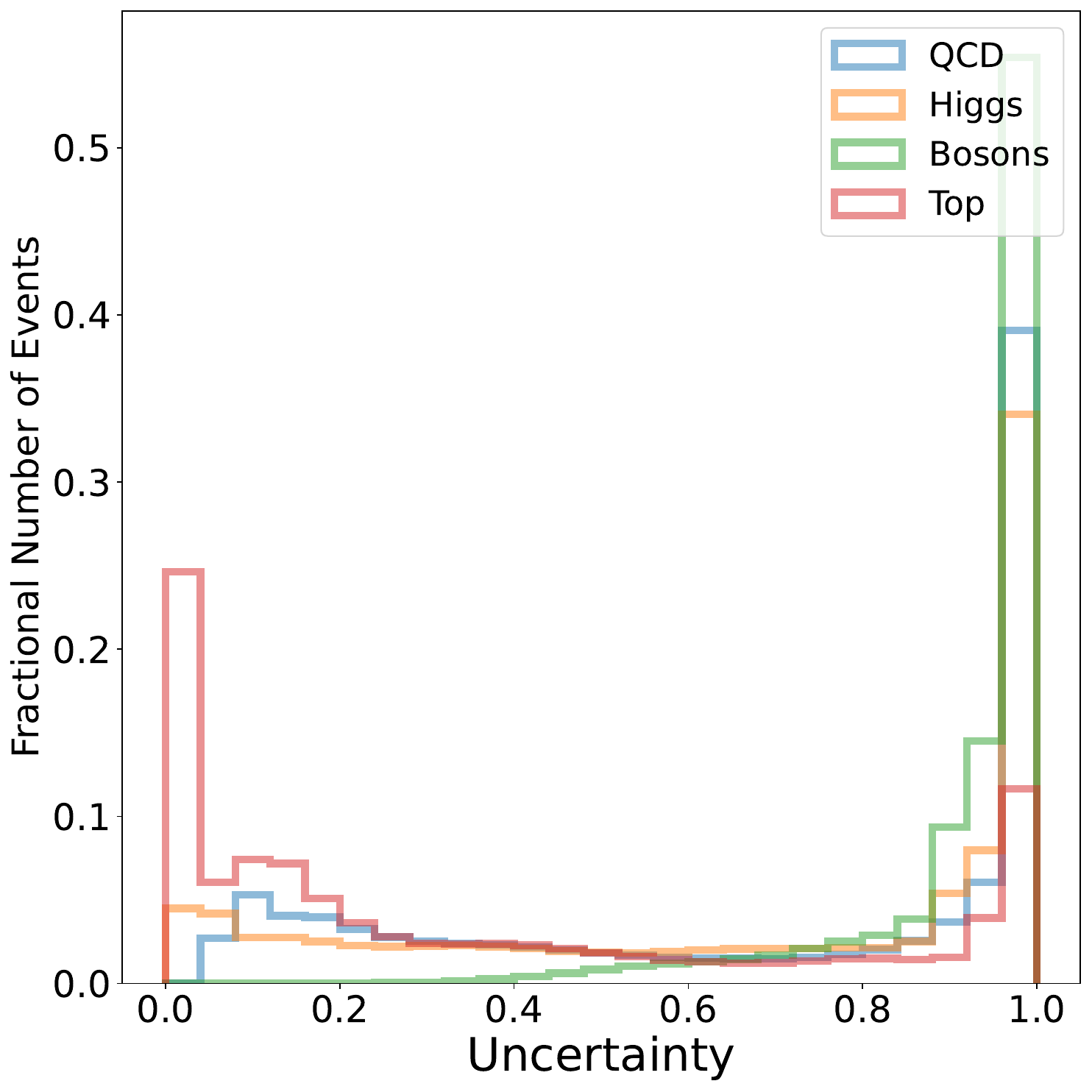}
        \label{fig:jetclass_1.0_baseline_unc_class}
    }
    \hspace*{-0.2cm}
    \subfloat[]{
        \raisebox{-0.15cm}{
        \includegraphics[width=0.3591\textwidth]{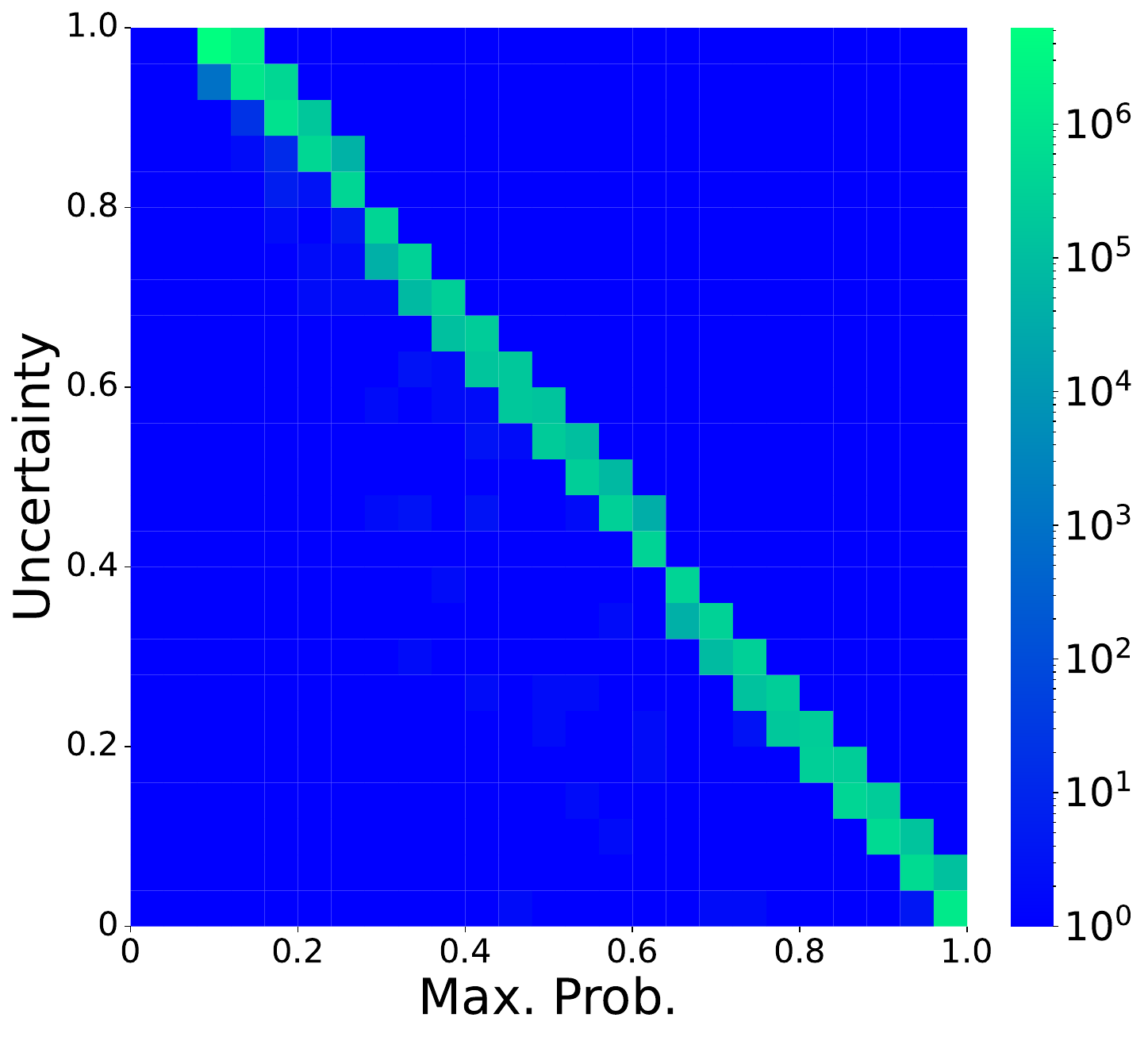}
        }
        \label{fig:jetclass_1.0_baseline_up}
    }

    \caption{For the \jetclass dataset, on each row, from left to right, \protect\subref{fig:jetclass_0.1_baseline_us},\protect\subref{fig:jetclass_1.0_baseline_us} uncertainty distribution, separated by correctly and incorrectly classified jets, \protect\subref{fig:jetclass_0.1_baseline_unc_class},\protect\subref{fig:jetclass_1.0_baseline_unc_class} uncertainty distribution for correctly classified jets, separated by initiating particle jet type, and \protect\subref{fig:jetclass_0.1_baseline_up},\protect\subref{fig:jetclass_1.0_baseline_up}
    logarithmic 2D histogram of maximum probability versus uncertainty for baseline EDL $\lambda_t (0.1) $ (top row) and $\lambda_t (1.0) $ (bottom row).
    }
    \label{fig:jetclass_baseline}
\end{figure}

The \JetClass column in Table~\ref{tab:baseline} shows an evaluation of the predictive performance of the \JetClass EDL models we studied. Similarly to the EDL models applied to the \JetNet dataset, as $\zeta$ increases, the classification accuracy decreases and AUROC increases, representing a tradeoff between predictive performance and conservative uncertainty quantification. The EDL model with $\zeta = 0.1$ has a marginally smaller accuracy but the AUROC improves significantly when compared with the $\zeta = 0$ model. The uncertainty distributions across different classes are shown in the UACM in Figure~\ref{fig:jetclass_uacm_0.1}. The uncertainty distributions show the desirable characteristics with large uncertainties being attributed to misclassified jets while the correctly classified jets typically have softer uncertainties. This is also evident from the uncertainty distributions given in Figure~\ref{fig:jetclass_0.1_baseline_us}. 

Figure~\ref{fig:jetclass_0.1_baseline_unc_class} provides a detailed overview of uncertainties associated with different jet classes, where jet classes are combined according to the originating particle. While most classes show a smoothly declining uncertainty profile, the bosons class, comprising the $Z \to q\bar{q}$ and $W \to qq'$ classes, show a bimodal distribution with the second peak close to the mode of the uncertainty distribution of the incorrectly classified jets. This is also seen in the UACM in Figure~\ref{fig:jetclass_uacm_0.1}. These two classes were also found to be most likely misclassified as one another, which can be attributed to the similarity in their invariant masses and final states. Two correctly classified jet categories have low uncertainties: $H \to \ell \nu q q'$ (4) and $t \to b \ell \nu$ (9). These are also the only jets that contain leptons, suggesting that the model has confidently learned to exploit final state characteristics such at the particle-type information and decay topology.

\begin{figure}[htbp]

\centering
    \includegraphics[width=1.0\textwidth]{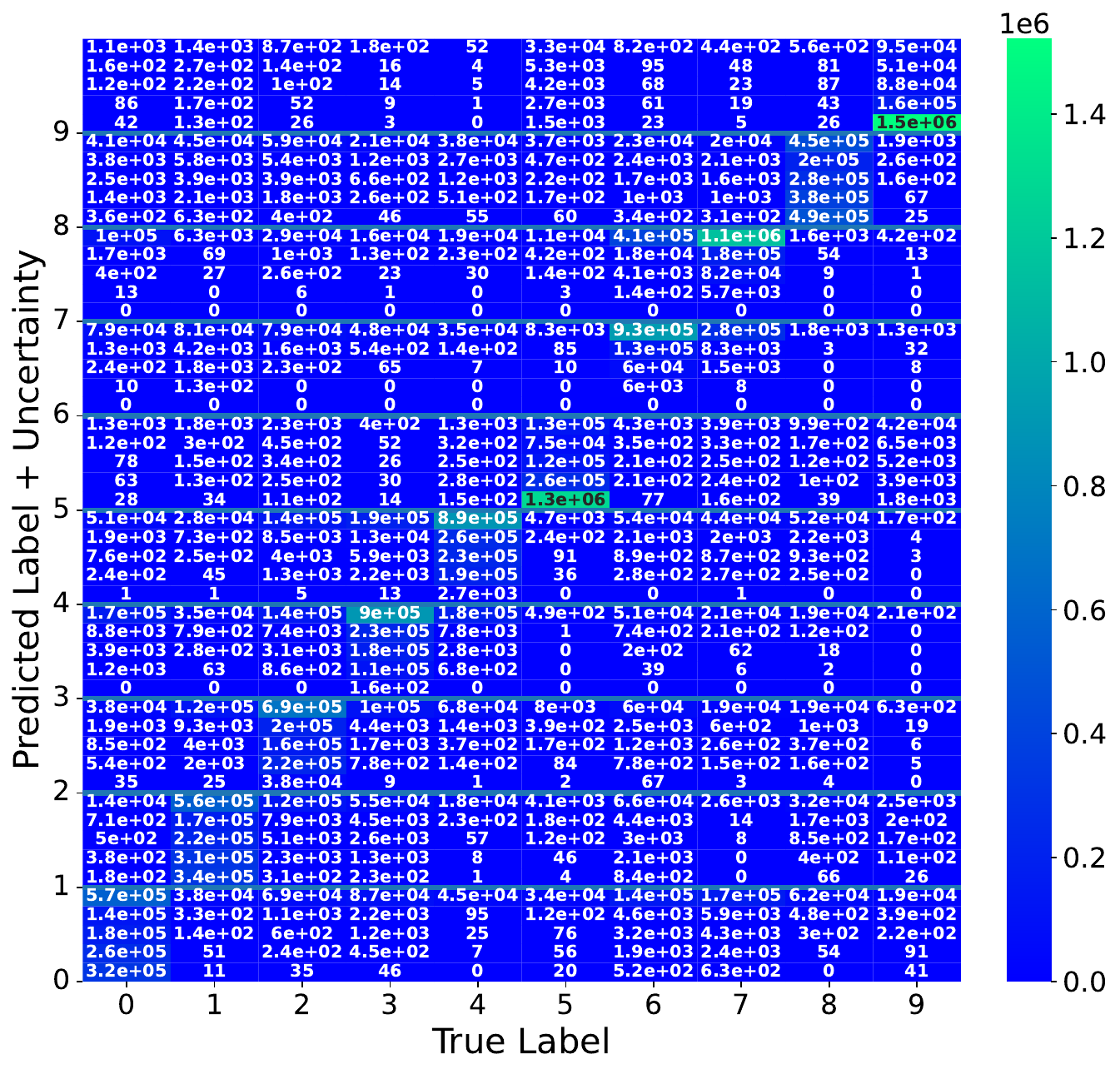}
    \label{fig:jetclass_1.0_baseline_lpu}
    
    \caption{Uncertainty Aware Confusion Matrix for baseline \jetclass EDL $\lambda_t (1.0)$.}
    \label{fig:jetclass_uacm_1.0}
\end{figure}

As $\zeta$ increases, we observe a significant difference in the uncertainty profile of the different jet classes. Both correctly and incorrectly classified jets tend to show very large uncertainties (Figure~\ref{fig:jetclass_1.0_baseline_us}) and all jet classes show strong bimodal distributions with a large peak near $u = 1.0$ (Figure~\ref{fig:jetclass_1.0_baseline_unc_class}). The uncertainty distributions can be further investigated from the UACM in Figure~\ref{fig:jetclass_uacm_1.0}. With increasing $\zeta$, the model leverages the larger contribution of the DL divergence term in the loss function to assign high uncertainties to most of the jets. This again relays the importance of considering the accuracy alongside AUROC to determine the performance of an EDL model.

As we conclude this section, we note that even EDL $\lambda_t(1.0)$ successfully distinguishes the jet classes with leptonic decay modes with low uncertainties. We also observe confusion of the model to distinguish the $W \to qq'$ and $Z \to q\bar{q}$ classes, misclassifying one into the other while assigning relatively large uncertainties on these class determinations.

\section{Comparison with Ensemble Methods for Uncertainty Quantification}
\label{sec:comp}

To determine the efficacy of EDL, we compare this method with two different Bayesian methods: Ensemble training and MC Dropout. Both Bayesian methods took ten times longer than EDL models on inference passes to estimate the uncertainty due to the nature of these methods.  As such, EDL models are preferable for systems with limited computational resources. However, the choice of model is subject to optimization and depends on the dataset and training set.

When benchmarking against the best-performing EDL model chosen from the optimal combination of accuracy and AUROC scores, we observe that the Ensemble methods typically provide better or comparable accuracies. On the other hand, MC Dropout performs similarly or worse than EDL in terms of accuracy. Both methods show worse performance in terms of AUROC. This indicates that a well-trained EDL model can provide similar performance in terms of accuracy but does a better job at assigning larger uncertainties to misclassified jets.

The results that compare EDL models with Bayesian methods for \topdata, \JetNet, and \JetClass datasets are summarized in Table~\ref{tab:baseline}. For \topdata, both the EDL and Ensemble models achieve the same ID accuracy, but MC Dropout has slightly worse prediction performance. The EDL $\lambda_t (0.7)$ and $\lambda_t (1.0)$ models outperform both Ensemble models on AUROC, suggesting that EDL is a better UQ method for the top tagging dataset. 

The performance of the baseline EDL models on the \JetNet dataset in exhibits a different trend. All EDL models with non-zero $\zeta$ perform better than the Bayesian methods in UQ at the expense of classification accuracy, although the degradation is modest. The best performing EDL model is the EDL-CT model with $\zeta = 0.1$ which has a slightly worse accuracy but a big improvement in AUROC compared to the Ensemble method. 

The performance of the baseline \JetClass models is similar in performance to the \JetNet models. We note that both accuracy and AUROC improve in the Ensemble model for \JetClass when compared with the benchmark of $\lambda_t(0)$. The best performing EDL model is the model with $\lambda_t(0.1)$ which has a slightly lower classification accuracy but a significantly larger AUROC.
\section{Interpretation of EDL Uncertainty Estimation}
\label{sec:interp}

Since the EDL model is a deterministic DNN that directly predicts a Dirichlet distribution, the model must also encode some information on the evidence gained for each class in the latent space. To understand how the model learns the uncertainty, we examine the distribution of variances in the latent space representation using Principal Component Analysis~\cite{jolliffe2016principal}. As shown in our previous work in Ref.~\cite{Khot_2023}, PCA reveals how the model reorganizes useful correlations with highly discriminative features. We perform similar studies on the PFIN latent space for all three datasets studied in this paper. We use the best performing model for each dataset, namely EDL $\lambda_t(1.0)$ for \topdata, EDL-CT with $\lambda_t^{\texttt{CT}}(0.1)$ for \JetNet, and EDL $\lambda_t(0.1)$ for the \JetClass dataset.

\begin{figure}[htbp]
    \centering
    
    \subfloat[]{
    \adjincludegraphics[width=0.48\textwidth]{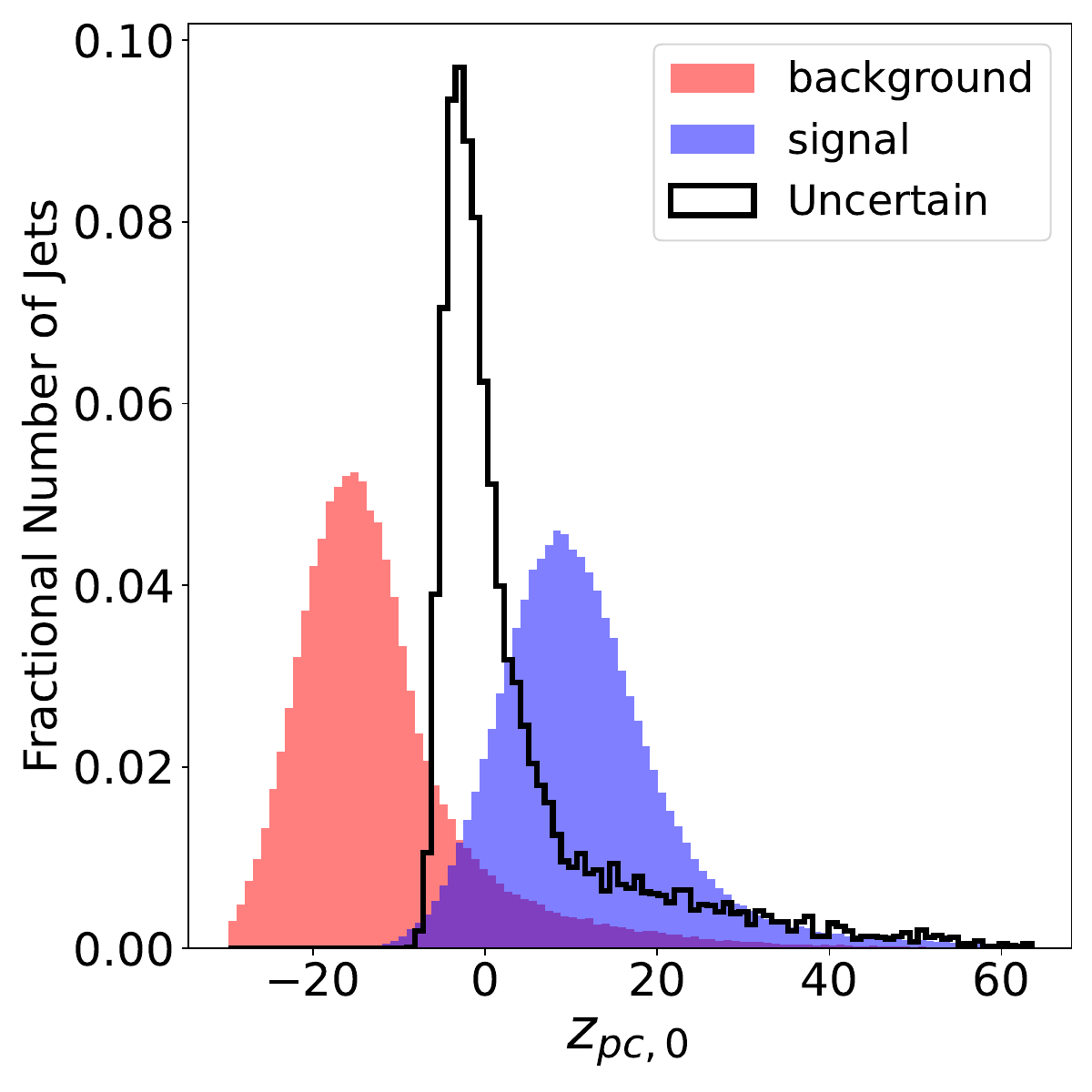}
    \label{fig:zpca_0}
    }
    \subfloat[]{
    \adjincludegraphics[width=0.48\textwidth]{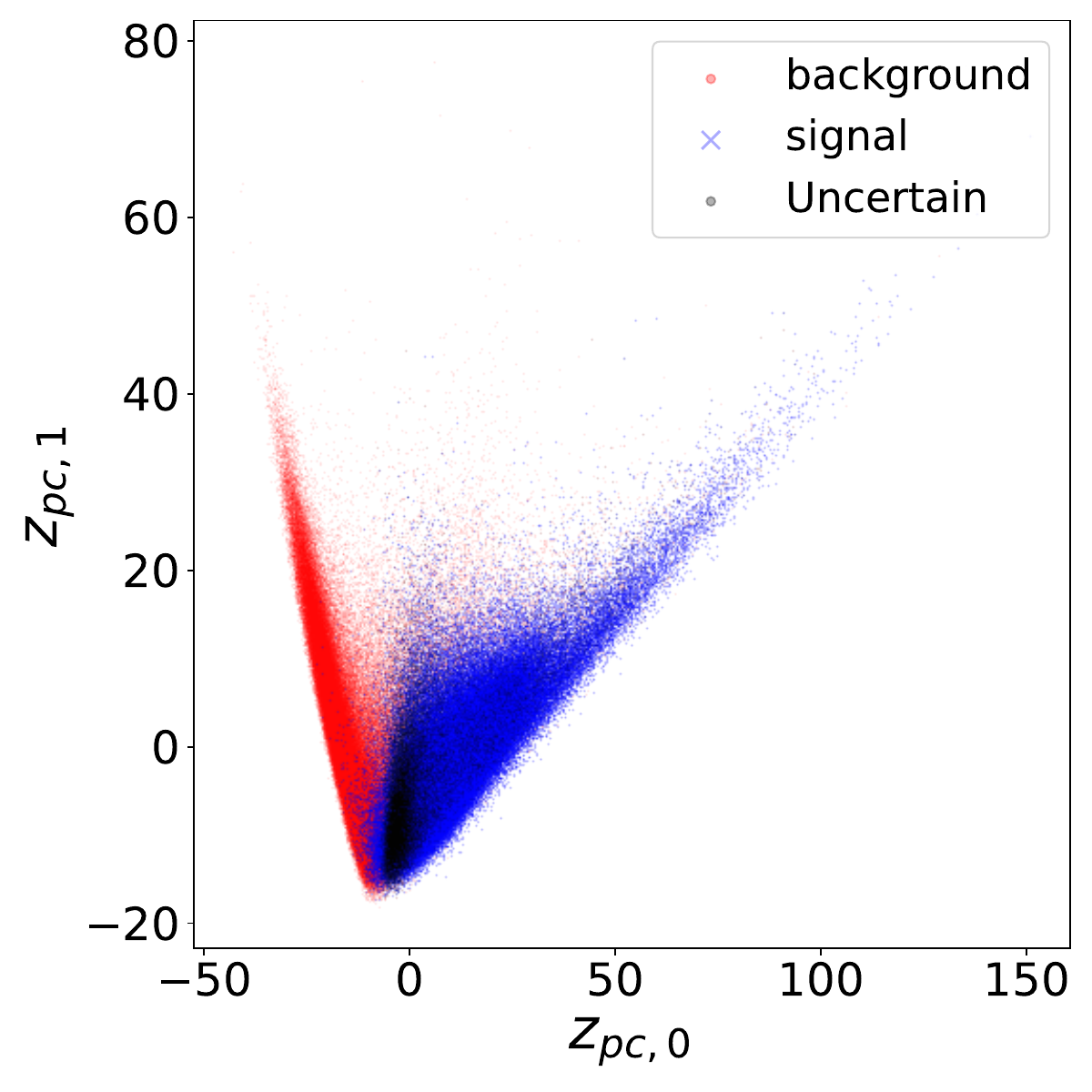}
    \label{fig:scatter_01}
    }
    
    \caption{For \topdata EDL $\lambda_t (1.0)$, \protect\subref{fig:zpca_0} distribution of first principal component for background QCD jets and signal top jets along with incorrectly predicted jets with uncertainties greater than 0.8, also known as "Uncertain" jets and \protect\subref{fig:scatter_01} pairwise distribution between first and second principal component for background, signal, and "Uncertain" jets.
    }
    \label{fig:latent_topdata}
\end{figure}

For the \topdata dataset, we found that 99\% of the observed variance in the test data was described by the top 37 principal components. Along with this, we set an uncertainty threshold at 0.8 and examine the distributions of the first principal component of the misclassified jets with an uncertainty higher than this threshold. We identify high-uncertainty misclassified jets as  \textit{uncertain} jets. 
Figure~\ref{fig:latent_topdata}(a) shows the distribution of the top principal component, $z_{pc,0}$ for the two jet classes along with the uncertain jets for EDL $\lambda_t (1.0)$. We can readily see how the large-uncertainty misclassified jets lie right at the overlap region, where discrimination is the hardest. We can also examine how the correlation between these PCA-transformed latent features further display large uncertainty at the intersection of the distributions in Figure~\ref{fig:latent_topdata}(b).

\begin{figure}[htbp]
    \centering
    
    \subfloat[]{
    \adjincludegraphics[width=0.4\textwidth]{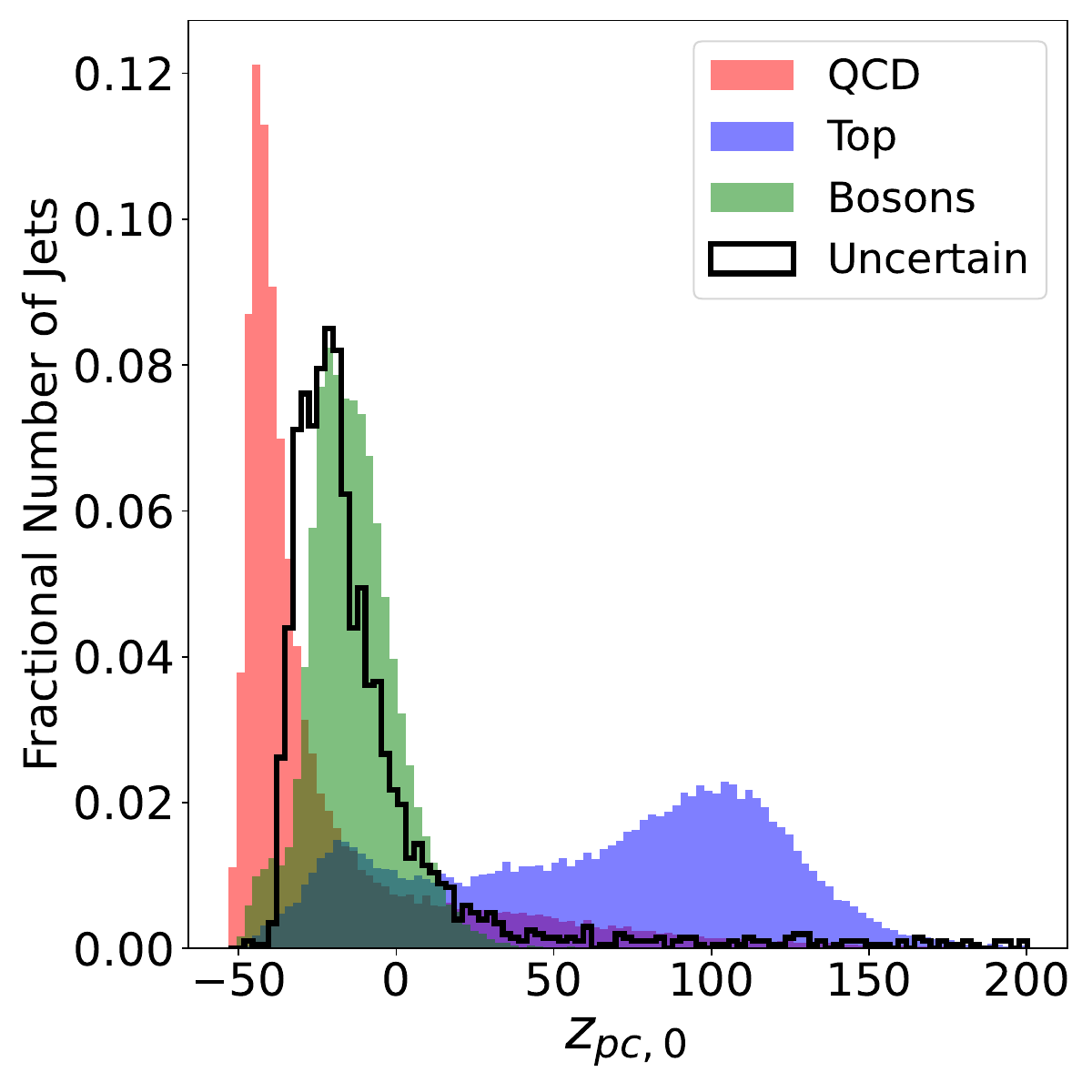}
    \label{fig:jetnet_zpca_0}
    }
    \subfloat[]{
    \adjincludegraphics[width=0.4\textwidth]{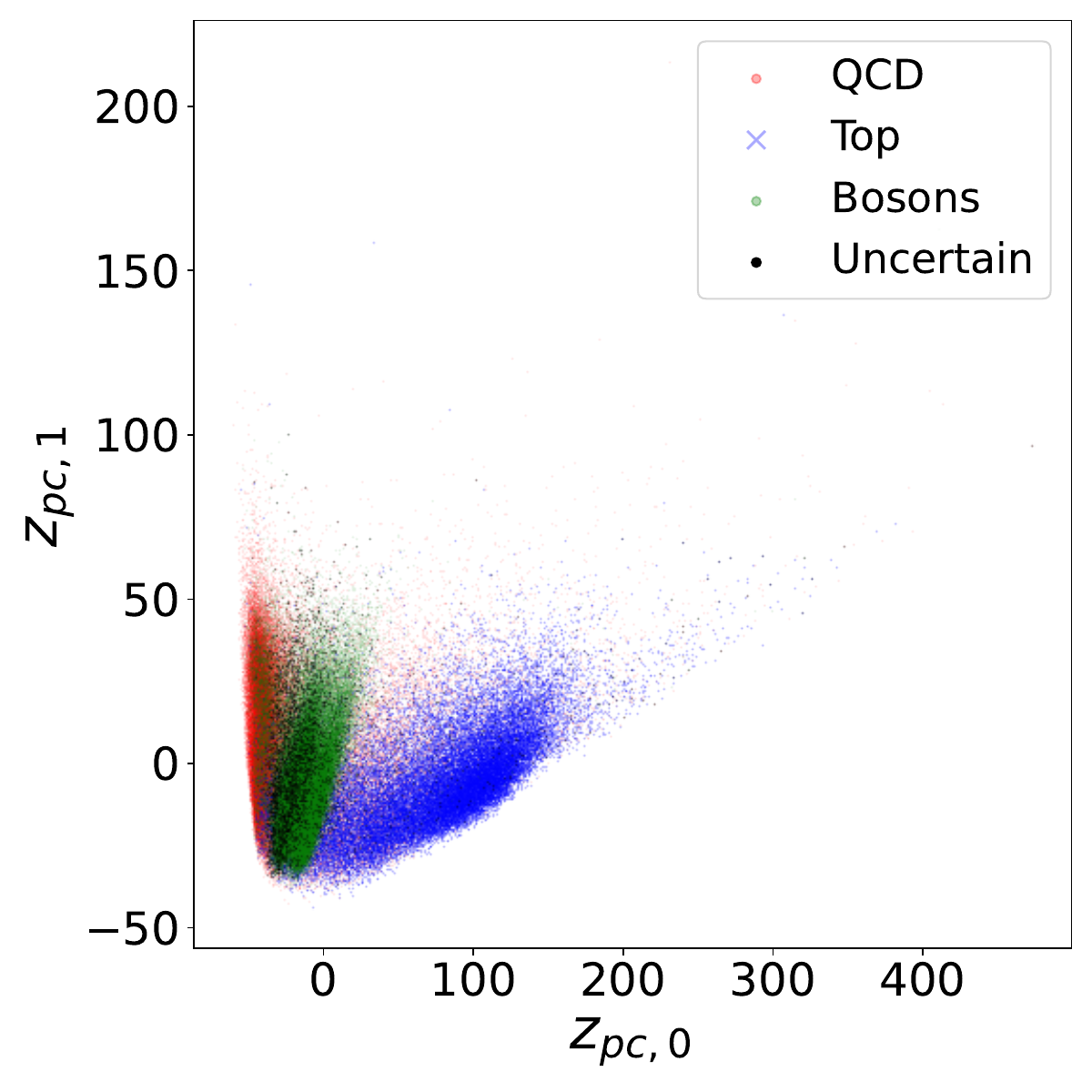}
    \label{fig:jetnet_scatter_zpca}
    }
    
    \subfloat[]{
    \adjincludegraphics[width=0.4\textwidth]{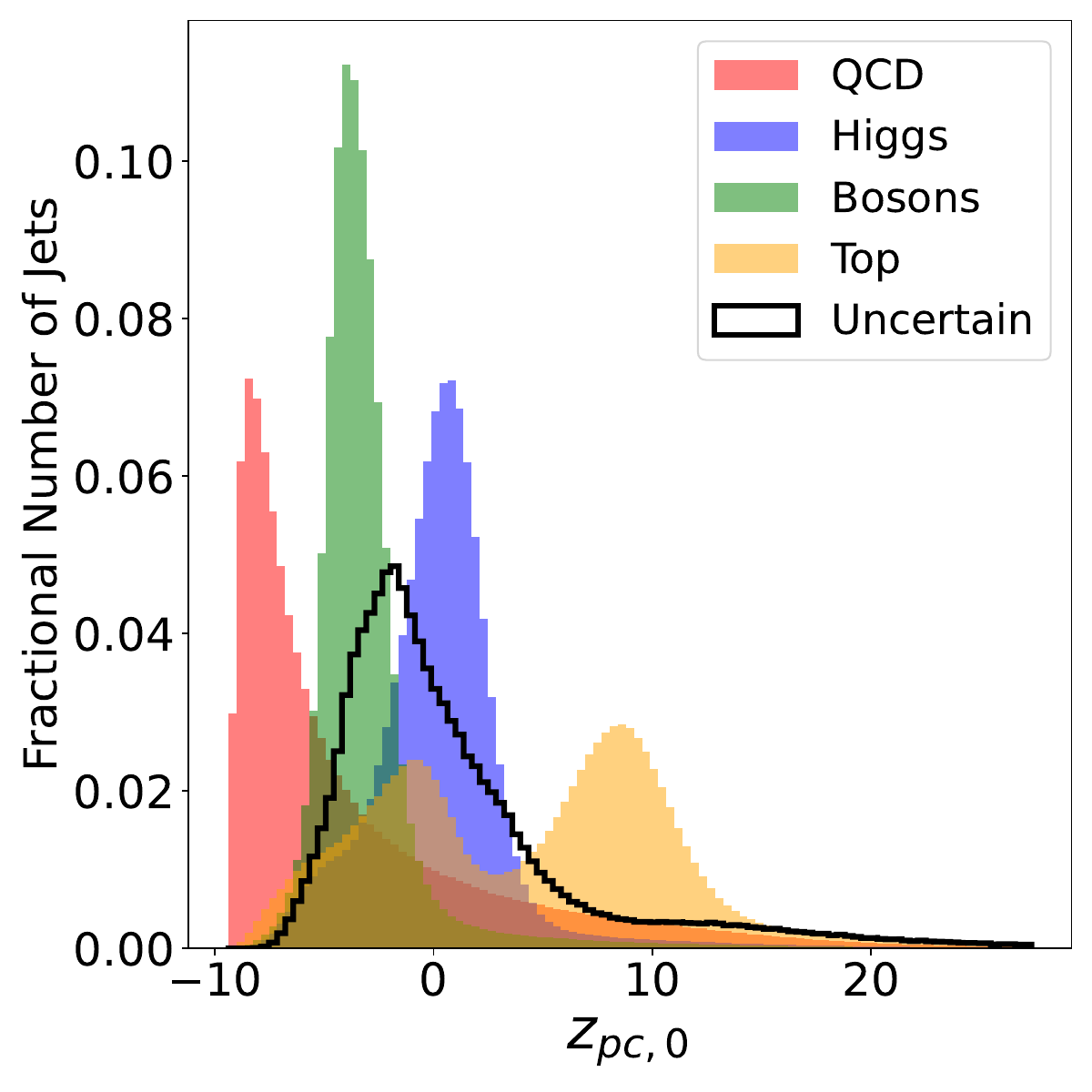}
    \label{fig:jetclass_zpca_0}
    }
    \subfloat[]{
    \adjincludegraphics[width=0.4\textwidth]{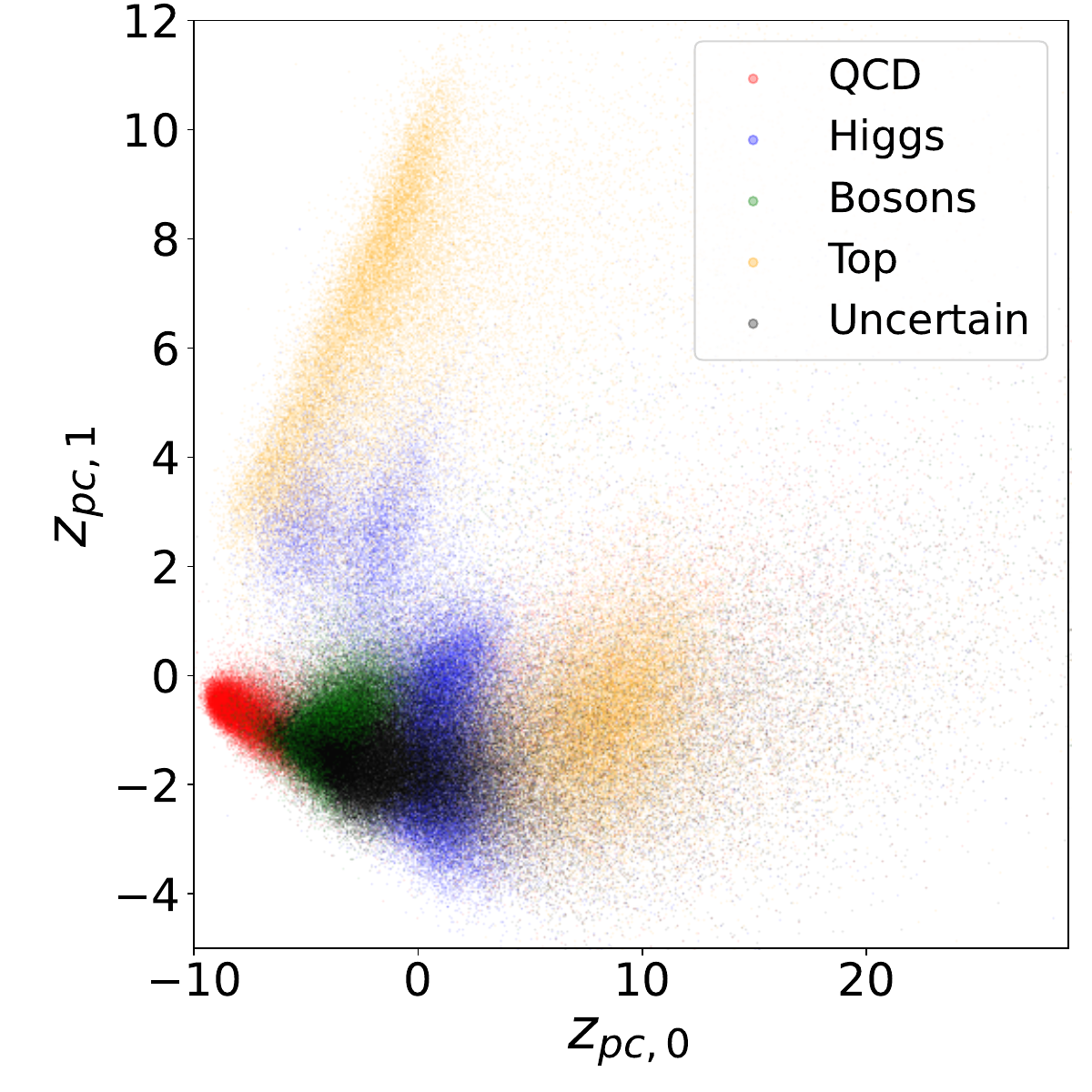}
    \label{fig:jetclass_scatter_zpca}
    }
    
    \caption{For \protect\subref{fig:jetnet_zpca_0} \JetNet EDL-CT $\lambda_t^{\texttt{CT}}(0.1)$ and \protect\subref{fig:jetclass_zpca_0} \JetClass EDL $\lambda_t (0.1)$, distribution of first principal components for jets grouped based on initiating particle types along with incorrectly predicted jets with uncertainties greater than 0.8, which we refer to as "Uncertain" jets and \protect\subref{fig:jetnet_scatter_zpca}, \protect\subref{fig:jetclass_scatter_zpca} pairwise distribution between first and second principal component for jets grouped on initiating particle types and "Uncertain" jets.}
    \label{fig:latent_jets}
\end{figure}

For the larger \JetNet and \JetClass datasets, we can group jet classes based on initiating particle types and analyze the latent space in Figure~\ref{fig:latent_jets}. They show similar patterns in how high-uncertainty misclassified jets are near the intersection of principal components and class types. The distribution of the principal components for top quarks are much further other class, which is likely why they usually have lower uncertainty as shown in Figures~\ref{fig:jetnet_unc} and~\ref{fig:jetclass_baseline}. 

Having examined how the uncertainty maps onto the principal components of the latent space, it is also instructive to investigate if learning of uncertainty impacts the ability of a model to incorporate information about physical jet characteristics. As stated in the original PFIN paper, jet-class information is found to be embodied in the distribution of correlations among latent space features~\cite{Khot_2023}. We repeated those studies in the context of our current experiments to find if the model still manages to embody jet class information in such correlations. We chose to examine jet features such as jet mass and the number of constituents which, as shown in Figures~\ref{fig:topdata-jet-feats}-\ref{fig:jetclass-jet-feats}, can have moderate-to-strong discriminative power and give estimates for uncertainty.

For the \topdata dataset, the first principal component, $z_{pc, 0}$, shows a strong correlation with jet mass for both jet categories with correlation coefficients of 0.9 and 0.8 for background and signal jets, respectively. Similarly, in the EDL models applied to the \JetNet dataset, the correlation coefficient between $z_{pc, 0}$ and QCD jets is 0.9 (with a similar level of correlation found for top jets) while for boson jets the correlation is weaker at 0.7. The first principal component shows comparable correlation with jet mass in the \JetClass dataset with correlation coefficients being 0.9 for QCD jets and 0.6 for all other jet categories. Despite the larger size of the \JetClass dataset as compared with \topdata and \JetNet datasets, EDL models do not diminish in their ability to construct expressive distributions in the latent space.

\begin{figure}[htbp]
\centering
\subfloat[]{
\includegraphics[width=0.4\textwidth]{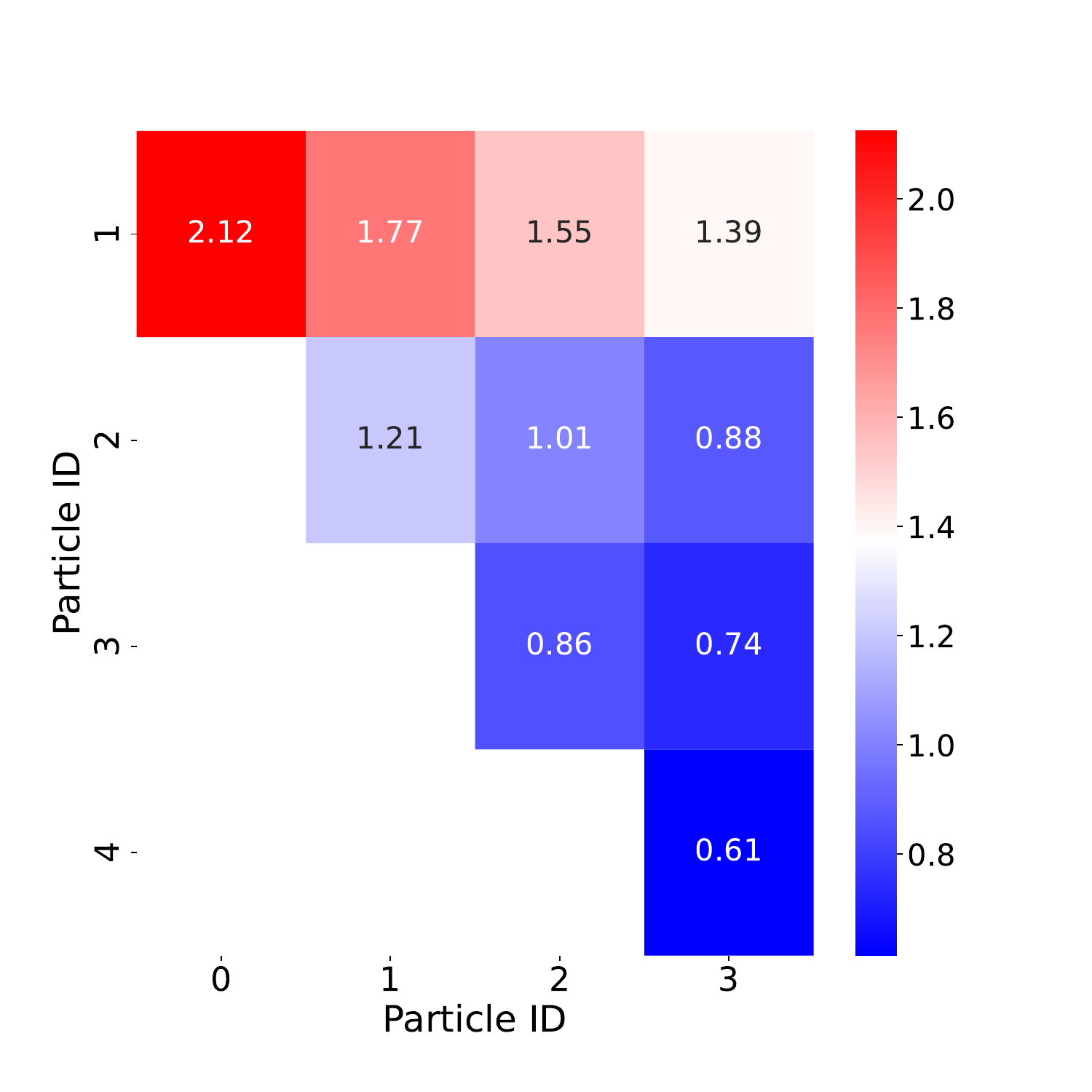}
\label{fig:dAUC-PFIN}            
}
\subfloat[]{
\includegraphics[width=0.4\textwidth]{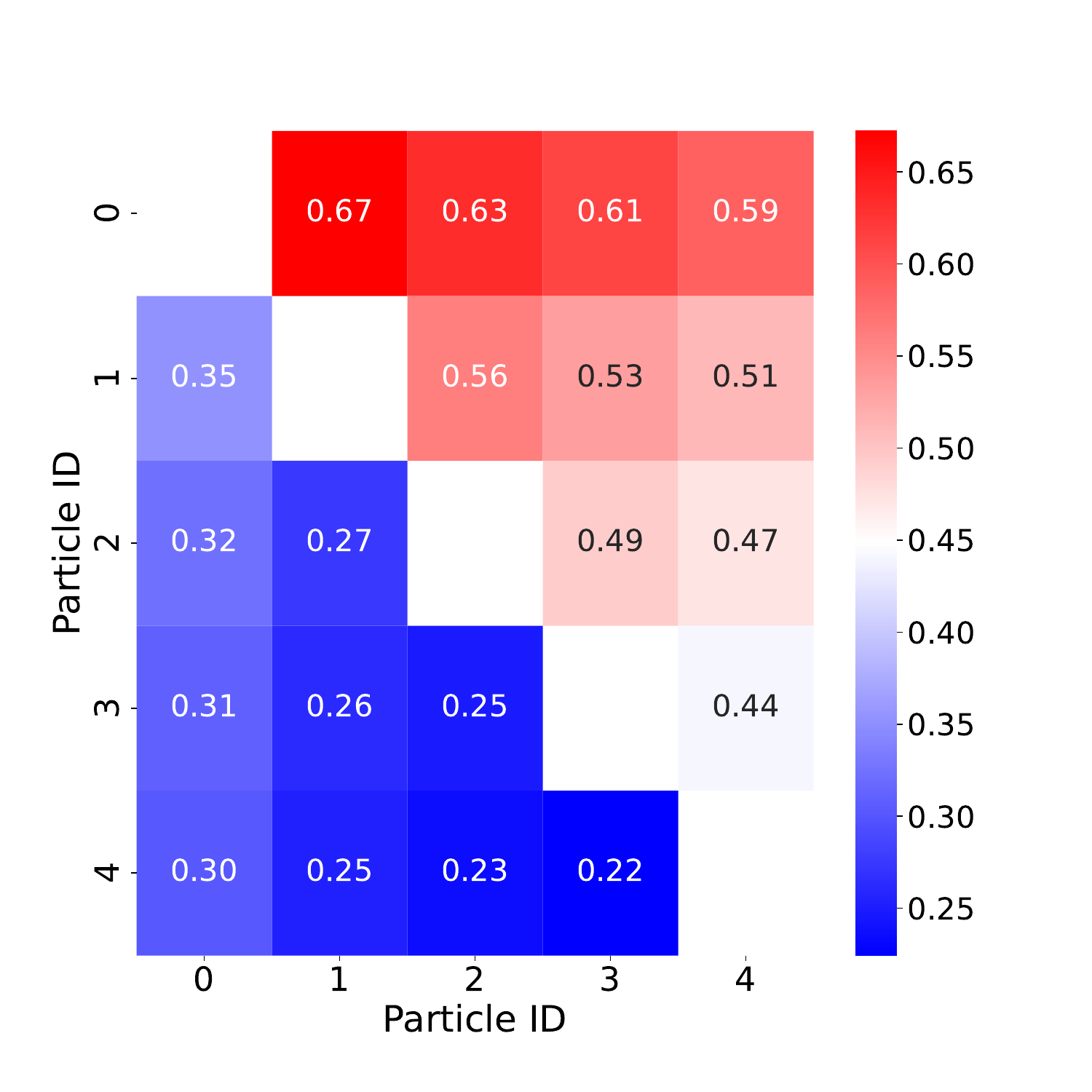}
\label{fig:MAD-PFIN}            
}

\subfloat[]{
\includegraphics[width=0.4\textwidth]{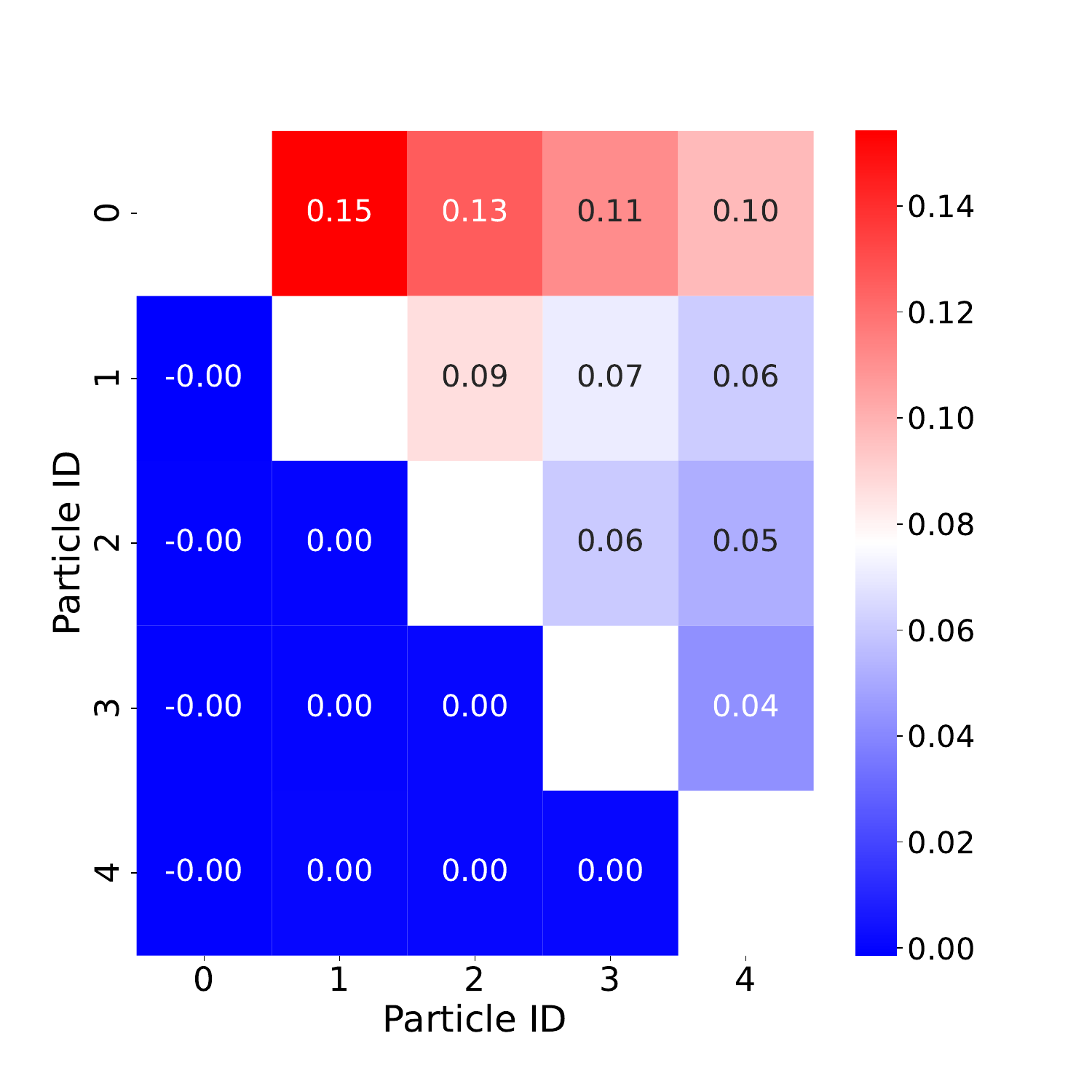}
\label{fig:dpred-PFIN}            
}
\subfloat[]{
\includegraphics[width=0.4\textwidth]{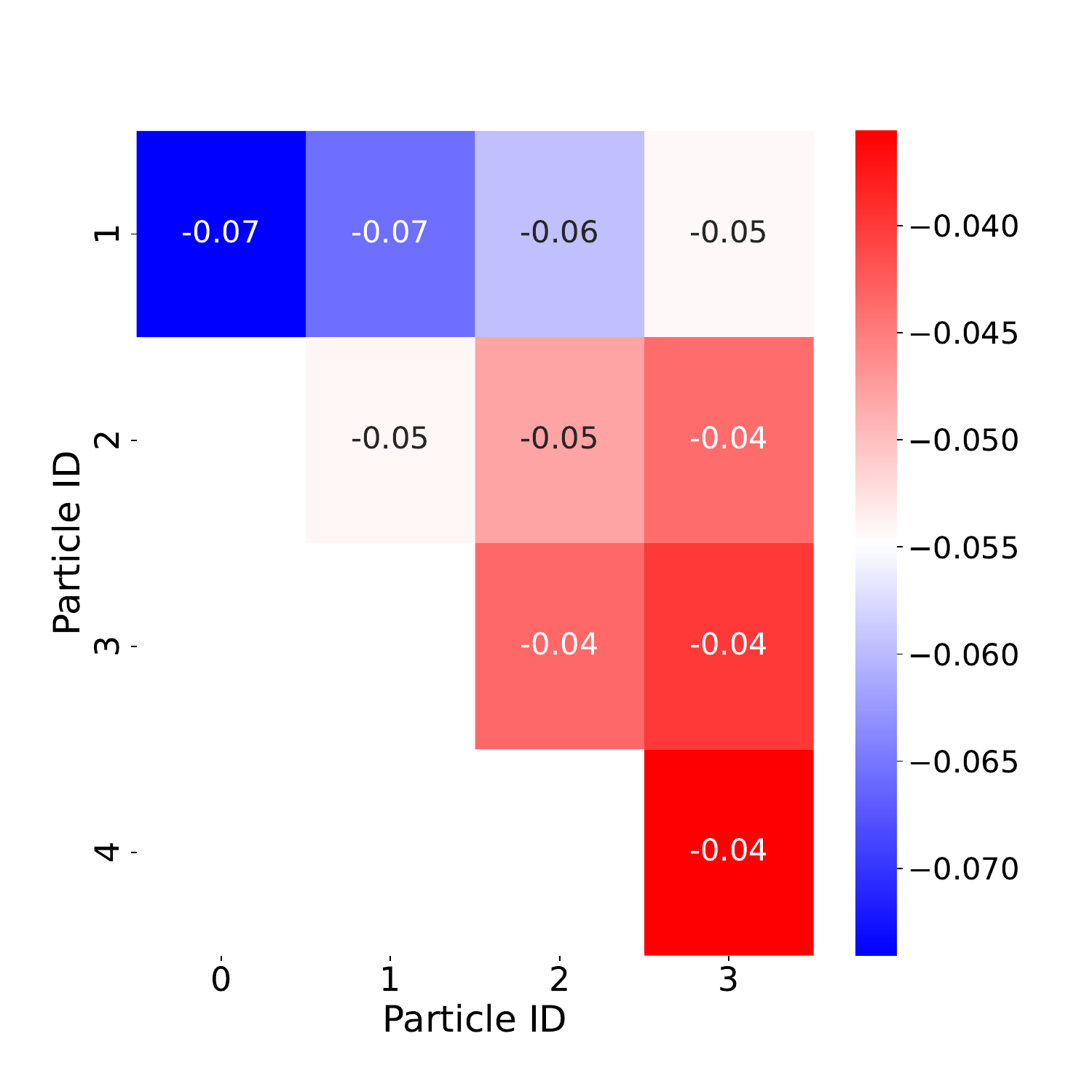}
\label{fig:dunc-PFIN}            
}
\caption{
For the \topdata baseline EDL $\lambda_t (1.0)$ model, \protect\subref{fig:dAUC-PFIN} \dAUC,
\protect\subref{fig:MAD-PFIN} normalized MAD relevance score,
\protect\subref{fig:dpred-PFIN} mean deviation in jet class prediction probabilities, and \protect\subref{fig:dunc-PFIN} mean deviation in uncertainties for pairwise masking of particle interaction features for the five most energetic constituents. Particle constituents are arranged in decreasing order of their energies. For \protect\subref{fig:MAD-PFIN} and \protect\subref{fig:dpred-PFIN}, the upper and lower diagonal entries represent the corresponding scores for the signal and background jet classes, respectively.
}
\label{fig:topdata-XAI}
\end{figure}

PFIN also allows us to explore the impact of pairwise particle interaction matrices on uncertainty quantification and jet classification. As explained in Ref.~\cite{Khot_2023}, we calculated the \dAUC and Mean Absolute Differential Relevance (MAD Relevance) score for each pair of particles by masking the corresponding input to the $\Phi_l$ network and calculating the deviation in model prediction with respect to the baseline model result. Additionally, we calculate the deviation in the model prediction probabilities and uncertainty using the \topdata dataset. These quantities are useful for evaluating the contribution of individual features by examining how the model performs when we mask a particle interaction. The results for the EDL baseline model with $\lambda_t (1.0)$ on the \topdata dataset are shown in Figure~\ref{fig:topdata-XAI}. 

The pairwise particle interactions play a particularly important role in identifying the signal jets. The mean deviations in the background jet class probabilities are barely impacted by masking interaction features. However, for the signal jets, this impact is found to be rather large, with the mean prediction probability reduced by almost 20\% when the interaction between the two most energetic jets is masked. In addition, the uncertainty slightly increases when masking interaction features, which is expected due to the removal of important information from the model.

\section{EDL for Anomaly Detection}
\label{sec:edl4ad}

There is a compelling and potentially powerful connection between UQ for ML models and detection of data anomalies having characteristics not seen in model training, such as under/overdensities or out-of-distribution (OOD) data. The foundational EDL paper~\cite{10.5555/3327144.3327239} demonstrated this capability using rotated handwritten numbers from the MNIST dataset~\cite{lecun-mnisthandwrittendigit-2010}. EDL has been applied to anomaly detection in numerous settings, for example the detection of maritime anomalies due to unusual vessel maneuvering~\cite{singh2022leveraginggraphdeeplearning}.

In this section, we examine how the EDL-based uncertainty behaves with OOD data. To create "anomalies" or OOD jets, we omit certain classes from the training dataset and analyze the uncertainties for both ID and OOD jets from the test dataset. We examine three different anomaly detection models for the \JetNet dataset shown in Table~\ref{tab:JetNetB}: \skiptop, \skipwz, \skiptwz. The models are evaluated based on the same metrics as the baseline models. 

\begin{table}[h]
    \centering
    \begin{tabular}{|c|c|c|c|c|}
    \hline
    \multicolumn{5}{|c|}{\jetnet Training Configurations} \\
    \hline
    Names & baseline & \skiptop & \skipwz & \skiptwz \\
    \hline
    In-distribution Jets & $g,q,t,W,Z$ & $g,q,W,Z$ & $g,q,t$ & $g,q$ \\
    \hline
    Out-of-distribution Jets &  & $t$ & $W,Z$ & $t,W,Z$ \\
    \hline
    \end{tabular}
    \caption{Experimental configurations within the \JetNet dataset highlighting different training scenarios by excluding specific particle classes.}
    \label{tab:JetNetB}
\end{table}

\begin{figure}[htbp]
\centering
\subfloat[]{
\includegraphics[width=0.33\textwidth]{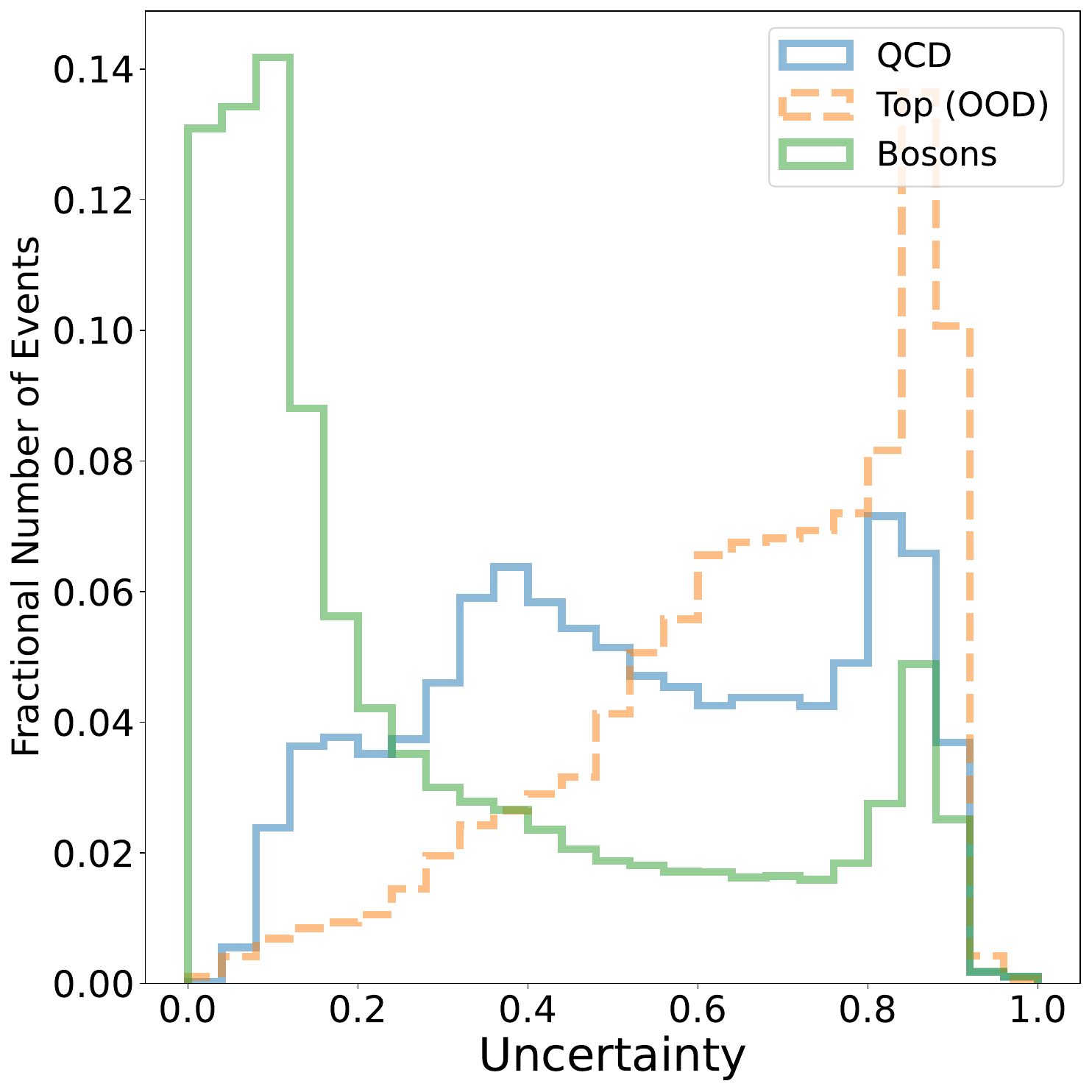}
\label{fig:skiptop-ad}            
}
\subfloat[]{
\includegraphics[width=0.33\textwidth]{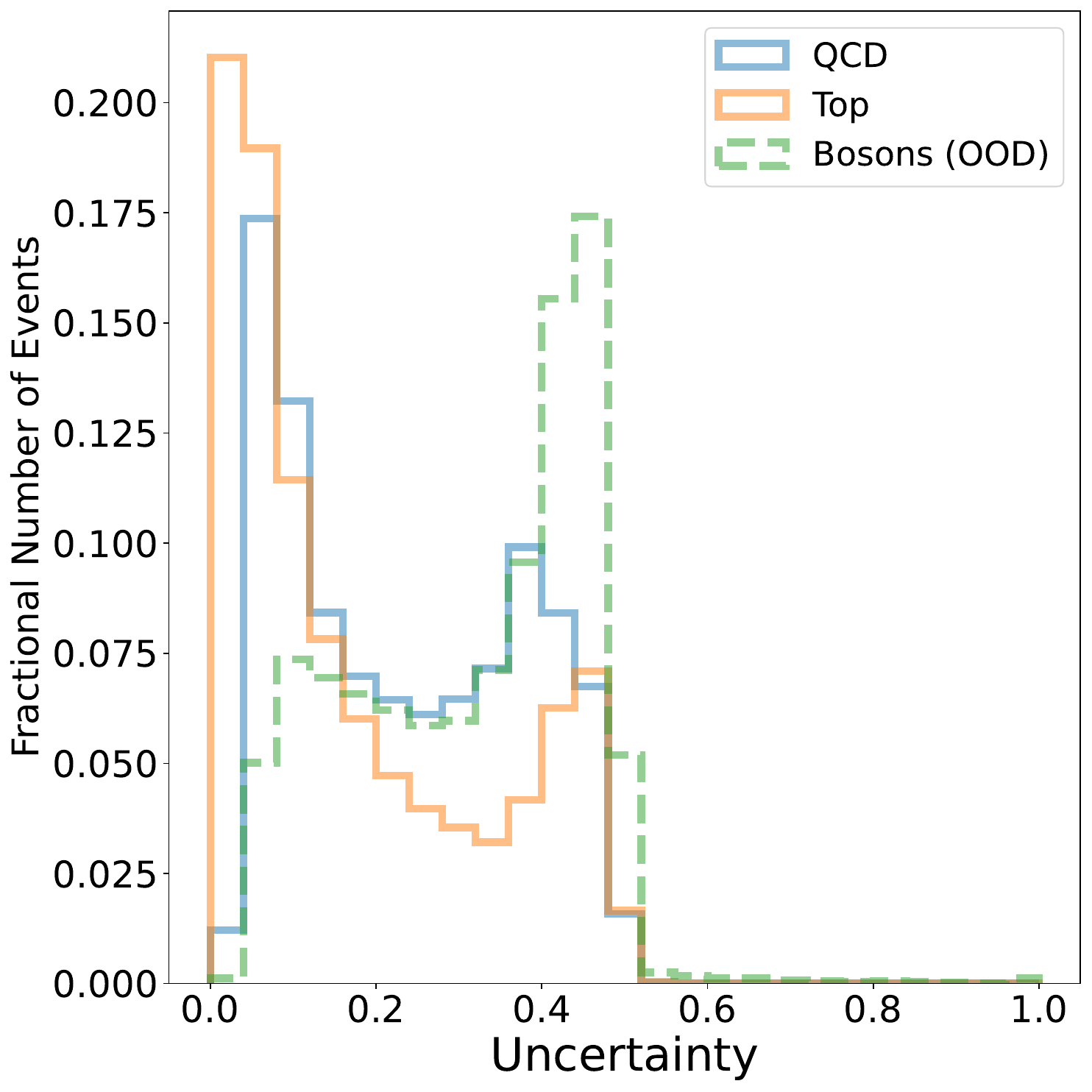}
\label{fig:skipwz-ad}            
}
\subfloat[]{
\includegraphics[width=0.33\textwidth]{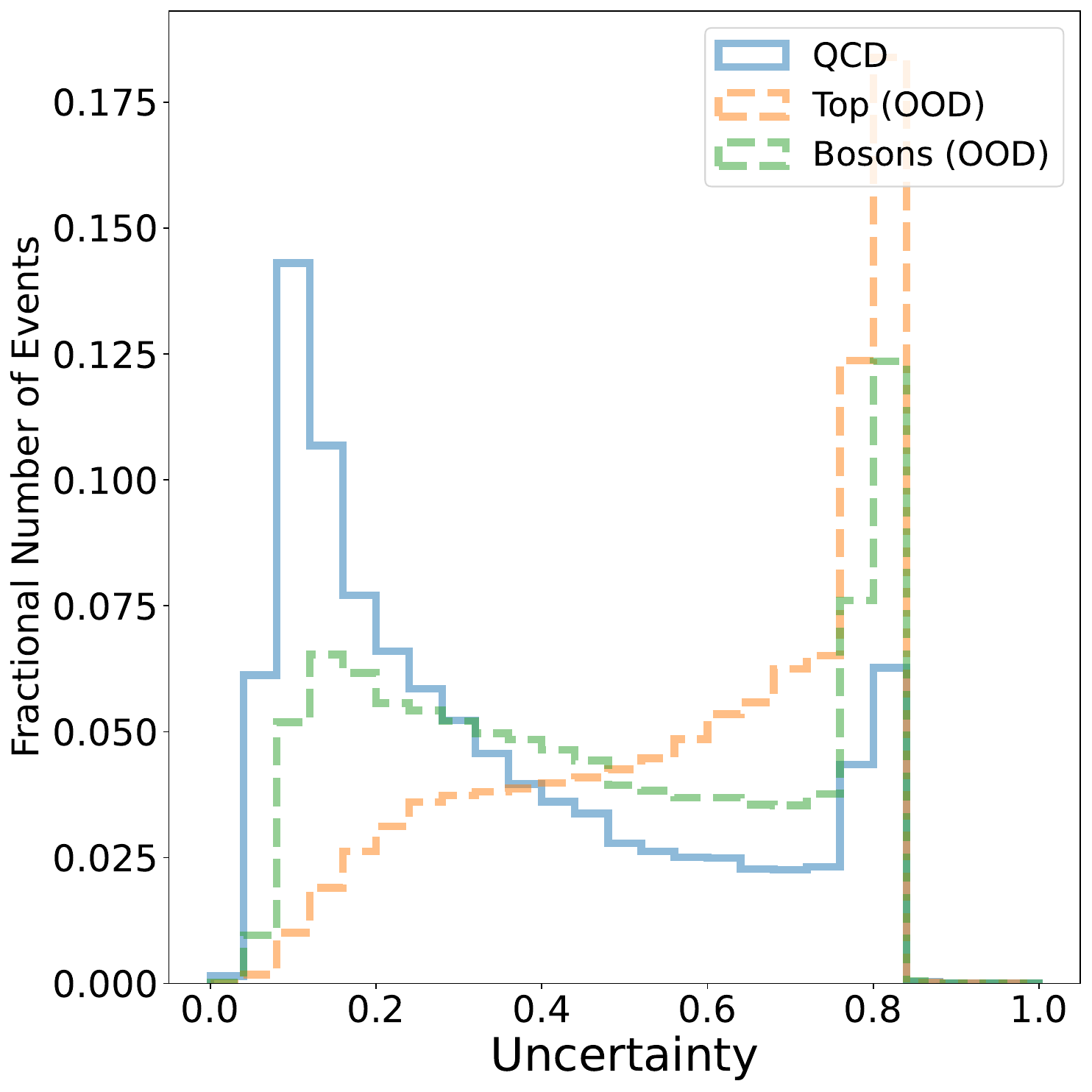}
\label{fig:skiptwz-ad}            
}
\caption{Uncertainty distributions, with jets separated by initiating particle jet type, for \protect\subref{fig:skiptop-ad} \jetnetskiptop EDL $\lambda_t (0.5) $, \protect\subref{fig:skipwz-ad} \jetnetskipwz EDL $\lambda_t (0.1) $, and \protect\subref{fig:skiptwz-ad} \jetnetskiptwz EDL $\lambda_t (0.6) $ models.}
\label{fig:unc-ad}
\end{figure}

In \jetnetskiptop EDL networks, we skip jets from top quarks during training and analyze how well the uncertainties identify these "anomalies" in the test set. The results are summarized in Table~\ref{tab:skipx-jetnet}. As shown in the \skiptop column of Table~\ref{tab:skipx-jetnet}, the best performing EDL model is EDL $\lambda_t (0.5)$ with an AUROC peaking at $0.754$. Increasing $\zeta$ further decreases both the ID accuracy and AUROC. As shown in Figure~\ref{fig:skiptop-ad}, \jetnetskiptop EDL $\lambda_t (0.5)$ assigns high uncertainties to most OOD jets, which serves as an indicator of the predictive limitations of the model on OOD data. There are many high-uncertainty QCD jets from misclassifications, making it difficult to differentiate between the ID QCD jets and OOD top jets. This points to a fundamental challenge in using EDL for OOD jet detection. The EDL uncertainty, in its simplest form, fails to distinguish between the jets that are hard to tell-apart and the jets that are unknown from the training data.

\begin{table}[!ht]
 \centering
 \resizebox{0.99\textwidth}{!}{
 \begin{tabular}{|l||c|c|c||c|c|c||c|c|c|}
    \hline
   & \multicolumn{3}{c||}{\skiptop} & \multicolumn{3}{c||}{\skipwz} & \multicolumn{3}{c|}{\skiptwz} \\ \hline \hline
    \textbf{Model} & Acc & AUC & STD & Acc & AUC & STD & Acc & AUC & STD  \\ \hline
    EDL $\lambda_t (0)$ & \textbf{0.815} & 0.380 & 0.511 & \textbf{0.818} & \textbf{0.824} & \textbf{0.753} & \textbf{0.837} & 0.386 & 0.614 \\ \hline
    EDL $\lambda_t (0.1)$ & 0.814 & 0.701 & 0.713 & 0.816 & 0.701 & 0.697 & 0.836 & 0.713 & \textbf{0.709} \\ \hline
    EDL $\lambda_t (0.5)$ & \textbf{0.815} & 0.754 & 0.754 & 0.816 & 0.697 & 0.696 & 0.833 & 0.682 & 0.682 \\ \hline
    EDL $\lambda_t (0.6)$ & 0.813 & \textbf{0.756} & \textbf{0.757} & 0.814 & 0.690 & 0.690 & \textbf{0.837} & \textbf{0.714} & 0.687 \\ \hline
    EDL $\lambda_t (1.0)$ & 0.811 & 0.743 & 0.744 & 0.815 & 0.666 & 0.666 & 0.836 & 0.690 & 0.690 \\ \hline
    EDL-CT $\lambda_t^{\texttt{CT}} (0.1)$ & 0.814 & 0.724 & 0.729 & 0.817 & 0.676 & 0.676 & 0.836 & 0.635 & 0.681 \\ \hline
    EDL-CT $\lambda_t^{\texttt{CT}} (0.5)$ & 0.808 & 0.746 & 0.745 & 0.816 & 0.681 & 0.680 & 0.835 & 0.694 & 0.694 \\ \hline
    EDL-CT $\lambda_t^{\texttt{CT}} (0.7)$ & 0.808 & 0.743 & 0.743 & 0.816 & 0.692 & 0.691 & 0.835 & 0.697 & 0.697 \\ \hline \hline
    Ensemble & 0.822 & 0.766 & - & 0.824 & 0.741 & - & 0.824 & 0.741 & -  \\ \hline
    MC Dropout & 0.810 & 0.656 & - & 0.817 & 0.717 & - & 0.833 & 0.693 & -  \\ 
    \hline
  \end{tabular}
  }
  \caption{ID Accuracy (Acc),  AUROC (AUC), and AUROC-STD (STD) of the EDL and Ensemble methods on variants of \JetNet: \skiptop, \skipwz, \skiptwz. The Ensemble model has 970k parameters, while all other models have 97k parameters. For each of the \JetNet variants, the entries marked in \textbf{bold} represent the EDL model with the highest central value for the corresponding metric. These measurements have uncertainties of $\mathcal{O}(0.001)$.}
  \label{tab:skipx-jetnet}
\end{table}

We observe similar situations when trying to detect OOD jets using EDL for the \skipwz and \skiptwz datasets, as shown in Figures~\ref{fig:skipwz-ad} and \ref{fig:skiptwz-ad} respectively. The \skipwz dataset considers the $W$ and $Z$ boson categories as anomalies while in the \skiptwz dataset, all jets coming from the heavy bosons and the top quark are considered OOD. In both cases, the uncertainty distribution of the QCD jets has a peak near the tail of the respective distribution, close to where the uncertainty assigned to most OOD jets is concentrated.
 
We note that choosing the best EDL model for the \skipwz dataset was trickier than the other cases. As shown in the \skipwz column of Table~\ref{tab:skipx-jetnet}, as $\zeta$ increases, both accuracy and AUROC decrease, with EDL $\lambda_t (0)$ having the highest AUROC. The EDL $\lambda_t (0)$ model performs better than even Ensemble and MC Dropout. This is much different from the previous \JetNet models where there was increased AUROC for non-zero $\zeta$. However, the physical range of uncertainties associated with the $\zeta = 0$ model is very narrow and close to zero, so uncertainty attributions are rather sporadic and noisy. Hence, uncertainty estimates are better characterized in the model with $\lambda = 0.1$. Both categories of ID jets show a strong peak near $u=0$, and a second peak close to the tail of the distribution. The peak near the larger uncertainties is much pronounced for QCD jets, which is a somewhat expected behavior based on our observations of the EDL model performance in the baseline case in Section~\ref{sec:jetnet}.

Since \skiptwz models skip both top quarks and bosons during training, the model is now a binary classifier for quarks and gluons. In terms of ID accuracy, the \skiptwz EDL models perform similarly to the binary top tagging models described in Section~\ref{sec:topdata}, with the accuracy barely decreasing as $\zeta$ increases. However, the AUROC score improves with non-zero $\zeta$, and peaks at $\zeta = 0.6$. In Figure~\ref{fig:skiptwz-ad}, there is a bimodal distribution for ID jets associated with low-uncertainty correctly-identified jets and high-uncertainty misclassified jets. The OOD jets have high uncertainties, but many of the ID jets are also assigned large uncertainties due to misclassification, which often occurs for hard-to-tell-apart jets.

Overall, it is difficult to differentiate between hard-to-tell-apart and OOD jets. In a way, this behavior is expected from EDL models. The EDL-assigned uncertainty to each jet instance reflects the level of confidence in the classification scores the model predicts. As a singular metric, we expect this quantity to be large whenever the model comes across a jet that is unlike anything it has seen before. We also expect this quantity to be large when this model encounters a jet with characteristics making it difficult to confidently place it in a single category. Misclassifications are likely to happen in such cases and though it is promising that OOD jets are associated with high uncertainties.

\section{Outlook on Model Selection and Limitations of the EDL Method}
\label{sec:outlook}

As we have discussed in Section~\ref{sec:review-edl} and demonstrated through our study of jet classification, the method of evidential deep learning provides a valuable method to faithfully assign epistemic uncertainties to deep classifiers. The model uncertainty defined in Eqn.~\ref{eqn:edl-3} provides a meaningful estimate of the level of confidence in model predictions. On the other hand, the uncertainties  associated with each class prediction is given by Eqn.~\ref{eqn:d-std}. Though both terms are ubiquitously termed as uncertainty in standard ML literature, their usage in the context of a physics analysis requires a proper examination of what these quantities represent. In the context of a jet classifier, the former would represent the quality of the classification, so a proper use-case of this uncertainty can be, for instance, using this quantity as a threshold for jet selection. The latter would be a more appropriate quantity to be assigned to physical distributions associated with jets, and can be incorporated as a systematic uncertainty in the context of likelihood optimization for a search or a precision measurement analysis.

As our analyses have demonstrated, the performance of the EDL mechanism is subject to (1) the choice of the hyperparameter $\zeta$ and (2) the choice of training methodology as illustrated in the differences between the standard EDL and EDL-CT methods. It is an artifact of the nature of the EDL loss function. Hence, it is important to define a systematic procedure to make the right choice of $\zeta$. The observations we made in Section~\ref{sec:results} suggest that model accuracy is typically the largest with $\zeta = 0$ with a small AUROC. A small increase in $\zeta$ yields in a significant increase in the AUROC with a marginal degradation in accuracy. Even with the EDL-CT models, smaller values of $\zeta$ tend to give better performance. These findings are also in line with the observations made in Ref.~\cite{shen2024uncertainty}. As a result, for the choice of right $\zeta$, model accuracy should be benchmarked against the accuracy obtained with $\zeta = 0$ while the AUROC should show a significant improvement over the $\zeta = 0$ case. In most use cases, a small, non-zero value of $\zeta$ for either EDL or the EDL-CT method would be most appropriate choice for the model.

Finally, we point out a potential limitation of the EDL method. The uncertainty that EDL assigns is an unbiased estimate of model uncertainty for a given choice of model parameters. However, as argued by some authors (e.g. in Ref.~\cite{shen2024uncertainty}), this is a conditional but incomplete estimate of model uncertainty as it does not take into account uncertainties arising from variations in model parameters. In that sense, the EDL uncertainties might be complementary to the uncertainties obtained from the Ensemble method since the latter attempts to capture the systematic variations in the model's predictions arising from variations in the model parameters. In the context of experimental analyses, a conservative account of both types of epistemic uncertainty could be made by incorporating both Ensemble-based variances with EDL-estimated uncertainties as independent and uncorrelated systematic uncertainties. However, the applicability and effectiveness of such a strategy might be dependent on the nature of the analysis itself. A more complete account of evidential uncertainties might require employing an ensemble of EDL models. That study is beyond the scope of this paper and leave it to future work.
\section{Conclusions and Outlook}
\label{sec:conclusion}

This paper presents a comprehensive study of evidential deep learning (EDL) in the context of uncertainty quantification (UQ) in jet tagging datasets. Our work has unveiled a number of important aspects regarding how the uncertainty and performance varies with the corresponding datasets. We have observed the EDL-based uncertainty and its comparable performance to Bayesian methods. The convergence and performance of the EDL method strongly depends on the choice of the annealing coefficient, $\zeta$. Larger annealing coefficients result in lower accuracy, higher AUROC, wider ranges of uncertainties, and a larger number of high-uncertainty jets. Hence, model selection of a robust EDL-based classifier relies on the proper choice of $\zeta$. Our empirical insights, as summarized in Section~\ref{sec:outlook}, suggest that in most use cases a model with a small nonzero $\zeta$ value would give a desirable AUROC while maintaining an accuracy close to the benchmark of $\zeta = 0$.

As a method of UQ, EDL provides unbiased estimates of uncertainties on class-wise predictions expressed as standard deviations of a parametric Dirichlet distribution. Given the physical range of uncertainties associated with EDL-based UQ varies with each choice of the annealing hyperparameter, the uncertainties predicted by this model must be regarded as \textit{post-hoc} uncertainties associated with a given instance of the model. In other words, EDL uncertainties express the confidence a model projects for a given choice of model parameters. These predictions should not be regarded as representative uncertainties distributed over a class of potential parameter and hyperparameter choices.

We also observe how the EDL-based uncertainty maps onto the latent space of the PFIN model. We demonstrate that high-uncertainty misclassified jets populate (based on the first principle component) at the intersection of jet distributions in latent space embeddings in all datasets. This bridges an important gap between our previous studies on model interpretability and the current work on UQ. Is is evident from our studies of the latent space embeddings that a well-tuned EDL model can show strong uncertainty associations for misclassified and hard-to-tell-apart jets. Finally, although the method of EDL shows promise leveraging UQ for the detection of OOD jets, anomaly detection (AD) using EDL can be limited in telling apart the OOD jets from the hard-to-tell-apart ID jets. Any attempt to reliably detect OOD jets can definitely benefit from additional degrees of freedom to identify anomalous jets from ID jets.

This work establishes a methodology to evaluate and optimize application of EDL for UQ and AD, using jet classification at the LHC as an important case study. While the results presented in this work rely exclusively on the PFIN model, the EDL method by itself remains model-agnostic. Post-hoc EDL uncertainties are reliable and unbiased estimates of model uncertainty, but it requires some effort to obtain performance optimization though hyperparameter tuning and training strategies. This paper also lays out the primary optimization criteria for selecting the best model for a given use case. As EDL uncertainties can be obtained in a single pass on the data during inference stage with minimal additions and modifications to a neural network model for classification or regression, it opens up potential applications for uncertainty-aware algorithms and hardware co-design for edge and low-latency applications, such as fast data reduction, detector triggering, and AD. In regard to AD, there is potential to leverage EDL to improve the performance and model independence of traditional approaches such as autoencoders, which we leave to future work.

\FloatBarrier


\section*{Authorship contribution statement}
\textbf{Ayush Khot}: Methodology, Analysis, Software, Visualization, Validation, Writing – original draft \& editing.
\textbf{Xiwei Wang}: Methodology, Analysis, Software, Visualization, Validation.
\textbf{Avik Roy}: Conceptualization, Methodology, Analysis, Software, Visualization, Validation, Writing – original draft, review \& editing.
\textbf{Volodymyr Kindratenko}: Conceptualization, Resources, Supervision, Writing – review \& editing.
\textbf{Mark S. Neubauer}: Conceptualization, Resources, Supervision, Writing – original draft, review \& editing.

\section*{Declaration of competing interest}
The authors declare that they have no known competing financial interests or personal relationships that could have appeared to influence the work reported in this paper.

\section*{Data availability}

The data that support the findings of this study are openly available at the following URL/DOI: \url{https://github.com/FAIR4HEP/PFIN4UQAD}

\section*{Acknowledgements}
The authors would like to thank the Center for Artificial Intelligence Innovation at the NCSA for support through our affiliation. This research is part of the Delta research computing project, which is supported by the National Science Foundation (award OCI 2005572), and the State of Illinois. Delta is a joint effort of the University of Illinois at Urbana-Champaign and its National Center for Supercomputing Applications. This work utilizes resources supported by the National Science Foundation’s Major Research Instrumentation program, grant 1725729, as well as the University of Illinois at Urbana-Champaign. This work was supported by the FAIR Data program of the U.S. Department of Energy, Office of Science, Advanced Scientific Computing Research, under contract number DE-SC0021258, the U.S. Department of Energy, Office of Science, High Energy Physics, under contract number DE-SC0023365, and the National Science Foundation Cooperative Agreement PHY-2117997.

\pagebreak

\newcommand{\newblock}{}
\bibliography{main2}
\bibliographystyle{JHEP}

\end{document}